\documentclass[preprint,preprintnumbers,amsmath,amssymb]{revtex4}

\usepackage{epsfig, color, ulem}
\usepackage{amsmath}

\begin{document}

\title{A modified Marchenko method to retrieve the wave field inside a layered metamaterial from reflection measurements at the surface}
\author{Kees Wapenaar}
\affiliation{Department of Geoscience and Engineering, Delft University of Technology, 2600 GA Delft, The Netherlands}
\date{\today}

\begin{abstract}
\noindent
With the Marchenko method it is possible to retrieve the wave field inside a medium from its reflection response at the surface.
To date, this method has predominantly been applied to naturally occurring materials. We extend the Marchenko method for applications in layered metamaterials 
with, in the low-frequency limit, effective negative constitutive parameters.  We illustrate the method with a numerical example, which confirms that the
method properly accounts for multiple scattering.
 The proposed method has potential applications for example in non-destructive testing of layered materials.

\end{abstract}

\maketitle

\section{Introduction}

Building on classical inverse scattering theory \cite{Marchenko55DAN, Lamb80Book, Chadan89Book, Budreck90IP},  recent research 
has opened new ways of retrieving the wave field inside a medium from its reflection response at the surface  
\cite{Rose2001PRA, Rose2002IP, Broggini2012EJP, Wapenaar2012GJI,  Neut2015GJI,  Elison2018GJI, Wapenaar2018SR, Dukalski2018GJI, Brackenhoff2019SE}
and using this for imaging \cite{Wapenaar2014GEO, Slob2014GEO, Broggini2014GEO, Meles2015GEO, Ravasi2016GJI, Neut2017GEO, Mildner2017GEO, Ravasi2017GEO,  Staring2018GEO}.
These methods, named after Marchenko \cite{Marchenko55DAN}, have in common that they account for multiple scattering inside the medium
and yet require only a single-sided reflection response as input, together with a background model of the medium. 

The Marchenko method has almost exclusively been applied to naturally occurring materials (with the exception of an application to non-reciprocal materials \cite{Wapenaar2019JASA}), 
but it has not yet been applied to metamaterials with, in the low-frequency limit, effective negative constitutive parameters.
A classical reference on wave propagation in materials with negative permittivity and permeability is the paper by Veselago \cite{Veselago68SPU}, in which it is shown that such materials exhibit negative refraction.
Since the discovery by Pendry \cite{Pendry2000PRL} that negative refraction makes a perfect lens 
 there has been a significant interest in electromagnetic wave propagation in metamaterials
\cite{Ziolkowski2001PRE, Marques2002PRB, Ziolkowski2003PRE, Ziolkowski2003OE, Engheta2006Book, Smith2006JOSA, Yu2011Science, Willis2011RS, Nemirovsky2012OE}.
Almost simultaneously, after the first fabrication of an elastic metamaterial with effective negative elastic parameters \cite{Liu2000Science},
much research has been directed to wave propagation in elastic  metamaterials 
\cite{Liu2005PRB, Guenneau2007NJP, Norris2009JASA, Willis2012CRM, Zhu2012PRB, Liu2012WM, Norris2012RS, Liang2012PRL, Haberman2016PhysicsToday, Nassar2017RS}.
 
Here we modify the Marchenko method for metamaterials. We start by formulating a wave equation that holds for elastodynamic and electromagnetic waves in natural materials and metamaterials. 
{We show that metamaterials, with negative phase slowness and positive group slowness, are by definition dispersive.
This implies that a modification of the standard Marchenko method is needed.}
Next,  we derive {wave field} representations for a layered medium, consisting of a mix of natural materials and metamaterials. Using these representations, we
derive the  Marchenko method {for such a medium. This method uses 
new window functions which better acknowledge the dispersive behaviour of waves propagating through metamaterial. We conclude by illustrating the modified Marchenko method} 
 with a numerical example.

\section{Wave equation for natural materials and metamaterials}\label{sec2}

\subsection{Basic equations}

Throughout this paper we consider scalar wave propagation in the 2D plane. This allows capturing
different wave phenomena by a unified wave equation. 
We define the Cartesian coordinate vector in the 2D plane as ${\bf x}=(x_1,x_3)$, where positive $x_3$ denotes depth in a horizontally layered medium. 
Quantities that are a function of space and time are denoted as $u({\bf x},t)$, where $t$ stands for time.
We define the temporal Fourier transform of $u({\bf x},t)$ as 
\begin{eqnarray}\label{eqA11}
&&u({\bf x},\omega)=\int_{-\infty}^\infty u({\bf x},t)\exp(i\omega t){\rm d}t,
\end{eqnarray}
where $\omega$ is the angular frequency and $i$ the imaginary unit. For convenience, quantities in the time and frequency domain are denoted by the same symbol (here $u$).
The inverse Fourier transform is defined as
\begin{eqnarray}\label{eqA11inva}
&&u({\bf x},t)=\frac{1}{2\pi}\int_{-\infty}^\infty u({\bf x},\omega)\exp(-i\omega t){\rm d}\omega.
\end{eqnarray}
Throughout this paper quantities in the time domain are real-valued, hence, equation (\ref{eqA11}) implies $u({\bf x},-\omega)=u^*({\bf x},\omega)$, where the asterisk denotes complex conjugation.
Using this property, the inverse Fourier transform can be rewritten as
\begin{eqnarray}\label{eqA11inv}
&&u({\bf x},t)=\frac{1}{\pi}\Re\int_{0}^\infty u({\bf x},\omega)\exp(-i\omega t){\rm d}\omega,
\end{eqnarray}
where $\Re$ denotes that the real part is taken.
Since the integral is taken over positive frequencies only, it is sufficient to 
restrict our derivations in the frequency domain to positive frequencies. 
This avoids complications related to the sign of the frequency.

In the space-frequency domain, we consider the following system of  equations in the low-frequency limit 
for 2D wave propagation in an inhomogeneous, transverse isotropic natural material or metamaterial
\begin{eqnarray}
-i\omega\alpha P+\partial_1Q_1+\partial_3Q_3&=&B,\label{eqA1}\\
-i\omega\beta_1 Q_1+\partial_1 P&=&C_1,\label{eqA2}\\
-i\omega\beta_3 Q_3+\partial_3 P&=&C_3.\label{eqA3}
\end{eqnarray}
These equations hold for acoustic (AC), horizontally polarised shear (SH), transverse-electric (TE) and transverse-magnetic (TM) wave fields.
Operator $\partial_i$ stands for the partial differential operator $\partial/\partial x_i$.
The wave fields [$P({\bf x},\omega)$ and $Q_i({\bf x},\omega)$] and sources [$B({\bf x},\omega)$ and $C_i({\bf x},\omega)$] are space- and frequency-dependent macroscopic quantities.
These are often denoted as $\langle P\rangle$, etc. \cite{Willis2011RS}, but  for notational convenience we will not use the brackets.
The medium parameters [$\alpha(x_3,\omega)$ and $\beta_i(x_3,\omega)$] are effective parameters (which, in a layered medium, are varying in the $x_3$-direction only).
At layer interfaces, where the medium parameters are discontinuous, the boundary conditions state that the wave field quantities $P$ and $Q_3$ are continuous. 
\begin{center}
{{\noindent \it Table 1: Quantities in equations (\ref{eqA1}) to (\ref{eqA3}). }
\begin{tabular}{||l|c|c|c|c|c|c|c|c|c|c|c|c|c|c||}
\hline\hline
& $P$ & $Q_1$ & $Q_3$ & $\alpha$ &$\beta_1$  & $\beta_3$ & $B$ & $C_1$ & $C_3$ \\
\hline
AC  & $p$ &$v_1$ & $v_3$  & $\kappa$ &${\rho}_{11}$  &  ${\rho}_{33} $   &$q$ & $F_1$ & $F_3$  \\
\hline
SH  & $v_2$ &$-\tau_{21}$ & $-\tau_{23}$  & ${\rho}_{22}$ &$4s_{1221}$ &  $4s_{3223}$&  $F_2$ & $2h_{21}$ & $2h_{23}$  \\
\hline
TE & $E_2$ &$H_3$ & $-H_1$ & ${\varepsilon}_{22}$ &$\mu_{33}$  & $\mu_{11}$ & $-J_2^{\rm e}$ & $-J_3^{\rm m}$& $J_1^{\rm m}$  \\
\hline
TM & $H_2$ &$-E_3$ & $E_1$ & $\mu_{22}$ &${\varepsilon}_{33}$  & ${\varepsilon}_{11}$  & $-J_2^{\rm m}$ & $J_3^{\rm e}$& $-J_1^{\rm e}$ \\
\hline
\hline
\end{tabular}
}
\end{center}
\mbox{}\\
The wave fields, sources and medium parameters are specified for the different wave phenomena in Table 1.
For AC and SH waves, $p$ is the acoustic pressure, $\tau_{ij}$ the stress, $v_i$ the particle velocity, $\kappa$ the compressibility, 
$\rho_{ij}$ the mass density, $s_{ijkl}$ the compliance, $q$ the volume injection-rate density, $F_i$ the external force density and $h_{ij}$ the external deformation-rate density.
For TE and TM waves, $E_i$ is the electric field strength, $H_i$ the magnetic field strength, $\varepsilon_{ij}$ the permittivity, $\mu_{ij}$ the permeability, 
$J_i^{\rm e}$ the external electric current density and $J_i^{\rm m}$ the external magnetic current density.

For natural materials the real parts of the medium parameters $\alpha$ and $\beta_i$ are positive (and the imaginary parts are positive or zero). 
Such a medium will be called a double-positive (DPS) medium \cite{Engheta2006Book}.
For metamaterials the real part of one or more of the medium parameters is negative.
When both $\alpha$ and $\beta_i$ have negative real parts, we speak of a double-negative (DNG) medium \cite{Engheta2006Book}.
{The phase slowness of a DNG medium is negative \cite{Veselago68SPU}, see also section \ref{sec1d}.}
To obey causality, {the group slowness should be positive. These opposite slownesses imply that}  the parameters of a DNG medium are frequency-dependent and complex-valued (with positive imaginary parts) \cite{Ziolkowski2003PRE}. {The inherent dispersive character of DNG media implies that the Marchenko method needs to be modified for such media, see section \ref{sec4}}.

In the following we separate the space- and frequency-dependency of the medium parameters, for DPS as well as DNG media, according to 
\begin{eqnarray}
\alpha(x_3,\omega)&=&\alpha_0(x_3) h_\alpha(\omega),\label{eqalpha}\\
\beta_i(x_3,\omega)&=&\beta_{i,0}(x_3) h_\beta(\omega),\label{eqbeta}
\end{eqnarray}
with positive real-valued $\alpha_0(x_3)$ and $\beta_{i,0}(x_3)$.
For DNG media, an often used model for the frequency-dependent functions is the Drude model \cite{Ziolkowski2001PRE}, where
\begin{eqnarray}
h_\alpha(\omega)&=&1-\frac{\omega_\alpha^2}{\omega(\omega+i\Gamma_\alpha)},\label{eqdrudea}\\
h_\beta(\omega)&=&1-\frac{\omega_\beta^2}{\omega(\omega+i\Gamma_\beta)},\label{eqdrudeb}
\end{eqnarray}
with small positive real-valued $\Gamma_\alpha$ and $\Gamma_\beta$. 
Note that $\Re(h_\alpha)<0$
 for $\omega^2<\omega_\alpha^2-\Gamma_\alpha^2$ and $\Im(h_\alpha)>0$ (where $\Im$ denotes the imaginary part) for all positive $\omega$. Similar properties hold for $h_\beta$.
 Hence, in the low-frequency limit these parameters obey the mentioned conditions for a DNG medium. 
 {In section \ref{sec1e} we confirm that the group slowness for this type of DNG medium is positive. }

In the following we assume that the medium parameters (for DPS and DNG media) are defined by the more general relations  (\ref{eqalpha}) and (\ref{eqbeta}). 
Whenever we use the Drude model for DNG media (equations (\ref{eqdrudea}) and (\ref{eqdrudeb})) we mention this explicitly.

\subsection{Matrix-vector wave equation}

We reorganise the basic equations (\ref{eqA1}) to (\ref{eqA3}) into a matrix-vector wave equation. This wave equation is a suited starting point for the derivation of representations
for the Marchenko method in section \ref{sec3}. 

\begin{figure}[t]
\centerline{\hspace{4.5cm}\epsfxsize=9cm \epsfbox{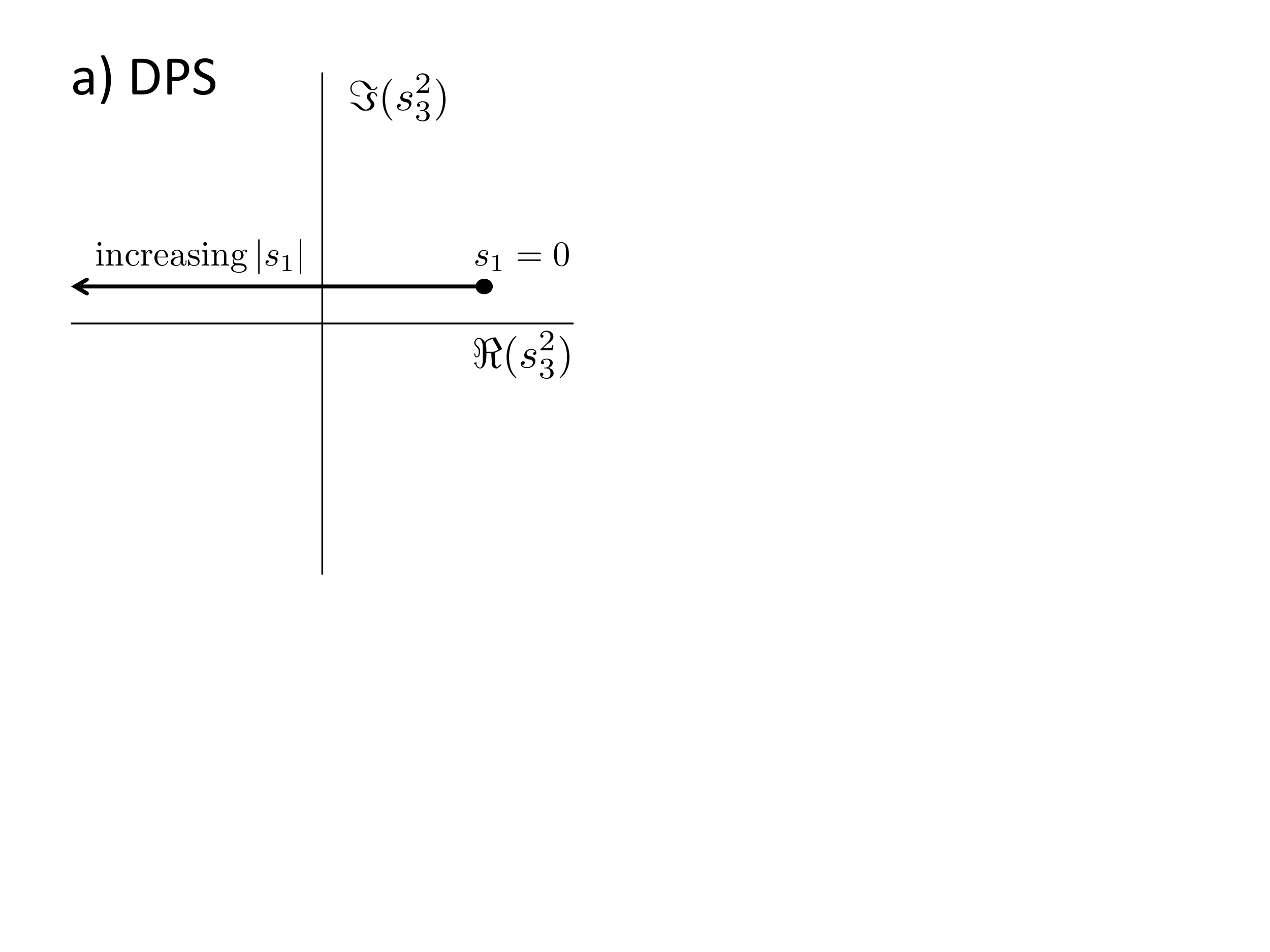}\hspace{-4.5cm}\epsfxsize=9cm \epsfbox{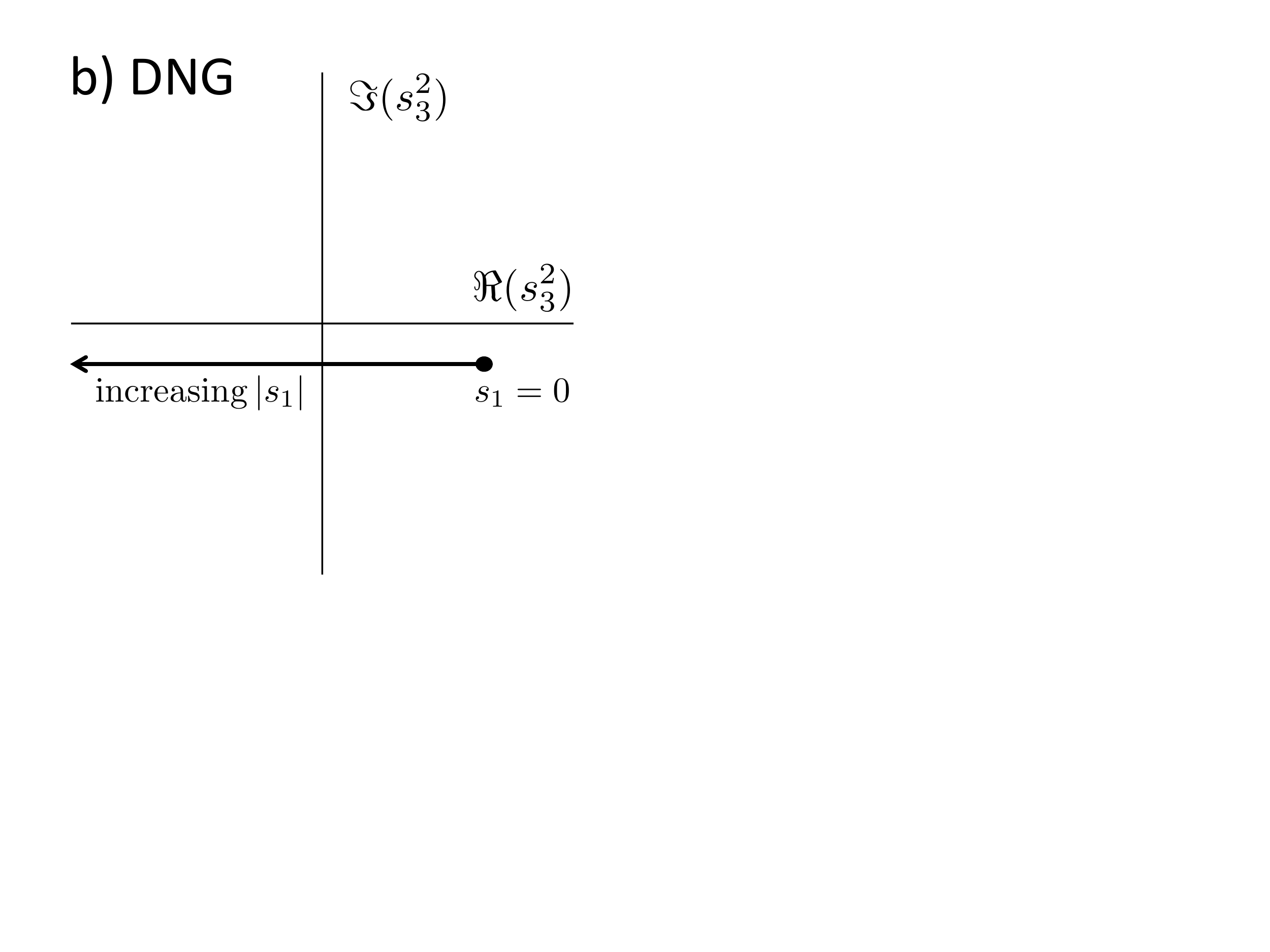}}
\vspace{-2.5cm}
\centerline{\hspace{4.5cm}\epsfxsize=9cm \epsfbox{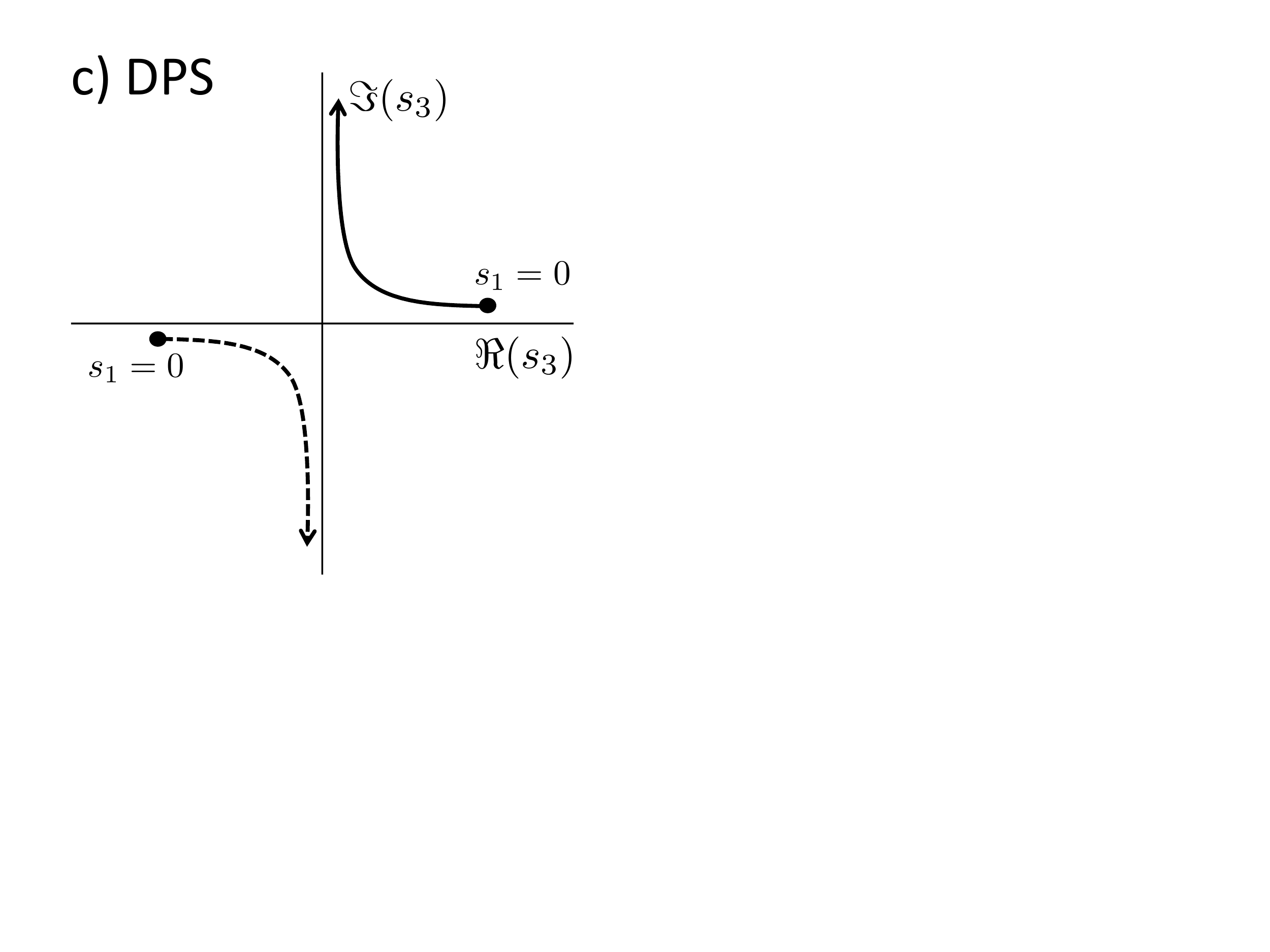}\hspace{-4.5cm}\epsfxsize=9cm \epsfbox{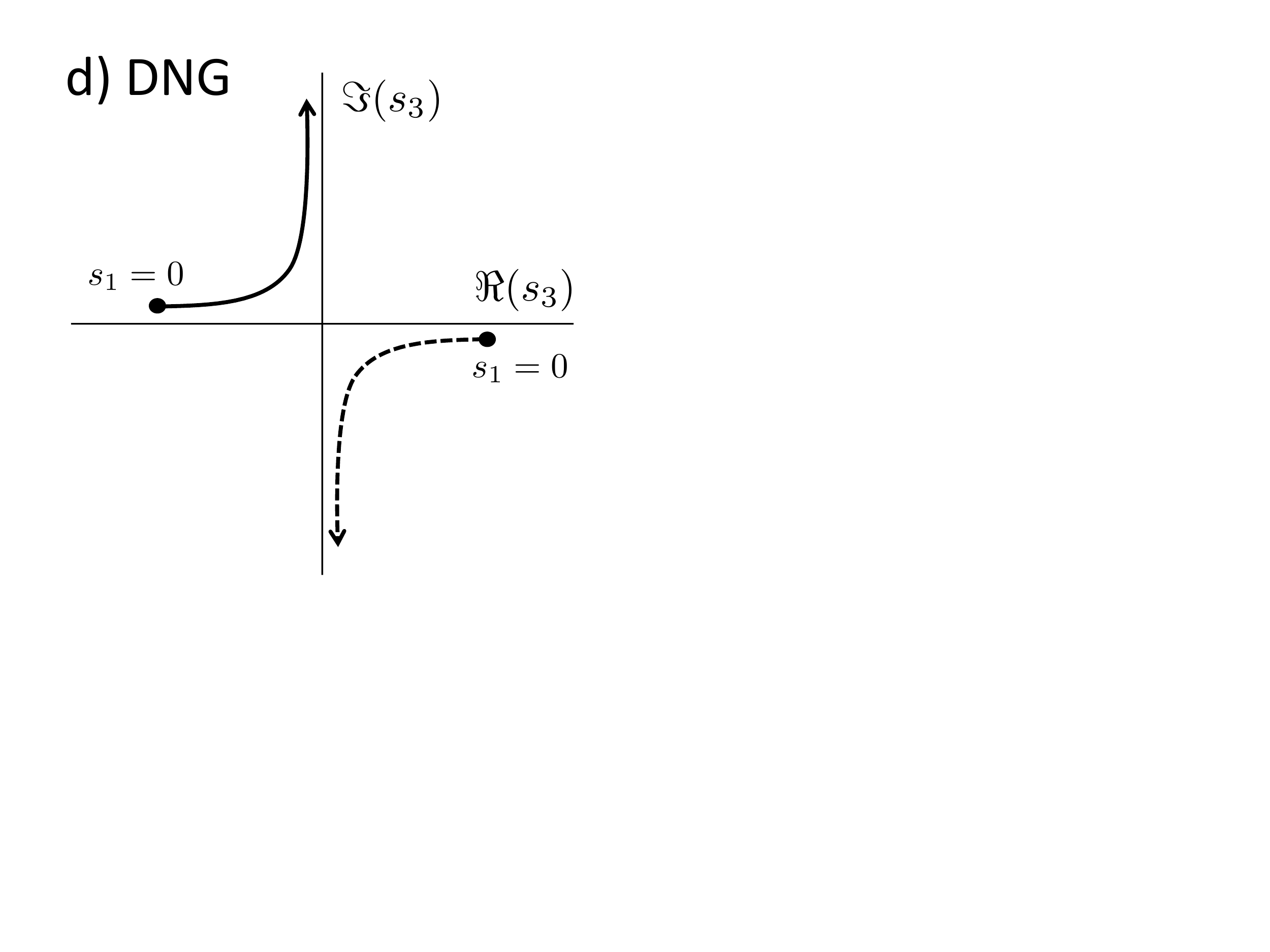}}
\vspace{-2.7cm}
\caption{Squared slowness $s_3^2$ in the complex plane for (a) DPS and (b) DNG medium.
Slowness $s_3$ in the complex plane for (c) DPS and (d) DNG medium. The solid curves in (c) and (d)  represent the proper square-roots of those in (a) and (b).}\label{Fig1}
\end{figure}

We define the spatial Fourier transform of a function $u(x_1,x_3,\omega)$ as
\begin{equation}\label{eqFTP}
\tilde u(s_1,x_3,\omega)=\int_{-\infty}^\infty u(x_1,x_3,\omega)\exp(-i\omega s_1 x_1){\rm d}x_1,
\end{equation}
with $s_1$ being the horizontal slowness. 
This transformation accomplishes a decomposition of the wave field $u(x_1,x_3,\omega)$ into plane-wave components $\tilde u(s_1,x_3,\omega)$.
We use equation (\ref{eqFTP})  to transform equations (\ref{eqA1}) to (\ref{eqA3}) from the space-frequency domain $(x_1,x_3,\omega)$  to the
slowness-depth-frequency  domain $(s_1,x_3,\omega)$. Differentiations with respect to $x_1$ thus become multiplications by $i\omega s_1$.
Eliminating $\tilde Q_1$ from the transformed equations, we obtain
the following matrix-vector wave equation  \cite{Thomson50JAP, Haskell53BSSA, Corones75JMAA, Ursin83GEO, Fishman84JMP, Hoop96JMP, Wapenaar2019GJI}
\begin{eqnarray}\label{eq22tilde}
\partial_3\tilde{\bf q} =\tilde{\bf A}\tilde{\bf q} +\tilde{\bf d},
\end{eqnarray}
with  wave vector $\tilde {\bf q} (s_1,x_3,\omega)$
and source vector $\tilde {\bf d} (s_1,x_3,\omega)$ 
defined as
\begin{eqnarray}\label{eq23}
\tilde {\bf q} =\begin{pmatrix} \tilde P \\ \tilde Q_3 \end{pmatrix}\quad\mbox{and}\quad
\tilde{\bf d} =\begin{pmatrix} \tilde C_3\\ \tilde B+ s_1\tilde C_1/\beta_1 \end{pmatrix}
\end{eqnarray}
and matrix $\tilde{\bf A}(s_1,x_3,\omega)$ defined as
\begin{eqnarray}\label{eq24pe}
\tilde{\bf A}=\begin{pmatrix}
 0 &i\omega\beta_3 \\  i\omega s_3^2/\beta_3 & 0
   \end{pmatrix},
\end{eqnarray}
with
\begin{eqnarray}
s_3^2=\alpha\beta_3 - \eta s_1^2, \quad\mbox{with}\quad \eta=\beta_3/\beta_1.\label{eqs3sq}
\end{eqnarray}
Note that  vector $\tilde {\bf q}$ defined in equation (\ref{eq23}) contains the wave field quantities that are continuous at interfaces between layers with different medium parameters.
Moreover, these quantities constitute the power flux density $j$ in the $x_3$-direction via $j=\frac{1}{2}\Re\{\tilde P^*\tilde Q_3\}$.
In the matrix-vector notation this can be written as
\begin{eqnarray}
j=\frac{1}{4} \tilde{\bf q}^\dagger{\bf K}\tilde{\bf q},\label{eqflux}
\end{eqnarray}
where  $\dagger$ denotes transposition and complex conjugation and where  matrix ${\bf K}$ is defined as
\begin{eqnarray}\label{Kmatrix}
{{\bf K}}=\begin{pmatrix} 0 & 1 \\ 1 & 0 \end{pmatrix}.
\end{eqnarray}
The quantity $s_3^2$ defined in equation (\ref{eqs3sq}) is the square of the vertical phase slowness.
Using equations (\ref{eqalpha}) and (\ref{eqbeta}) it can be written as 
\begin{eqnarray}
s_3^2=\frac{1}{c_0^2} h_\alpha h_\beta - \eta_0 s_1^2, \label{eqs3sqag}
\end{eqnarray}
with
\begin{eqnarray}
c_0=(\alpha_0\beta_{3,0})^{-\frac{1}{2}} \quad\mbox{and}\quad \eta_0=\beta_{3,0}/\beta_{1,0}.
\end{eqnarray}
Since $h_\alpha$ and $h_\beta$ are complex-valued functions, $s_3^2$ is complex-valued as well.
Defining $h_\alpha=h_\alpha^{\rm r}+i h_\alpha^{\rm i}$ and $h_\beta=h_\beta^{\rm r}+i h_\beta^{\rm i}$ we may write
\begin{eqnarray}
\Re (s_3^2)&=&\frac{1}{c_0^2} (h_\alpha^{\rm r} h_\beta^{\rm r}-h_\alpha^{\rm i} h_\beta^{\rm i}) - \eta_0 s_1^2,\\
\Im (s_3^2)&=&\frac{1}{c_0^2} (h_\alpha^{\rm r} h_\beta^{\rm i}+h_\alpha^{\rm i} h_\beta^{\rm r}).
\end{eqnarray}
For DPS media, with $h_\alpha^{\rm r}$, $h_\beta^{\rm r}$, $h_\alpha^{\rm i}$, $h_\beta^{\rm i}$ all positive (or zero), we have $\Im (s_3^2)\ge 0$.
For this situation, Figure \ref{Fig1}(a) illustrates $s_3^2$ in the complex plane for a fixed frequency $\omega$ and variable $s_1$. 
For DNG media, with $h_\alpha^{\rm r}$, $h_\beta^{\rm r}$ both negative and $h_\alpha^{\rm i}$, $h_\beta^{\rm i}$ both positive, we have $\Im (s_3^2)<0$.
For this situation $s_3^2$ is illustrated in the complex plane in Figure \ref{Fig1}(b).

\subsection{Decomposition of matrix-vector wave equation}

We reorganise the matrix-vector wave equation into an equation for downgoing and upgoing waves.
The eigenvalue decomposition of matrix $\tilde{\bf A} (s_1 ,x_3,\omega) $ reads
\begin{eqnarray}\label{eqALHLtilde}
\tilde{\bf A}=\tilde{\bf L}\tilde{\bf\Lambda}\tilde{\bf L}^{-1},
\end{eqnarray}
with $\tilde{\bf\Lambda}(s_1 ,x_3,\omega)$, $\tilde{\bf L}(s_1 ,x_3,\omega)$ and $\{\tilde{\bf L}(s_1 ,x_3,\omega)\}^{-1}$ defined as
\begin{eqnarray}
\tilde{\bf\Lambda}
&=&\begin{pmatrix}i\omega s_3 & 0 \\ 0 & -i\omega s_3\end{pmatrix},\label{eqAH}\\
\tilde{\bf L}&=&\begin{pmatrix}1 & 1\\  s_3/\beta_3 &  -s_3/\beta_3  \end{pmatrix},\label{eqAL}\\
\tilde{\bf L}^{-1}&=&\frac{1}{2}\begin{pmatrix}1 & \beta_3/s_3\\ 1 &  - \beta_3/s_3   \end{pmatrix}.\label{eqALinv}
\end{eqnarray}
The vertical phase slowness $s_3$ is defined as the square-root of $s_3^2$, {i.e., 
\begin{eqnarray}
s_3=\pm\sqrt{\frac{1}{c_0^2} h_\alpha h_\beta - \eta_0 s_1^2}.
\end{eqnarray}
It} is illustrated in Figures \ref{Fig1}(c) and \ref{Fig1}(d) for DPS and DNG
media, respectively. In both cases there are two square-roots, indicated by the two curves in these figures. 
In section \ref{sec1d}  we discuss how to choose the proper square-roots.

We introduce a decomposed field vector $\tilde{\bf p}$ and a decomposed source vector $\tilde{\bf s}$ via
\begin{eqnarray}
&&\tilde{\bf q}=\tilde{\bf L} \tilde{\bf p},\label{eqdecom2tilde}\\
&&\tilde{\bf d}=\tilde{\bf L}\tilde{\bf s}, \label{eqdecom3tilde}
\end{eqnarray}
with
\begin{eqnarray}
&& \tilde{\bf p}=\begin{pmatrix} \tilde P^+ \\ \tilde P^- \end{pmatrix},\quad \tilde{\bf s}=\begin{pmatrix} \tilde S^+ \\ \tilde S^- \end{pmatrix}.\label{eqdecom4tilde}
\end{eqnarray}
Substitution of equations  (\ref{eqALHLtilde}),  (\ref{eqdecom2tilde}) and  (\ref{eqdecom3tilde}) into the matrix-vector wave equation (\ref{eq22tilde}) yields
\begin{eqnarray}
&&\partial_3\tilde{\bf p}=\bigl(\tilde{\bf\Lambda}-\tilde{\bf L}^{-1}\partial_3\tilde{\bf L}\bigr)\tilde{\bf p}+\tilde{\bf s}.\label{eqonewaytilde}
\end{eqnarray}
This is a coupled system of equations for the wave field {components} $\tilde P^+$ and $\tilde P^-$, respectively. 
{}
According to equations (\ref{eq23}), (\ref{eqAL}), (\ref{eqdecom2tilde}) and (\ref{eqdecom4tilde}), we have 
\begin{eqnarray}\label{eq51aa}
\tilde P = \tilde P^+ + \tilde P^-,
\end{eqnarray}
hence, the wave field {components} $\tilde P^+$ and $\tilde P^-$ have the same physical dimension as the field quantity $\tilde P$.
Therefore we speak of field-normalised decomposition (opposed to flux-normalized decomposition). 

{From the theory for lossless DPS media it is known that the components $\tilde P^+$ and $\tilde P^-$ represent downgoing and upgoing wave fields, respectively 
\cite{Corones75JMAA, Ursin83GEO, Fishman84JMP, Wapenaar89Book, Hoop96JMP}. 
This still holds true for DPS and DNG media with or without losses, provided the proper choices are made for the sign of the vertical phase slowness $s_3$.
This is discussed in the next section.}

\subsection{Phase slowness}\label{sec1d}

We can express the power flux density $j$ in the $x_3$-direction in terms of  {downgoing and upgoing} wave fields by
substituting $\tilde{\bf q}=\tilde{\bf L} \tilde{\bf p}$ into equation (\ref{eqflux}). Using equations (\ref{Kmatrix}) and (\ref{eqAL}) we thus obtain
\begin{eqnarray}\label{eq511}
j&=&\frac{1}{4} \tilde{\bf q}^\dagger{\bf K}\tilde{\bf q}=\frac{1}{4}\tilde {\bf p}^\dagger\tilde{\bf L}^\dagger{\bf K}\tilde{\bf L}\tilde {\bf p}\nonumber\\
&=&\frac{1}{2} \Re\bigl(s_3/\beta_3\bigr)\Bigl(|\tilde P^+|^2-|\tilde P^-|^2\Bigr)
 + \Im\bigl(s_3/\beta_3\bigr)\Im\Bigl((\tilde P^+)^*\tilde P^-\Bigr).
\end{eqnarray}
For the discussion on the sign of the vertical phase slowness $s_3$, consider an independent downgoing wave field  $\tilde P^+$ in a homogeneous medium. 
For this situation the power flux density can be written as 
\begin{eqnarray}
j&=&\frac{1}{2} \Re(s_3/\beta_3)|\tilde P^+|^2
=\frac{h_\beta^{\rm r}\Re(s_3)+h_\beta^{\rm i}\Im(s_3)}{2\beta_{3,0}|h_\beta|^2}|\tilde P^+|^2,
\end{eqnarray}
where we used equation (\ref{eqbeta}) to express $\beta_3$ in terms of the positive quantity $\beta_{3,0}$ and  $h_\beta=h_\beta^{\rm r}+i h_\beta^{\rm i}$.
We now determine the signs of $\Re(s_3)$ and $\Im(s_3)$ such that  $\tilde P^+$ has a positive power flux density  in the positive $x_3$-direction \cite{Veselago68SPU}.
For DPS media, with $h_\beta^{\rm r}$ and $h_\beta^{\rm i}$ both positive, we find that this condition is fulfilled when $\Re(s_3)> 0$ and $\Im(s_3)> 0$.
Hence, the solid curve in  Figure \ref{Fig1}(c) represents the proper square-root of $s_3^2$. 
We write this square-root as
\begin{eqnarray}
s_3=+\sqrt{\frac{1}{c_0^2} h_\alpha h_\beta - \eta_0 s_1^2}, \label{eqs3sqagsqr}
\end{eqnarray}
with the $+$ sign in front of the square-root denoting that $\Re(s_3)> 0$.
For DNG media, with $h_\beta^{\rm r}$ negative and $h_\beta^{\rm i}$ positive,
we find that $j$ is positive when $\Re(s_3)< 0$ and $\Im(s_3)> 0$.
Hence, for this situation the solid curve in  Figure \ref{Fig1}(d) represents the proper square-root of $s_3^2$.
We write this square-root as
\begin{eqnarray}\label{eq34}
s_3=-\sqrt{\frac{1}{c_0^2} h_\alpha h_\beta - \eta_0 s_1^2}, \label{eqs3sqagsqrm}
\end{eqnarray}
with the $-$ sign in front of the square-root denoting that $\Re(s_3)< 0$.
Equations (\ref{eqs3sqagsqr}) and (\ref{eqs3sqagsqrm}) express the fact that the {(real part of the)} vertical phase slowness is positive for DPS media and negative for DNG media.
Given these square-roots, we find in the same way that $j$ is negative for  an independent upgoing wave field  $\tilde P^-$ in a homogeneous medium.

\begin{figure}[h]
\centerline{\hspace{4.5cm}\epsfxsize=9cm \epsfbox{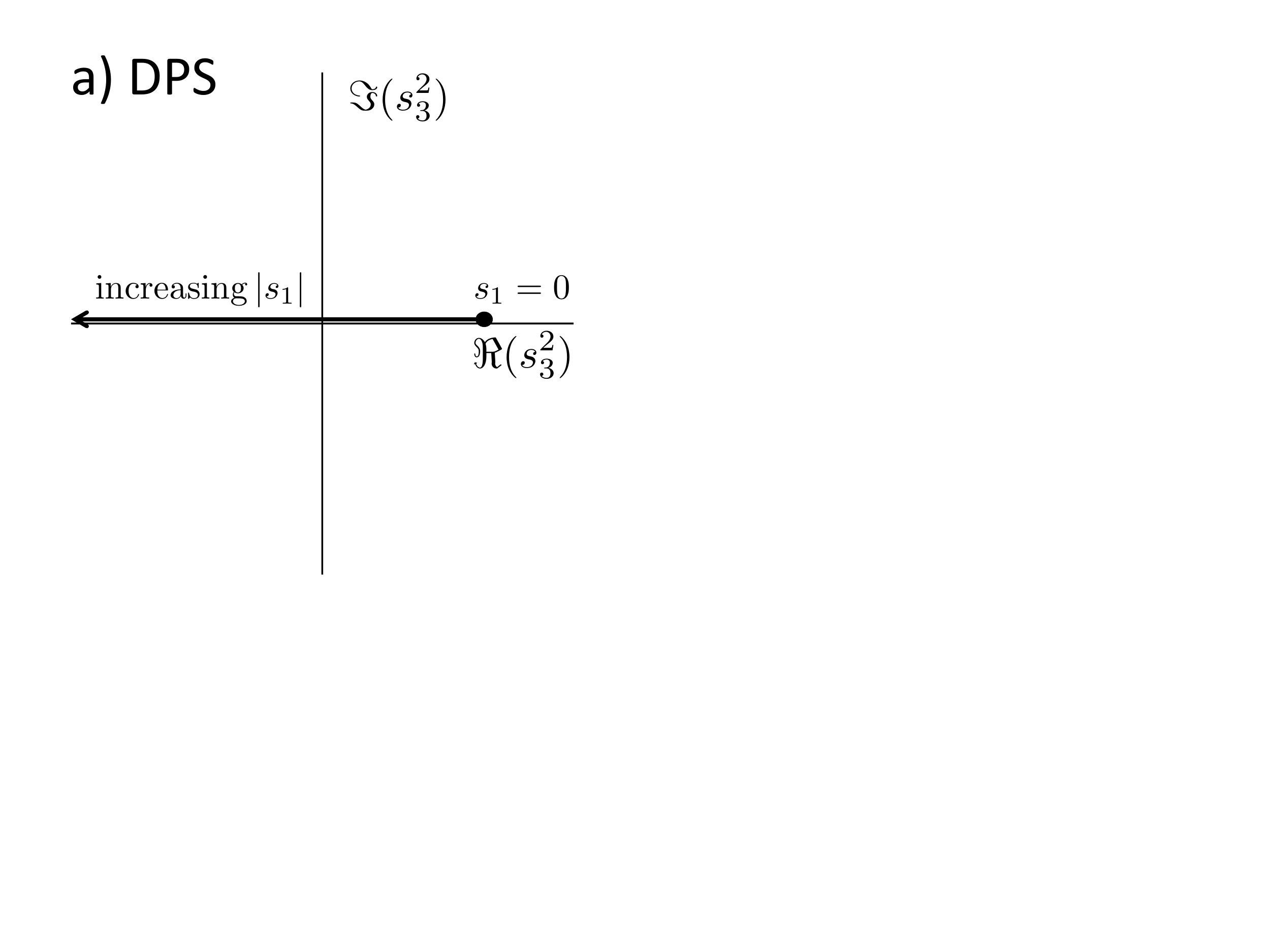}\hspace{-4.5cm}\epsfxsize=9cm \epsfbox{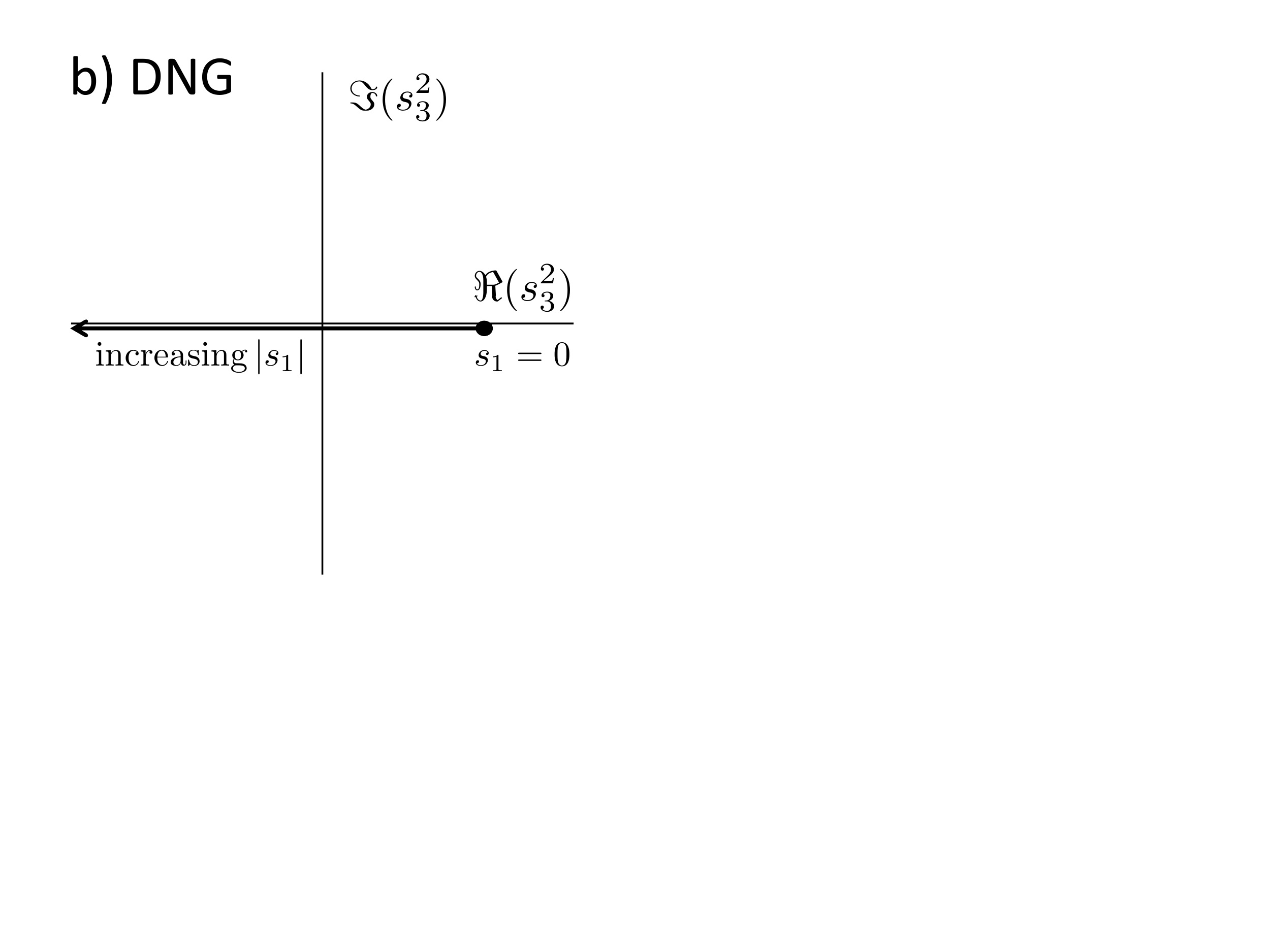}}
\vspace{-2.5cm}
\centerline{\hspace{4.5cm}\epsfxsize=9cm \epsfbox{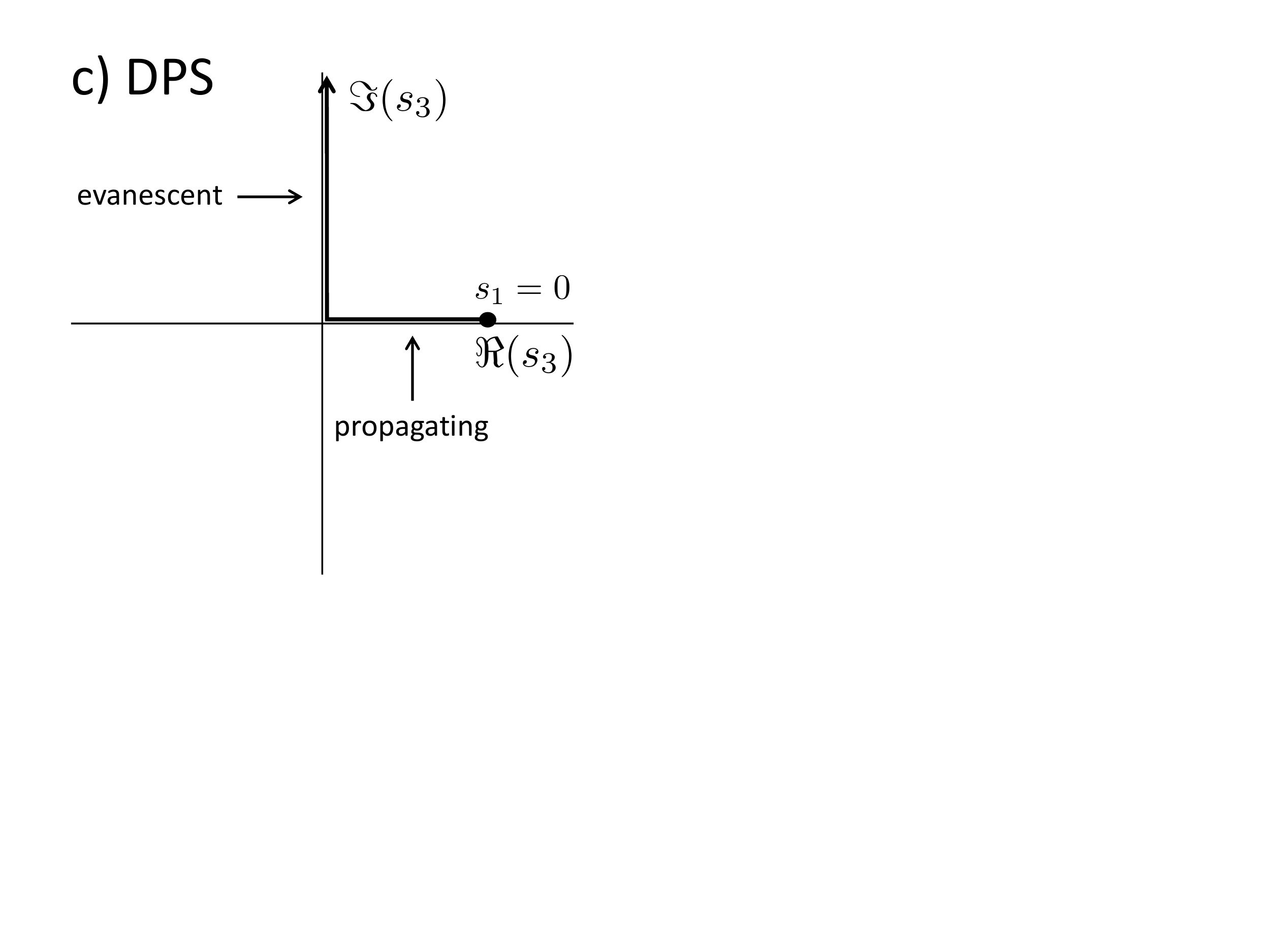}\hspace{-4.5cm}\epsfxsize=9cm \epsfbox{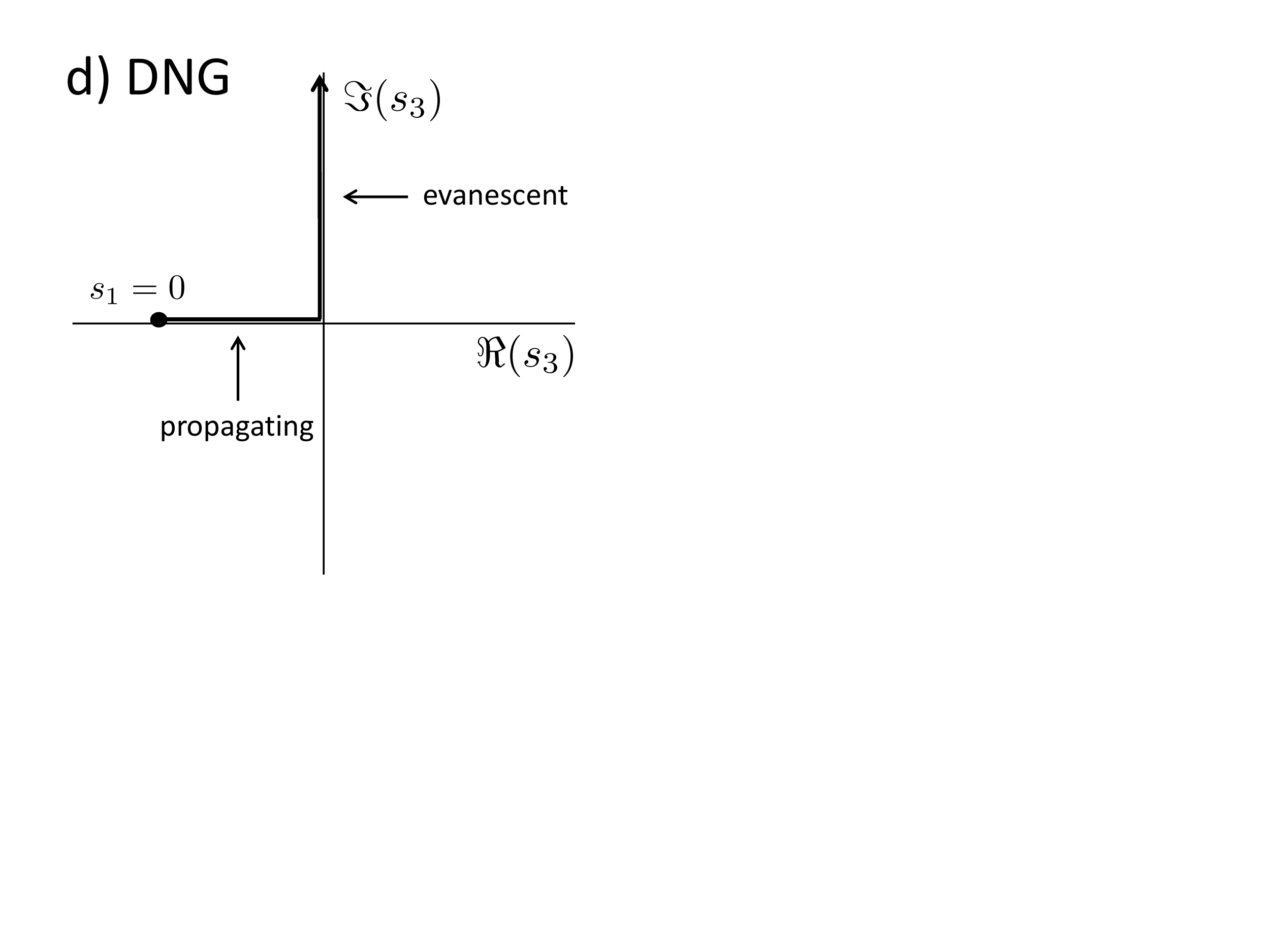}}
\vspace{-2.7cm}
\caption{As Figure \ref{Fig1}, for the limiting case of vanishing loss parameters. The curves in (c) and (d) are the proper square-roots of those in (a) and (b).}\label{Fig2}
\end{figure}

Figure \ref{Fig2} shows $s_3^2$ and $s_3$ in the complex plane for the limiting situation of vanishing losses, i.e., vanishing imaginary parts of the medium parameters.
Note that the real and imaginary branches of $s_3$  correspond to propagating and evanescent waves, respectively.

\subsection{Group slowness}\label{sec1e}

Despite the fact that the vertical phase slowness in a DNG medium is negative, the vertical group slowness should be positive. 
This restricts the choice of models for the functions 
$h_\alpha(\omega)$ and $h_\beta(\omega)$. We define the vertical group slowness as
\begin{eqnarray}
s_3^{\rm gr}&=&\Re\biggl(\frac{\partial (\omega s_3)}{\partial\omega}\biggr).
\end{eqnarray}
Substituting equation (\ref{eqs3sqagsqrm}), taking for convenience $h_\alpha(\omega)=h_\beta(\omega)=h(\omega)$, we obtain
\begin{eqnarray}\label{eq36}
s_3^{\rm gr}&=&\Re\biggl(\frac{h(h+\omega \frac{\partial h}{\partial\omega})-\eta_0c_0^2s_1^2}{c_0^2s_3}\biggl).
\end{eqnarray}
We analyse this expression for the Drude model of equations (\ref{eqdrudea}) and (\ref{eqdrudeb}), 
 with $\omega_\alpha=\omega_\beta=\omega_0$ and $\Gamma_\alpha=\Gamma_\beta =\Gamma$, hence
 \begin{eqnarray}\label{eq37}
h(\omega)=1-\frac{\omega_0^2}{\omega(\omega+i\Gamma)}
\end{eqnarray}
and
\begin{eqnarray}\label{eq38}
h+\omega \partial h/\partial\omega=1+\frac{\omega_0^2}{(\omega+i\Gamma)^2}.
\end{eqnarray}
The condition for a DNG medium,  $\Re(h)<0$, requires  $\omega^2<\omega_0^2-\Gamma^2$.

We evaluate the sign of   $s_3^{\rm gr}$ for two special situations.
First we consider  vertically propagating waves, i.e.,  $s_1=0$. From equation (\ref{eq34}) we find $s_3=h/c_0$ for 
  $\omega^2<\omega_0^2-\Gamma^2$. Using this in equation (\ref{eq36}) we obtain  \cite{Ziolkowski2001PRE}
 \begin{eqnarray}
s_3^{\rm gr}
&=&\frac{\Re(h+\omega \partial h/\partial\omega)}{c_0}
=\frac{1}{c_0}\Re\biggl(1+\frac{\omega_0^2}{(\omega+i\Gamma)^2}\biggr).\label{eqgrsl}
\end{eqnarray}
Assuming small $\Gamma$ (a sufficient condition is $\Gamma<\omega$), we find indeed that the vertical group slowness $s_3^{\rm gr}$ is positive.
 
Next we  consider non-zero $s_1$ and analyse equation (\ref{eq36}) for  the limit $\Gamma\to 0$.
From equations  (\ref{eq37}) and (\ref{eq38}) we find $h=1-\omega_0^2/\omega^2$ and $h(h+\omega \partial h/\partial\omega)=1-\omega_0^4/\omega^4$.
Using this in equation (\ref{eq36}) we obtain
\begin{eqnarray}
s_3^{\rm gr}&=&\frac{\eta_0c_0^2s_1^2+\omega_0^4/\omega^4-1}{c_0^2\sqrt{(\omega_0^2/\omega^2-1)^2/c_0^2 - \eta_0 s_1^2}}.
\end{eqnarray}
For $\omega<\omega_0$ the nominator is positive for all $s_1$. 
For propagating waves the denominator is real-valued and positive as well, hence
 $s_3^{\rm gr}$ is positive for this situation.

\section{Representations for the Marchenko method}\label{sec3}

\subsection{Propagation invariants for DPS and DNG media}

We consider a medium configuration consisting of a homogeneous DPS upper half-space $x_3\le x_{3,0}$ and a horizontally layered
lower half-space $x_3>x_{3,0}$, which may consist of an arbitrary mix of DPS and DNG layers.
The effective medium parameters in this configuration are $\alpha(x_3,\omega)$ and $\beta_i(x_3,\omega)$. 
These parameters  may vary continuously as a function of $x_3$ within each layer and jump by a finite amount at layer interfaces. 
 We assume that the losses are small and for the derivation of the Marchenko method we ignore the imaginary parts of $\alpha(x_3,\omega)$ and $\beta_i(x_3,\omega)$  
(however, in the numerical example in section \ref{sec5} we model the input data with complex-valued medium parameters).
Assuming that the sources are restricted to the upper half-space $x_3 \le x_{3,0}$, the wave field inside the layers  is governed by  wave equation (\ref{eq22tilde}) with $\tilde{\bf d}={\bf 0}$. 
Moreover, $\tilde{\bf q}$ is continuous at layer interfaces.
We derive propagation invariants \cite{Haines88GJI, Kennett90GJI, Koketsu91GJI, Takenaka93WM},
which we will use for the derivation of the representations for the Marchenko method in the next section. 
We consider two independent wave vectors $\tilde {\bf q}_A$ and $\tilde {\bf q}_B$
and  will show that 
$\tilde{\bf q}_A^t{\bf N}\tilde{\bf q}_B$ and $\tilde{\bf q}_A^\dagger{\bf K}\tilde{\bf q}_B$ 
are propagation invariants (i.e., that they are independent of the coordinate $x_3$ for $x_3>x_{3,0}$). Here superscript $t$ denotes transposition and matrix ${\bf N}$ is defined as
\begin{eqnarray}\label{eq4.3K}
&&{{\bf N}}=\begin{pmatrix} 0 & 1 \\ -1 & 0 \end{pmatrix}.
\end{eqnarray}
Obviously the quantities $\tilde{\bf q}_A^t{\bf N}\tilde{\bf q}_B$ and $\tilde{\bf q}_A^\dagger{\bf K}\tilde{\bf q}_B$ are continuous at layer interfaces. Hence, to show that these quantities are propagation 
invariants for the layered medium, it suffices to show that they are propagation invariants inside a layer.
Evaluating $\partial_3\{\tilde{\bf q}_A^t{\bf N}\tilde{\bf q}_B\}$ and $\partial_3\{\tilde{\bf q}_A^\dagger{\bf K}\tilde{\bf q}_B\}$, using equation (\ref{eq22tilde})  with $\tilde{\bf d}={\bf 0}$, we obtain
\begin{eqnarray}
&&\hspace{-.4cm}\partial_3\{\tilde{\bf q}_A^t{\bf N}\tilde{\bf q}_B\}=
\tilde{\bf q}_A^t\tilde{\bf A}^t{\bf N}\tilde{\bf q}_B+\tilde{\bf q}_A^t{\bf N}\tilde{\bf A}\tilde{\bf q}_B,\label{eqPI1}\\
&&\hspace{-.4cm}\partial_3\{\tilde{\bf q}_A^\dagger{\bf K}\tilde{\bf q}_B\}=
\tilde{\bf q}_A^\dagger{\bf A}^\dagger{\bf K}\tilde{\bf q}_B+\tilde{\bf q}_A^\dagger{\bf K}\tilde{\bf A}\tilde{\bf q}_B.\label{eqPI2}
\end{eqnarray}
Matrix $\tilde{\bf A}$, defined in equation (\ref{eq24pe}), obeys for real-valued medium parameters the following symmetry relations
\begin{eqnarray}
\tilde{\bf A}^t{\bf N}&=&-{\bf N}\tilde{\bf A},\label{eq46}\\
\tilde{\bf A}^\dagger{\bf K}&=&-{\bf K}\tilde{\bf A}.\label{eq47}
\end{eqnarray}
Hence, the right-hand sides of equations (\ref{eqPI1}) and (\ref{eqPI2}) are equal to zero, which confirms that 
$\tilde{\bf q}_A^t{\bf N}\tilde{\bf q}_B$ and $\tilde{\bf q}_A^\dagger{\bf K}\tilde{\bf q}_B$ are propagation invariants.

Next, we derive propagation invariants for decomposed wave fields.
Consider two independent decomposed wave vectors $\tilde {\bf p}_A$ and $\tilde {\bf p}_B$, which are related to  $\tilde {\bf q}_A$ and $\tilde {\bf q}_B$, respectively, via equation (\ref{eqdecom2tilde}).
We obtain propagation invariants for these decomposed wave vectors
by substituting $\tilde{\bf q}_A=\tilde{\bf L} \tilde{\bf p}_A$ and $\tilde{\bf q}_B=\tilde{\bf L} \tilde{\bf p}_B$ 
into the propagation invariants $\tilde{\bf q}_A^t{\bf N}\tilde{\bf q}_B$ and $\tilde{\bf q}_A^\dagger{\bf K}\tilde{\bf q}_B$. 
Using equations (\ref{Kmatrix}), (\ref{eqAL}) and  (\ref{eq4.3K}), we obtain
\begin{eqnarray}
\tilde{\bf q}_A^t{\bf N}\tilde{\bf q}_B=\tilde{\bf p}_A^t\tilde{\bf L}^t{\bf N}\tilde{\bf L}\tilde{\bf p}_B
=-2(s_3/\beta_3) \tilde{\bf p}_A^t{\bf N}\tilde{\bf p}_B\label{eqPI3}
\end{eqnarray}
and
\begin{eqnarray}
&&\tilde{\bf q}_A^\dagger{\bf K}\tilde{\bf q}_B=\tilde{\bf p}_A^\dagger\tilde{\bf L}^\dagger{\bf K}\tilde{\bf L}\tilde{\bf p}_B
=2 \tilde{\bf p}_A^\dagger\bigl(\Re(s_3/\beta_3){\bf J}-i\Im(s_3/\beta_3){\bf N}\bigr)\tilde{\bf p}_B,\label{eqPI4}
\end{eqnarray}
with
\begin{eqnarray}\label{eq4.3J}
&&{{\bf J}}=\begin{pmatrix} 1 & 0 \\ 0 & -1 \end{pmatrix}.
\end{eqnarray}
From equations (\ref{eqdecom4tilde}), (\ref{eq4.3K}) and (\ref{eqPI3}) we obtain the propagation invariant
\begin{eqnarray}
(s_3/\beta_3)\bigl(\tilde P_A^+\tilde P_B^- - \tilde P_A^-\tilde P_B^+\bigr).\label{eq49}
\end{eqnarray}
From equations (\ref{eqdecom4tilde}), (\ref{eqPI4}) and (\ref{eq4.3J}) we obtain for propagating waves (i.e., for real-valued $s_3$)
the propagation invariant
\begin{eqnarray}
(s_3/\beta_3)\bigl((\tilde P_A^+)^*\tilde P_B^+ - (\tilde P_A^-)^*\tilde P_B^-\bigr).\label{eq50}
\end{eqnarray}

\begin{figure}[t]
\centerline{\epsfxsize=10.cm \epsfbox{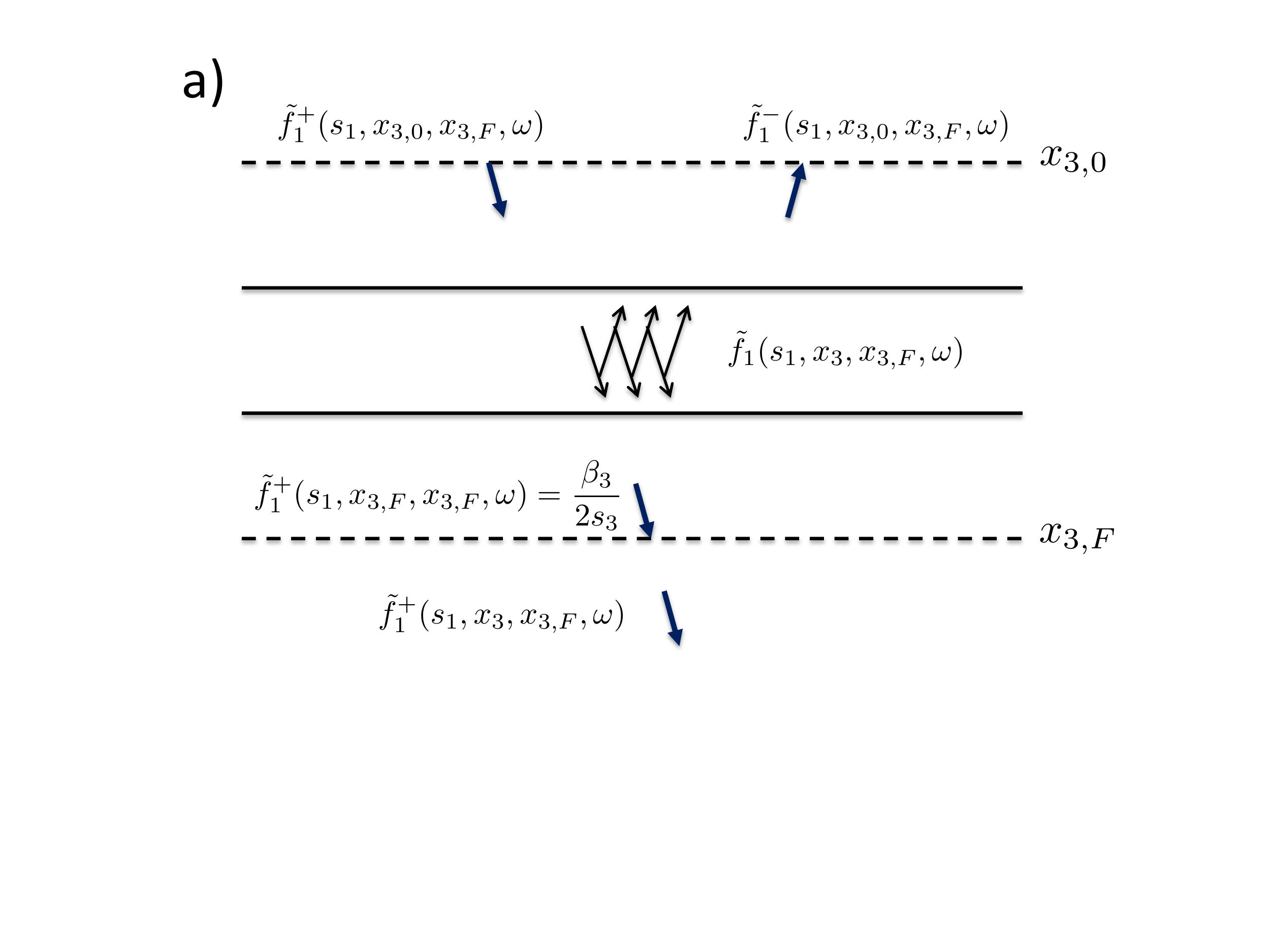}}
\vspace{-1cm}
\centerline{\epsfxsize=10.cm \epsfbox{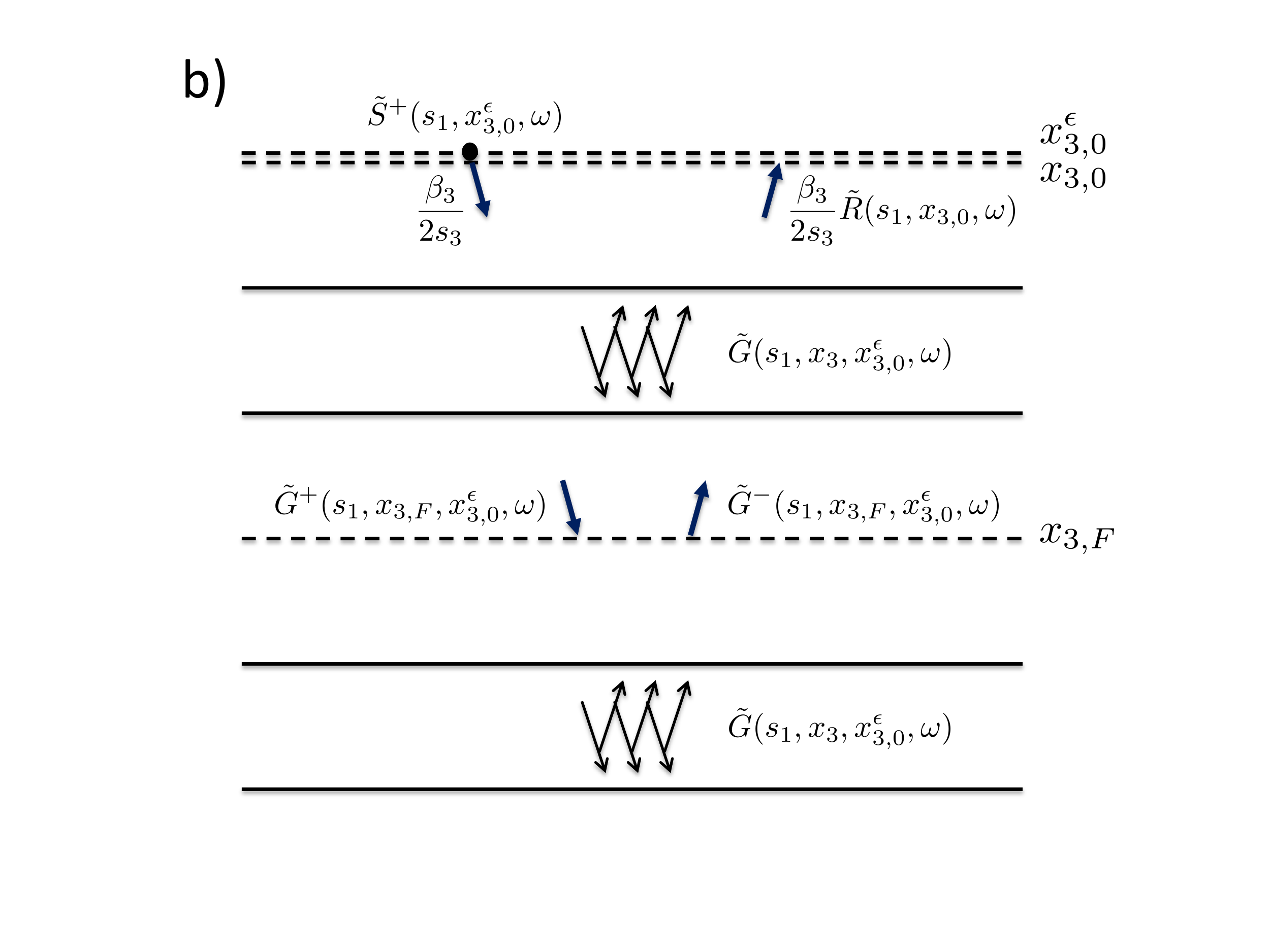}}
\vspace{-1cm}
\caption{(a) The focusing function $\tilde f_1=\tilde f_1^++\tilde f_1^-$, defined in a truncated version of the actual medium. (b) The Green's function $\tilde G=\tilde G^++\tilde G^-$, defined in the actual medium.}\label{Figg3}
\end{figure}

\subsection{Representations}

We use the propagation invariants of equations (\ref{eq49}) and (\ref{eq50}) to derive representations for the Marchenko method. 
We  introduce decomposed  focusing functions (Figure \ref{Figg3}(a)) and Green's functions (Figure \ref{Figg3}(b)) and derive relations between them using equations (\ref{eq49}) and (\ref{eq50}), 
with $\tilde P_A^\pm$ and $\tilde P_B^\pm$ replaced by the focusing functions and Green's functions, respectively \cite{Slob2014GEO, Wapenaar2014GEO}.

First we discuss the Green's functions. For the source quantities in equation (\ref{eq23}) we take $\tilde B(s_1,x_3,\omega)=\delta(x_3-x_{3,0}^\epsilon)$ and  $\tilde C_i(s_1,x_3,\omega)=0$,
where $x_{3,0}^\epsilon=x_{3,0}-\epsilon$, with $\epsilon$ a vanishing positive constant, so that the source of the Green's function is located in the homogeneous upper half-space, just above $x_{3,0}$ (Figure \ref{Figg3}(b)).
For the wave field $\tilde P$ in equation (\ref{eq23}) we take $\tilde P=\tilde G(s_1,x_3,x_{3,0}^\epsilon,\omega)$, with  $x_{3,0}^\epsilon$ and $x_3$ denoting the 
source and receiver coordinates of the Green's function. We decompose the Green's function at the receiver position $x_3$ into downgoing and upgoing components
$\tilde P^+=\tilde G^+(s_1,x_3,x_{3,0}^\epsilon,\omega)$ and $\tilde P^-=\tilde G^-(s_1,x_3,x_{3,0}^\epsilon,\omega)$, respectively. 
Analogous to equation (\ref{eq51aa}), 
these components are related to the total Green's function via
\begin{eqnarray}
\tilde G=\tilde G^++\tilde G^-.\label{eq51}
\end{eqnarray}
Furthermore, according to equations (\ref{eq23}), (\ref{eqALinv}), (\ref{eqdecom3tilde}) and (\ref{eqdecom4tilde}), the decomposed Green's sources are related to the total Green's source
$\tilde B(s_1,x_3,\omega)=\delta(x_3-x_{3,0}^\epsilon)$ via
\begin{eqnarray}
\tilde S^\pm(s_1,x_3,\omega)=\pm (\beta_3/2s_3)\delta(x_3-x_{3,0}^\epsilon).\label{eq52}
\end{eqnarray}
The source $\tilde S^-$ radiates upgoing waves into the homogeneous half-space above $x_{3,0}^\epsilon$, which will not return into the layered medium and will therefore not be further considered.
The source $\tilde S^+$ radiates downgoing waves into the  medium below $x_{3,0}^\epsilon$. Due to {propagation and} scattering in the layered medium, the field at any depth $x_3>x_{3,0}^\epsilon$ consists of 
the downgoing and upgoing components $\tilde G^+(s_1,x_3,x_{3,0}^\epsilon,\omega)$ and $\tilde G^-(s_1,x_3,x_{3,0}^\epsilon,\omega)$. At $x_3=x_{3,0}$, i.e., at a vanishing distance $\epsilon$ 
below the source, the downgoing component reads
\begin{eqnarray}
\tilde G^+(s_1,x_{3,0},x_{3,0}^\epsilon,\omega)&=&(\beta_3/2s_3)\lim_{\epsilon\to 0}\exp\{i\omega s_3\epsilon\}
=\frac{\beta_3(x_{3,0},\omega)}{2s_3 (s_1,x_{3,0},\omega)}.\label{eq53}
\end{eqnarray}
This follows from equations  (\ref{eqAH}), (\ref{eqdecom4tilde}), (\ref{eqonewaytilde}) and (\ref{eq52}), taking into account that the medium between $x_{3,0}^\epsilon$ and $x_{3,0}$ is homogeneous.
At the same depth level ($x_3=x_{3,0}$) we relate the upgoing component to the reflection response $\tilde R(s_1,x_{3,0},\omega)$ of the layered medium, via
\begin{eqnarray}
\tilde G^-(s_1,x_{3,0},x_{3,0}^\epsilon,\omega)&=&\frac{\beta_3(x_{3,0},\omega)\tilde R(s_1,x_{3,0},\omega)}{2s_3 (s_1,x_{3,0},\omega)},\label{eq54}
\end{eqnarray}
where the factor $\beta_3/2s_3$ is introduced to compensate for the source properties expressed by equation (\ref{eq52}).
The decomposed Green's functions $\tilde G^+$ and $\tilde G^-$ will be substituted for $\tilde P_B^+$ and $\tilde P_B^-$ in the  propagation invariants of equations (\ref{eq49}) and (\ref{eq50}).
Table 2 shows these functions at depth level $x_3=x_{3,0}$ (just below the source) and at an arbitrary depth level $x_3=x_{3,F}$ (with $x_{3,F}>x_{3,0}$) inside the layered medium.
Note that for convenience  we dropped the superscript $\epsilon$ from $x_{3,0}^\epsilon$.

\begin{center}
{\small
{\noindent \it Table 2: Quantities used in the propagation invariants of equations (\ref{eq49}) and (\ref{eq50}).}
\begin{tabular}{||l|c|c|c|c||}
\hline\hline
&&&&\\
& $\tilde P_A^+ (s_1,x_3,\omega) $ & $\tilde P_A^-(s_1,x_3,\omega) $ & $\tilde P_B^+ (s_1,x_3,\omega) $ & $\tilde P_B^-(s_1,x_3,\omega) $  \\
&&&&\\
\hline
&&&&\\
$x_3=x_{3,0}$  & $\tilde f_1^+(s_1,x_{3,0},x_{3,F},\omega)$ &$\tilde f_1^-(s_1,x_{3,0},x_{3,F},\omega)$ &  $\frac{\beta_3(x_{3,0},\omega)}{2s_3 (s_1,x_{3,0},\omega)}$ & 
$\frac{\beta_3(x_{3,0},\omega)\tilde R (s_1,x_{3,0},\omega)} {2s_3 (s_1,x_{3,0},\omega)} $\\
&&&&\\
\hline
&&&&\\
$x_3=x_{3,F}$  &  $\frac{\beta_3(x_{3,F},\omega)}{2s_3 (s_1,x_{3,F},\omega)}$  &$0$ & $\tilde G^+ (s_1,x_{3,F},x_{3,0},\omega) $ & $\tilde G^-(s_1,x_{3,F},x_{3,0},\omega) $ \\
&&&&\\
\hline\hline
\end{tabular}
}
\end{center}
\mbox{}\\

Next we discuss the focusing functions \cite{Wapenaar2013PRL}. We define these functions in a truncated version of the actual medium (Figure \ref{Figg3}(a)). This truncated medium is taken identical to the actual medium above $x_3=x_{3,F}$ and 
homogeneous below this depth level. We call $x_{3,F}$ the focal depth. Analogous to equation (\ref{eq51}) we define the focusing function $\tilde f_1(s_1,x_3,x_{3,F},\omega)$ as a superposition of downgoing and upgoing components 
$\tilde f_1^+(s_1,x_3,x_{3,F},\omega)$ and $\tilde f_1^-(s_1,x_3,x_{3,F},\omega)$, respectively, according to
\begin{eqnarray}
\tilde f_1=\tilde f_1^++\tilde f_1^-.\label{eq55}
\end{eqnarray}

The downgoing focusing function $\tilde f_1^+(s_1,x_{3,0},x_{3,F},\omega)$ is incident to the truncated layered medium from the upper boundary $x_{3,0}$ and is designed such that 
$\tilde f_1^+(s_1,x_3,x_{3,F},\omega)$ focuses at the focal depth $x_3=x_{3,F}$.
Inside the medium {propagation and} scattering takes place and the upgoing focusing function $\tilde f_1^-(s_1,x_3,x_{3,F},\omega)$  {eventually} reaches the upper boundary $x_3=x_{3,0}$.
Below the focal depth $x_{3,F}$ the focusing function continues propagating downward into the homogeneous lower half-space of the truncated medium.  We define $\tilde T(s_1,x_{3,F},x_{3,0},\omega)$ as the transmission
response of the truncated medium between $x_{3,0}$ and $x_{3,F}$. Hence, the propagation of the focusing function from $x_{3,0}$ to $x_{3,F}$ is described by
\begin{eqnarray}
&&\tilde f_1^+(s_1,x_{3,F},x_{3,F},\omega)=
\tilde T(s_1,x_{3,F},x_{3,0},\omega)\tilde f_1^+(s_1,x_{3,0},x_{3,F},\omega).\label{eq56}
\end{eqnarray}
The left-hand side describes the focused field at $x_{3,F}$. We could define this as $\tilde f_1^+(s_1,x_{3,F},x_{3,F},\omega)=1$ (with the inverse Fourier transform of 1 being a temporal delta function).
However, in analogy with the Green's function at the source depth in equation (\ref{eq53}), we define  the focused field at the focal depth as 
\begin{eqnarray}
\tilde f_1^+(s_1,x_{3,F},x_{3,F},\omega)=\frac{\beta_3(x_{3,F},\omega)}{2s_3 (s_1,x_{3,F},\omega)}.\label{eq57}
\end{eqnarray}
From equations (\ref{eq56}) and (\ref{eq57}) it follows that the downgoing focusing function $\tilde f_1^+(s_1,x_3,x_{3,F},\omega)$ for $x_3=x_{3,0}$ is related to the transmission response of the truncated medium via
\begin{eqnarray}
&&\hspace{-.4cm}\tilde f_1^+(s_1,x_{3,0},x_{3,F},\omega)=\frac{\beta_3(x_{3,F},\omega)}{2s_3 (s_1,x_{3,F},\omega)\tilde T(s_1,x_{3,F},x_{3,0},\omega)}.\nonumber\\
\label{eq58}
\end{eqnarray}
Since the truncated medium is homogeneous below the focal depth there is no upgoing field at the focal depth, hence
\begin{eqnarray}
\tilde f_1^-(s_1,x_{3,F},x_{3,F},\omega)=0.\label{eq59}
\end{eqnarray}
The decomposed focusing functions $\tilde f_1^+$ and $\tilde f_1^-$ will be substituted for $\tilde P_A^+$ and $\tilde P_A^-$ in the  propagation invariants of equations (\ref{eq49}) and (\ref{eq50}).
Table 2 shows these functions at depth levels $x_3=x_{3,0}$  and $x_3=x_{3,F}$.
An underlying assumption for the propagation invariants is that the
fields $\tilde P_A^\pm$ and $\tilde P_B^\pm$ are defined in the same source-free medium.
This condition is fulfilled in the region between $x_{3,0}$ and $x_{3,F}$. Substituting the quantities of Table 2 into the propagation invariant of equation (\ref{eq49}) and equating the results for  $x_{3,0}$ and $x_{3,F}$ yields
\begin{eqnarray}
\tilde G^-(s_1,x_{3,F},x_{3,0},\omega) + \tilde f_1^-(s_1,x_{3,0},x_{3,F},\omega)
=\tilde R (s_1,x_{3,0},\omega)\tilde f_1^+(s_1,x_{3,0},x_{3,F},\omega).\label{eq145}
\end{eqnarray}
In a similar way we obtain from the propagation invariant of equation (\ref{eq50}) for propagating waves
\begin{eqnarray}
\tilde G^+(s_1,x_{3,F},x_{3,0},\omega) - \{\tilde f_1^+(s_1,x_{3,0},x_{3,F},\omega)\}^*
=-\tilde R (s_1,x_{3,0},\omega)\{\tilde f_1^-(s_1,x_{3,0},x_{3,F},\omega)\}^*.\label{eq146}
\end{eqnarray}
These representations express the downgoing and upgoing components of the Green's function at an arbitrarily chosen depth level 
$x_3=x_{3,F}$ in terms of the reflection response at the surface $x_3=x_{3,0}$ and decomposed focusing functions.
These representations hold for a layered medium {consisting of} an arbitrary mix of DPS and DNG layers. 
The reflection response $\tilde R (s_1,x_{3,0},\omega)$ can be obtained from measurements at the surface $x_{3,0}$.
According to equation (\ref{eq58}), the focusing function $\tilde f_1^+(s_1,x_{3,0},x_{3,F},\omega)$ could in principle be obtained from the transmission response of the truncated medium.
However, this would require detailed knowledge of the medium between $x_{3,0}$ and $x_{3,F}$. 
In the next section we discuss the Marchenko method, which enables retrieving the focusing functions from the reflection response
at the surface and  a background model of the medium.

\section{The Marchenko method}\label{sec4}

We start by transforming the representations of equations (\ref{eq145}) and (\ref{eq146}) to the time domain.
Analogous to equation (\ref{eqA11inv}), we define the following inverse Fourier transform 
\begin{eqnarray}\label{eqA11invb}
u(s_1,x_3,\tau)=\frac{1}{\pi}\Re\int_{0}^\infty \tilde u(s_1,x_3,\omega)\exp(-i\omega \tau){\rm d}\omega,
\end{eqnarray}
where $\tau$ is the so-called intercept time \citep{Stoffa89Book}.
Applying this inverse transform to equations (\ref{eq145}) and (\ref{eq146}), we obtain 
\begin{eqnarray}
&&G^-(s_1,x_{3,F},x_{3,0},\tau) + f_1^-(s_1,x_{3,0},x_{3,F},\tau)\label{eq145tau}\\
&&=\int_{-\infty}^\tau R (s_1,x_{3,0},\tau-\tau')f_1^+(s_1,x_{3,0},x_{3,F},\tau'){\rm d}\tau'\nonumber
\end{eqnarray}
and
\begin{eqnarray}
&&G^+(s_1,x_{3,F},x_{3,0},\tau) - f_1^+(s_1,x_{3,0},x_{3,F},-\tau)\label{eq146tau}\\
&&=-\int_{-\infty}^\tau R (s_1,x_{3,0},\tau-\tau')f_1^-(s_1,x_{3,0},x_{3,F},-\tau'){\rm d}\tau'.\nonumber
\end{eqnarray}

\begin{figure}[b]
\centerline{\epsfxsize=7.5cm \epsfbox{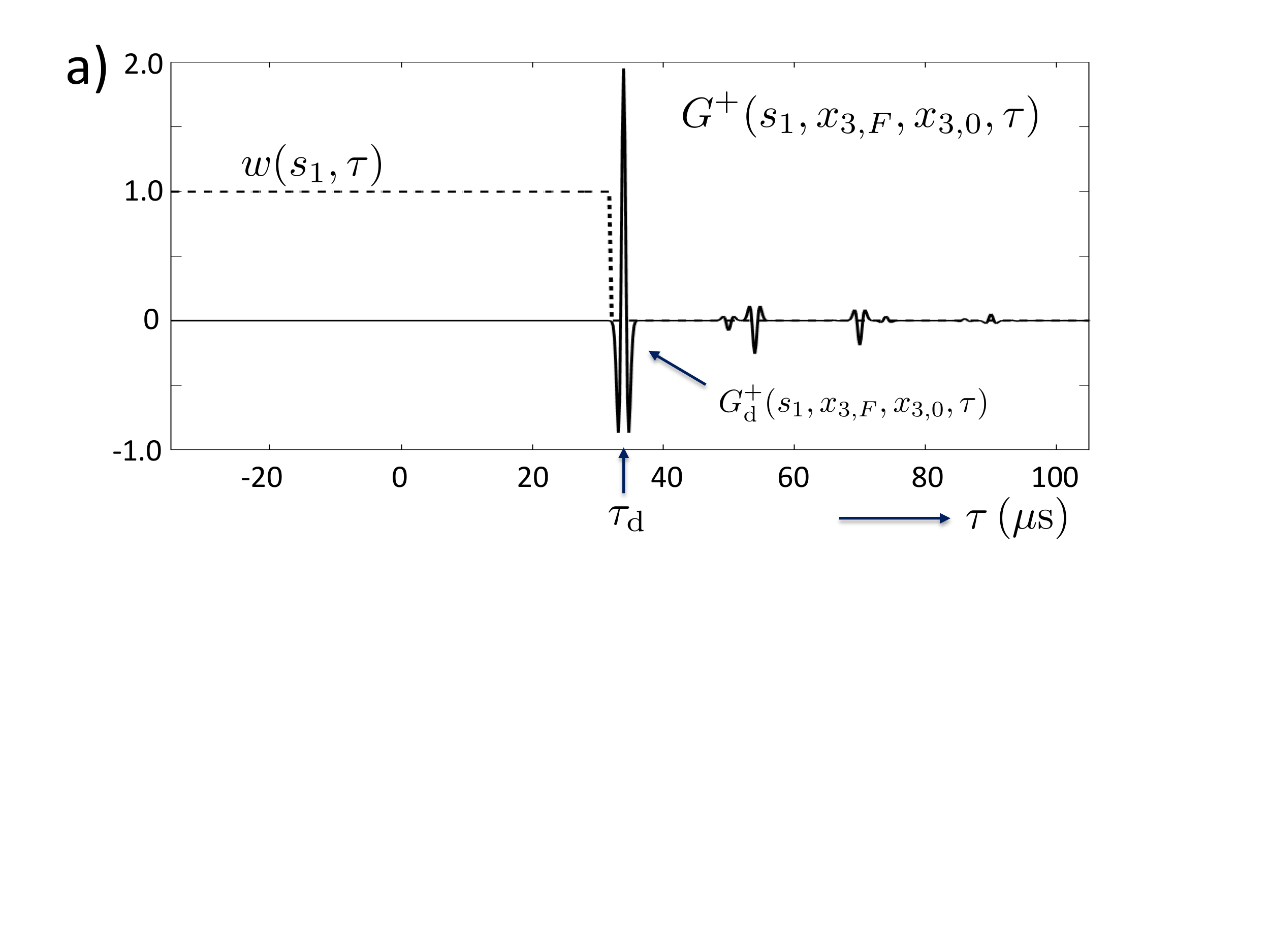}}
\vspace{-2.3cm}
\centerline{\epsfxsize=7.5cm \epsfbox{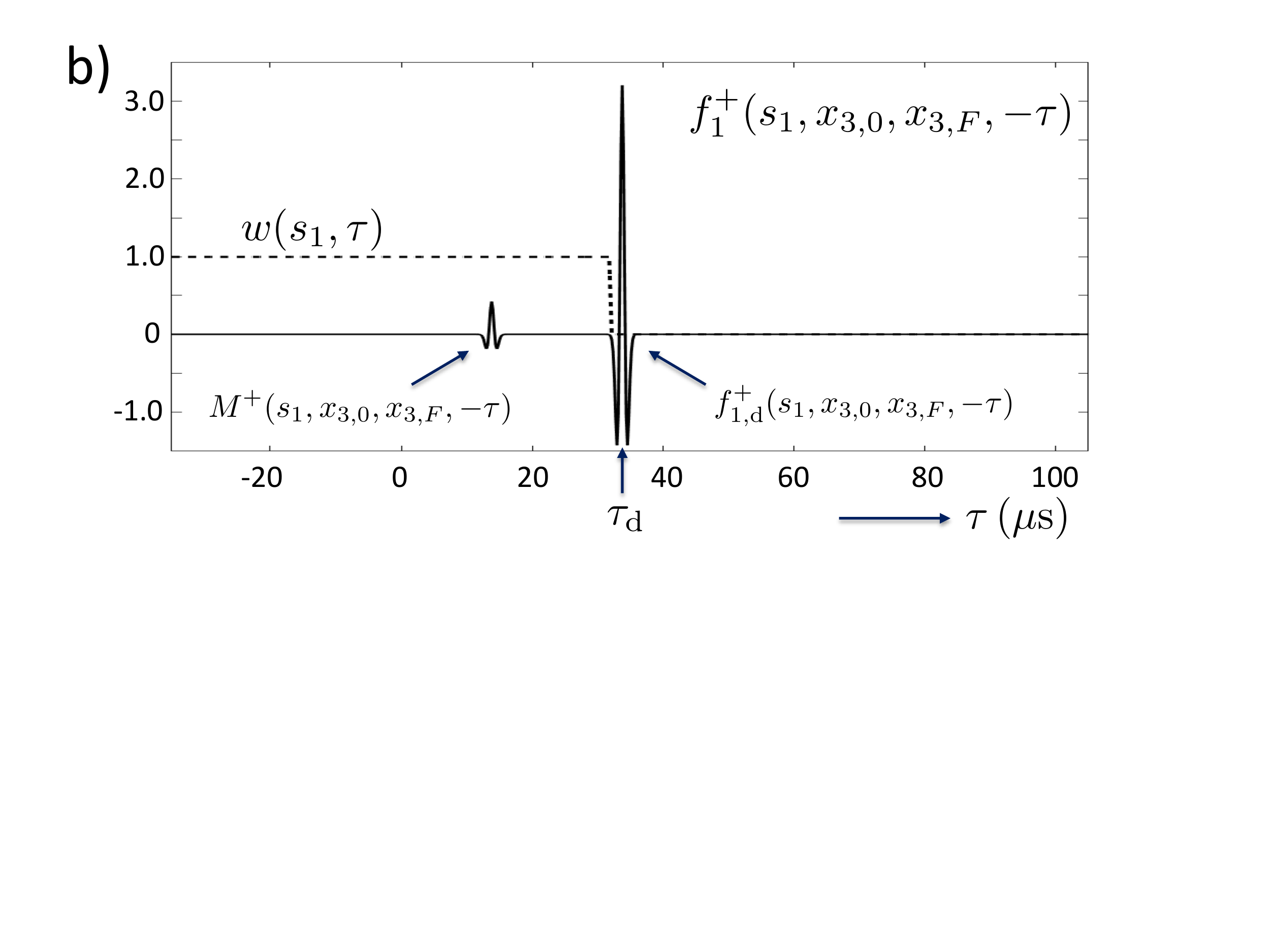}}
\vspace{-2.3cm}
\centerline{\epsfxsize=7.5cm \epsfbox{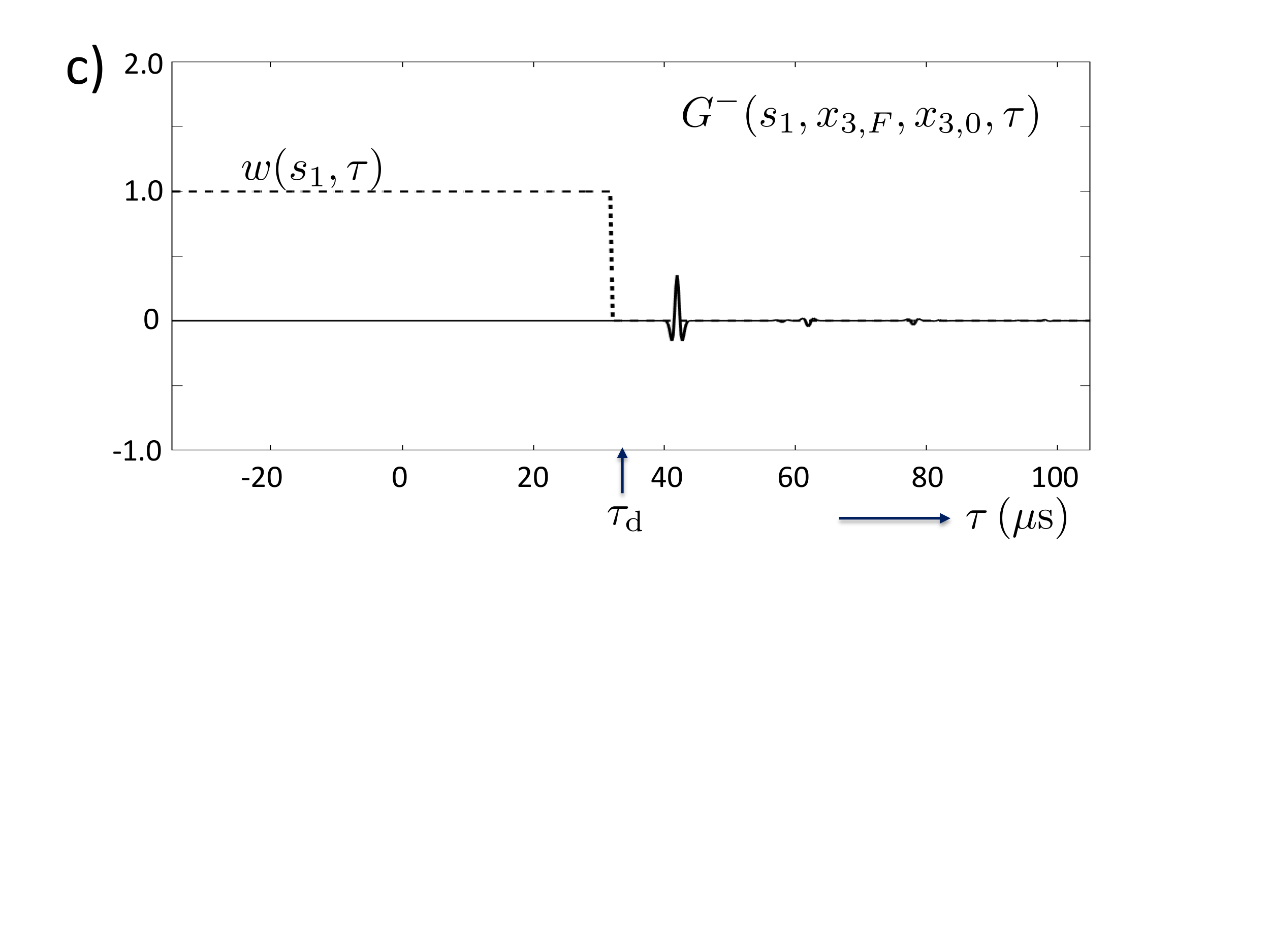}}
\vspace{-2.3cm}
\centerline{\epsfxsize=7.5cm \epsfbox{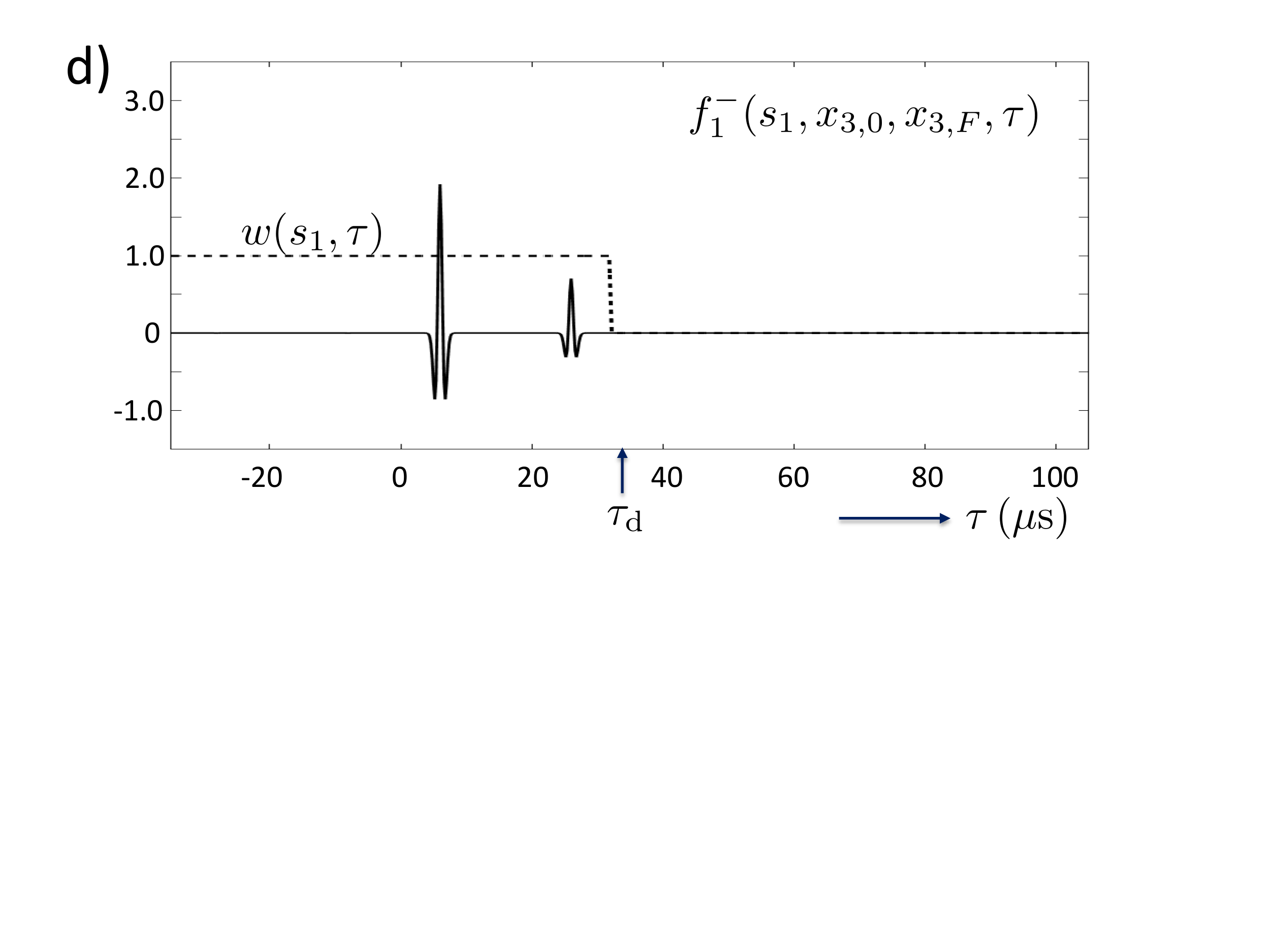}}
\vspace{-2.3cm}
\caption{The Green's functions and focusing functions in the left-hand sides of equations (\ref{eq145tau}) and (\ref{eq146tau}), for the situation of a 
layered medium consisting of DPS layers only. The time window $w(s_1,\tau)$, indicated by the dashed lines, is the same for all functions.}\label{Figure3}
\end{figure}

For the derivation of the Marchenko method we need time windows that suppress the 
Green's functions on the left-hand sides of  representations (\ref{eq145tau}) and (\ref{eq146tau}), so that we are left with
two equations for the two focusing functions. First we briefly review these windows for the situation of a layered medium consisting of DPS layers only.
Figure \ref{Figure3} shows an example of the functions in the left-hand sides of equations (\ref{eq145tau}) and (\ref{eq146tau}) for such a medium.
Note that these functions (represented by the solid lines) have been convolved with a symmetric band-limited wavelet.
{For convenience they have also been multiplied by a factor $2s_3 / \beta_3$ to compensate for the source properties defined in equation (\ref{eq52}).}
Figures \ref{Figure3}(a) and (b) show the functions in the left-hand side of equation (\ref{eq146tau}).
The direct arrival of the downgoing Green's function $G^+(s_1,x_{3,F},x_{3,0},\tau)$ in Figure \ref{Figure3}(a)
coincides with the time-reversed direct arrival of the focusing function $f_1^+(s_1,x_{3,0},x_{3,F},\tau)$ in Figure \ref{Figure3}(b).
For this focusing function  we write 
\begin{eqnarray}
 f_1^+(s_1,x_{3,0},x_{3,F},\tau)&=&f_{1,{\rm d}}^+(s_1,x_{3,0},x_{3,F},\tau)+M^+(s_1,x_{3,0},x_{3,F},\tau),\label{eq65}
\end{eqnarray}
where $f_{1,{\rm d}}^+$ is the direct arrival  and $M^+$  a coda, following the direct arrival
(in Figure \ref{Figure3}(b) this coda is time-reversed and consists of a single event only, but more generally it consists of multiple events).
We define a time window $w(s_1,\tau)=\theta(\tau_{\rm d}(s_1)-\tau_\epsilon-\tau)$, where $\theta(\tau)$ is the Heaviside step function,
 $\tau_{\rm d}(s_1)$ the traveltime of the direct arrival of the downgoing Green's function and $\tau_\epsilon$ is half the duration of the symmetric wavelet.
 This window  is indicated by the dashed lines in Figure \ref{Figure3}. It suppresses the downgoing Green's function in Figure \ref{Figure3}(a)
and  the time-reversed direct arrival of the focusing function in Figure \ref{Figure3}(b). It passes the time-reversed coda $M^+(s_1,x_{3,0},x_{3,F},-\tau)$ in Figure \ref{Figure3}(b).
Figures \ref{Figure3}(c) and (d) show the functions in the left-hand side of equation (\ref{eq145tau}). Since the first arrival of the upgoing Green's function $G^-(s_1,x_{3,F},x_{3,0},\tau)$
arrives later than that of the downgoing Green's function, the time window $w(s_1,\tau)$ suppresses the upgoing Green's function in Figure \ref{Figure3}(c). 
It passes the focusing function $f_1^-(s_1,x_{3,0},x_{3,F},\tau)$ in Figure \ref{Figure3}(d).
Applying the window to both sides of equations (\ref{eq145tau}) and  (\ref{eq146tau}), the windowed equations 
can be solved for $f_1^-(s_1,x_{3,0},x_{3,F},\tau)$ and $M^+(s_1,x_{3,0},x_{3,F},\tau)$, after which the Green's functions follow 
from the unwindowed equations \cite{Slob2014GEO, Wapenaar2014GEO}.

\begin{figure}[b]
\centerline{\epsfxsize=7.5cm \epsfbox{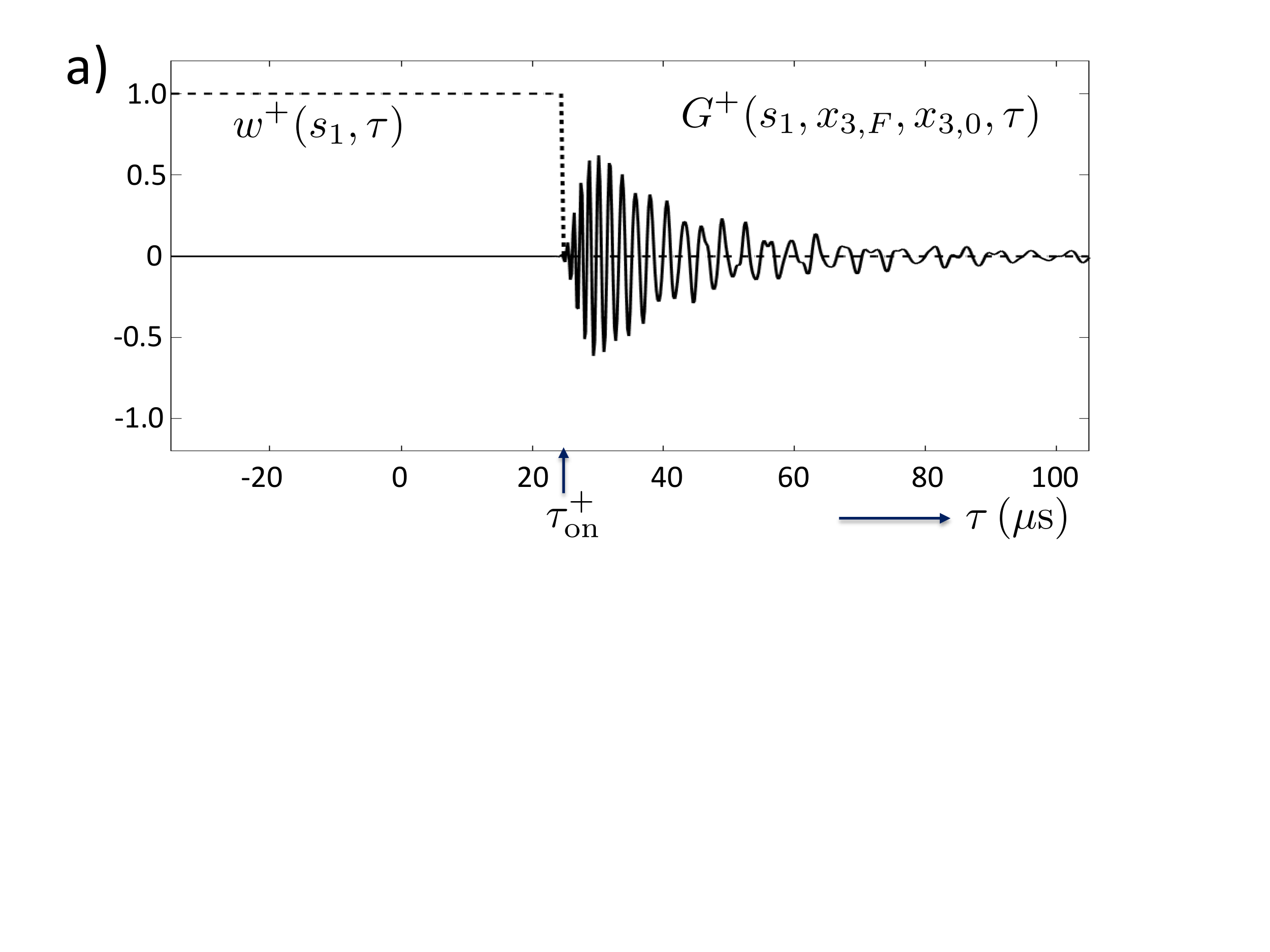}}
\vspace{-2.3cm}
\centerline{\epsfxsize=7.5cm \epsfbox{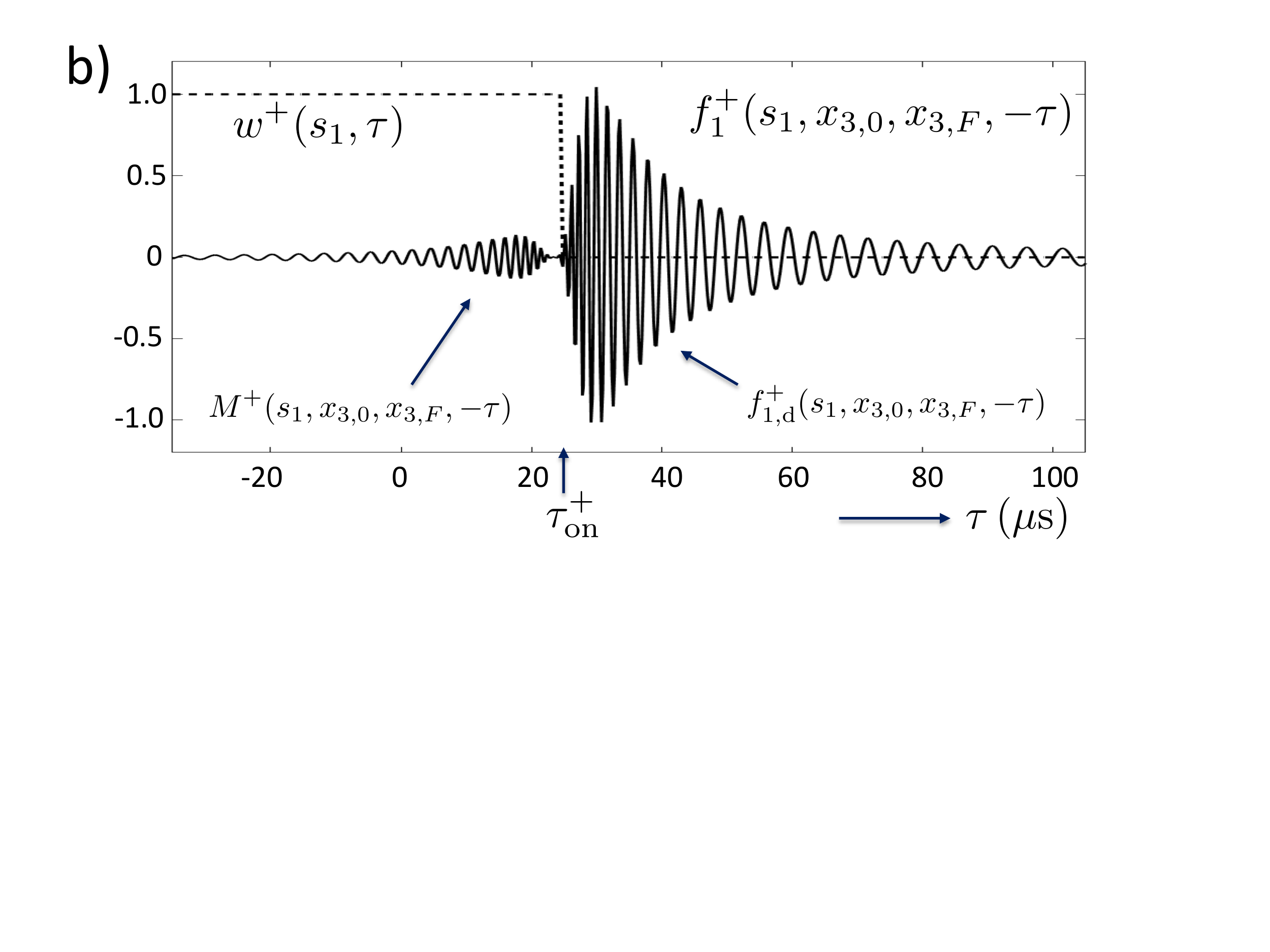}}
\vspace{-2.3cm}
\centerline{\epsfxsize=7.5cm \epsfbox{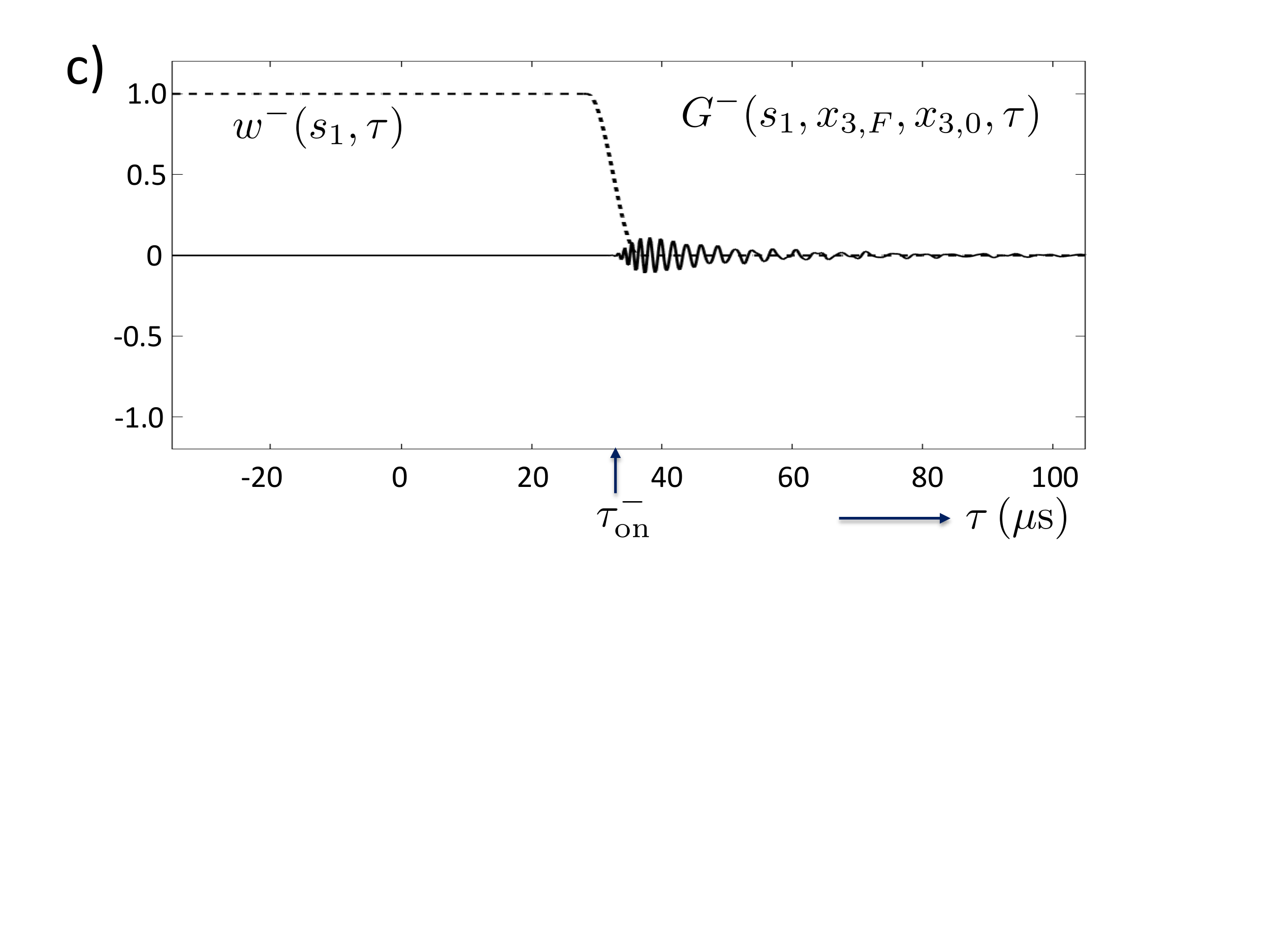}}
\vspace{-2.3cm}
\centerline{\epsfxsize=7.5cm \epsfbox{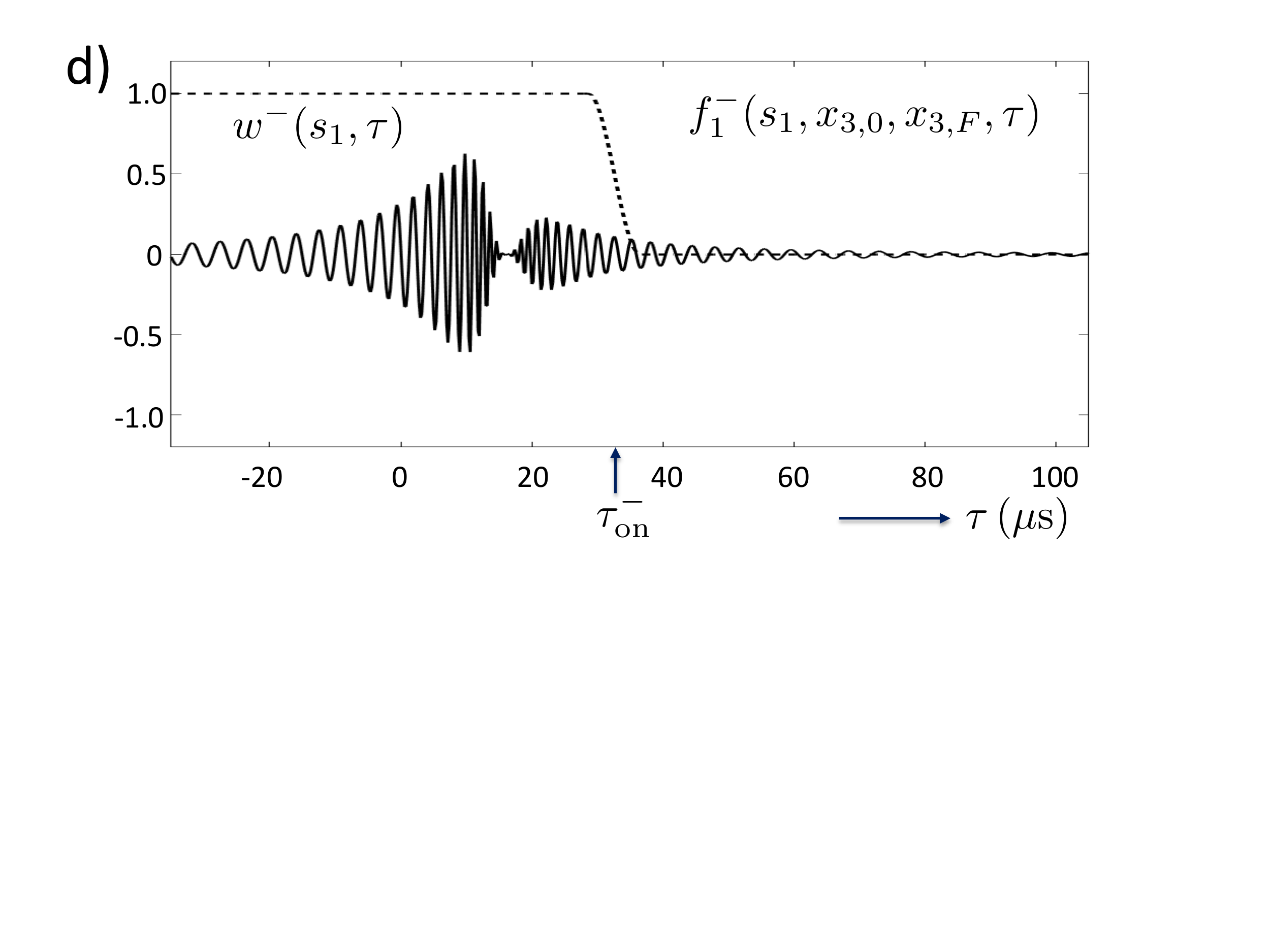}}
\vspace{-2.3cm}
\caption{As Figure \ref{Figure3}, but for the situation of a 
layered medium consisting of DPS and DNG layers. Note the different time windows $w^+(s_1,\tau)$ and $w^-(s_1,\tau)$, indicated by the dashed lines.}\label{Figure4}
\end{figure}

Next we discuss the time windows for the situation of a layered medium consisting of a mix of DPS and DNG layers. 
Figures \ref{Figure4}(a) and (b) show an example of the functions in the left-hand side of equation (\ref{eq146tau}) (again convolved with a symmetric band-limited wavelet
{and multiplied by a factor $2s_3 / \beta_3$}). 
Due to the highly dispersive character of the DNG layers, these functions
are very different from their counterparts in Figures \ref{Figure3}(a) and (b). Nevertheless, for the focusing function $f_1^+(s_1,x_{3,0},x_{3,F},\tau)$ we can again distinguish
between a direct arrival $f_{1,{\rm d}}^+$ and a coda $M^+$, see Figure \ref{Figure4}(b) and equation (\ref{eq65}).
We define a time window 
\begin{eqnarray}
w^+(s_1,\tau)=\theta(\tau_{\rm on}^+(s_1)-\tau),
\end{eqnarray}
where $\tau_{\rm on}^+(s_1)$ is the traveltime of the onset of the downgoing Green's function. 
{This window is indicated by the dashed lines in Figures \ref{Figure4}(a) and \ref{Figure4}(b).}
This window suppresses the downgoing Green's function in Figure \ref{Figure4}(a)
and  the time-reversed direct arrival of the focusing function in Figure \ref{Figure4}(b); it passes the time-reversed coda $M^+(s_1,x_{3,0},x_{3,F},-\tau)$ in Figure \ref{Figure4}(b).
Figures \ref{Figure4}(c) and (d) show the functions in the left-hand side of equation (\ref{eq145tau}). 
The onset time $\tau_{\rm on}^-(s_1)$ of the upgoing Green's function in Figure \ref{Figure4}(c) is larger than that of the downgoing Green's function.
Note, however, that {the dispersive tail of} $f_1^-(s_1,x_{3,0},x_{3,F},\tau)$ in Figure \ref{Figure4}(d) 
exceeds not only the onset time of the downgoing Green's function but also that of the upgoing Green's function.
Hence, the upgoing Green's function in Figure \ref{Figure4}(c) and the focusing function in Figure \ref{Figure4}(d) cannot be uniquely separated by a time window.
We define a time window 
\begin{eqnarray}\label{eq67}
w^-(s_1,\tau)=\theta_{\rm tap}(\tau_{\rm on}^-(s_1)-\tau),
\end{eqnarray}
where $\theta_{\rm tap}(\tau)$ is a tapered step function. It is indicated by the dashed lines in Figures
\ref{Figure4}(c) and \ref{Figure4}(d). The taper should be chosen such that this window  suppresses the upgoing Green's function in Figure \ref{Figure4}(c) as good as possible and 
leaves the focusing function $f_1^-(s_1,x_{3,0},x_{3,F},\tau)$ in Figure \ref{Figure4}(d) as much as possible intact. It is unavoidable that this approach leads to approximations, particularly in the situation of thin layers.

Assuming a proper window function $w^-(s_1,\tau)$ can be found, the application of this window to both sides of
equation (\ref{eq145tau}), and window  $w^+(s_1,\tau)$ to both sides of equation (\ref{eq146tau}), yields the following system of coupled Marchenko equations
for  $f_1^-(s_1,x_{3,0},x_{3,F},\tau)$ and $M^+(s_1,x_{3,0},x_{3,F},\tau)$
\begin{eqnarray}
 f_1^-(s_1,x_{3,0},x_{3,F},\tau)=
w^-(s_1,\tau)\int_{-\infty}^\tau R (s_1,x_{3,0},\tau-\tau')f_1^+(s_1,x_{3,0},x_{3,F},\tau'){\rm d}\tau'\label{eq145tauwin}
\end{eqnarray}
and
\begin{eqnarray}
M^+(s_1,x_{3,0},x_{3,F},-\tau)=
w^+(s_1,\tau)\int_{-\infty}^\tau R (s_1,x_{3,0},\tau-\tau')f_1^-(s_1,x_{3,0},x_{3,F},-\tau'){\rm d}\tau',\label{eq146tauwin}
\end{eqnarray}
with $f_1^+(s_1,x_{3,0},x_{3,F},\tau)$ defined in equation (\ref{eq65}).
This system of equations can be solved by the following iterative scheme
\begin{eqnarray}
 f_{1,k}^-(s_1,x_{3,0},x_{3,F},\tau)=
w^-(s_1,\tau)\int_{-\infty}^\tau R (s_1,x_{3,0},\tau-\tau')f_{1,k}^+(s_1,x_{3,0},x_{3,F},\tau'){\rm d}\tau'\label{eq145tauwiniter}
\end{eqnarray}
and
\begin{eqnarray}
 M_{k+1}^+(s_1,x_{3,0},x_{3,F},-\tau)=
w^+(s_1,\tau)\int_{-\infty}^\tau R (s_1,x_{3,0},\tau-\tau')f_{1,k}^-(s_1,x_{3,0},x_{3,F},-\tau'){\rm d}\tau',\label{eq146tauwiniter}
\end{eqnarray}
with 
\begin{eqnarray}
 f_{1,k}^+(s_1,x_{3,0},x_{3,F},\tau)&=&f_{1,{\rm d}}^+(s_1,x_{3,0},x_{3,F},\tau)
 +M_k^+(s_1,x_{3,0},x_{3,F},\tau),\label{eq147tauwiniter}
\end{eqnarray}
starting with $M_1^+(s_1,x_{3,0},x_{3,F},\tau)=0$. 
Note that this scheme requires the measured reflection response $R (s_1,x_{3,0},\tau)$ of the layered medium, estimates of the onset times $\tau_{\rm on}^+(s_1)$ and  $\tau_{\rm on}^-(s_1)$, and 
an estimate of the direct arrival of the focusing function, $f_{1,{\rm d}}^+(s_1,x_{3,0},x_{3,F},\tau)$.
Assuming a background model of the medium is available, the primary downgoing and upgoing waves can be modelled, from which the onset times can be retrieved.
Moreover, the direct arrival of the focusing function can be obtained, analogous to equation (\ref{eq58}),
 from the direct arrival of the transmission response $\tilde T_{\rm d}(s_1,x_{3,F},x_{3,0},\omega)$ (i.e., the modelled primary downgoing wave), according to
\begin{eqnarray}
\tilde f_{1,{\rm d}}^+(s_1,x_{3,0},x_{3,F},\omega)=
\frac{\beta_3(x_{3,F},\omega)}{2s_3 (s_1,x_{3,F},\omega)\tilde T_{\rm d}(s_1,x_{3,F},x_{3,0},\omega)},\label{eq71c}
\end{eqnarray}
followed by an inverse Fourier transform.

Once the iterative scheme has converged, the retrieved focusing functions can be used in 
representations   (\ref{eq145tau}) and (\ref{eq146tau}) to obtain the decomposed Green's functions
$G^-(s_1,x_{3,F},x_{3,0},\tau)$ and $G^+(s_1,x_{3,F},x_{3,0},\tau)$ and, finally, the total Green's function 
$G(s_1,x_{3,F},x_{3,0},\tau)=G^+(s_1,x_{3,F},x_{3,0},\tau)+G^-(s_1,x_{3,F},x_{3,0},\tau)$. Note that the latter can be interpreted as the response to a source at the surface $x_{3,0}$,
observed by a virtual receiver at $x_{3,F}$ inside the medium. Using reciprocity, 
$G(s_1,x_{3,0},x_{3,F},\tau)$ can be interpreted as the response to a virtual source at  $x_{3,F}$ inside the medium, observed by a receiver at $x_{3,0}$.
The retrieved Green's function contains the direct arrival and the primary and multiple reflections of the layered medium. 
The direct arrival comes from the background model whereas the primary and multiple reflections come from the measured reflection response at the surface.

Next, we show how to obtain the response between a virtual source and a virtual receiver, both inside the medium. To this end, note that 
the decomposed Green's functions are mutually related via \citep{Wapenaar2000SEG, Amundsen2001GEO}
\begin{eqnarray}
G^-(s_1,x_{3,F},x_{3,0},\tau)=
 \int_{-\infty}^\tau R(s_1,x_{3,F},\tau-\tau')G^+(s_1,x_{3,F},x_{3,0},\tau'){\rm d}\tau',\label{eq888}
\end{eqnarray}
where $R(s_1,x_{3,F},\tau)$ is the reflection response at depth level $x_{3,F}$ of the medium below this depth level, assuming a homogeneous medium above this depth level.
By inverting equation (\ref{eq888}), which is done by deconvolution, $R(s_1,x_{3,F},\tau)$ is obtained from $G^-(s_1,x_{3,F},x_{3,0},\tau)$ and $G^+(s_1,x_{3,F},x_{3,0},\tau)$.
This deconvolution process removes all the multiple reflections occurring in the medium above  $x_{3,F}$.
The retrieved reflection response $R(s_1,x_{3,F},\tau)$ can be interpreted as the response to a virtual source for downgoing waves at $x_{3,F}$, observed by a virtual receiver for upgoing waves at $x_{3,F}$.

\section{Numerical example}\label{sec5}

\begin{figure}[t]
\vspace{-.5cm}
\centerline{\epsfxsize=9cm \epsfbox{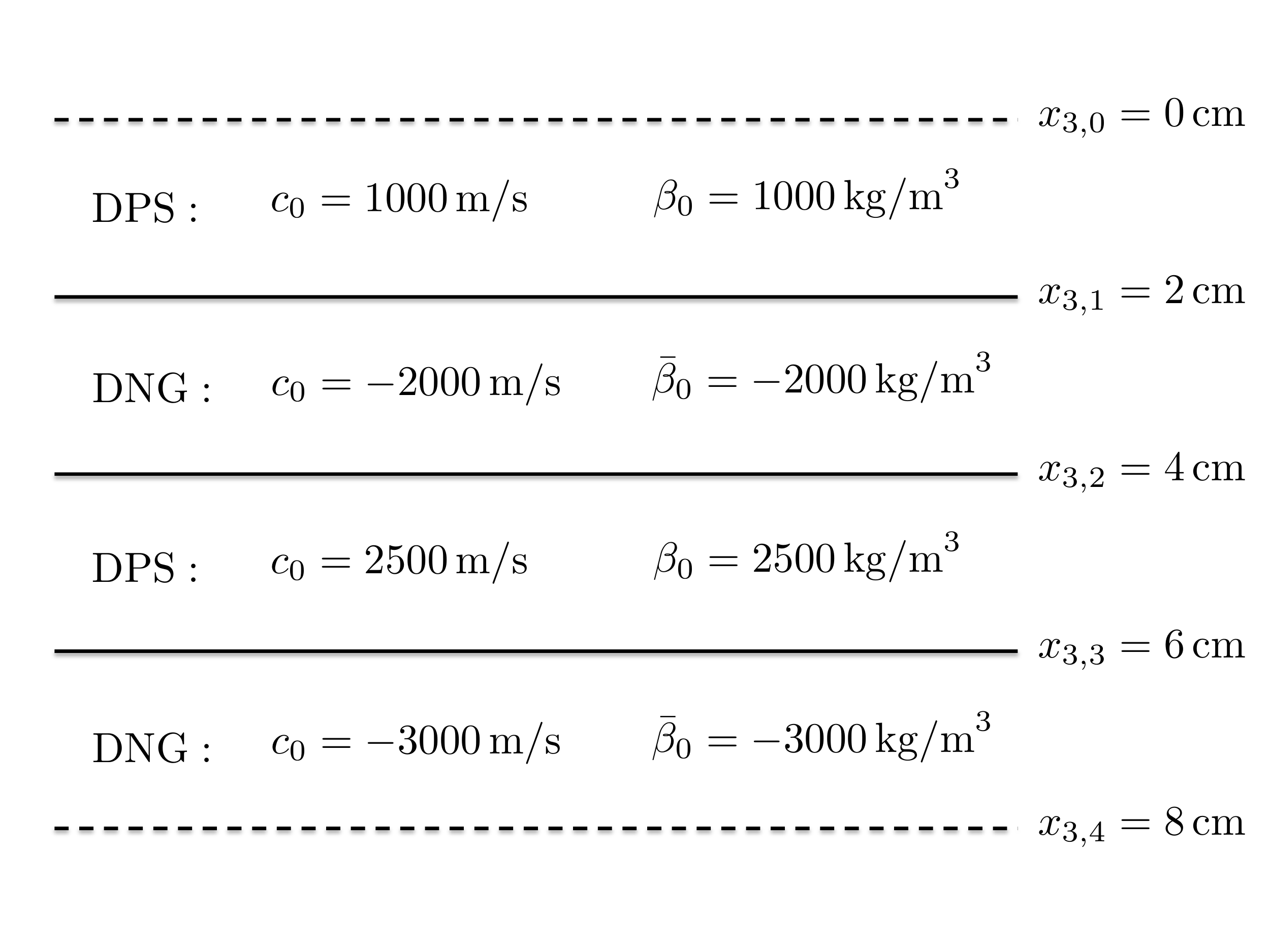}}
\vspace{-.5cm}
\caption{Layered acoustic medium, consisting of DPS and DNG layers.}\label{Fig3}
\end{figure}

We illustrate the Marchenko method with a numerical example for a horizontally layered acoustic medium consisting of a mix of DPS and DNG layers, see Figure \ref{Fig3}.
All layers are homogeneous and isotropic, with  $\beta_1=\beta_3=\beta$, where $\beta$ stands for the mass density (see Table 1).

The DPS layers consist of natural non-dispersive materials, with $h_\alpha(\omega)=h_\beta(\omega)=1$. Hence, according to equations (\ref{eqalpha}) and (\ref{eqbeta}),
the layer parameters simplify to $\alpha(\omega)=\alpha_0$ and $\beta(\omega)=\beta_0$. In Figure \ref{Fig3}, the parameters of the DPS layers are the mass density $\beta_0$ and the
phase velocity $c_0=(\alpha_0\beta_0)^{-\frac{1}{2}}$. 

The DNG layers consist of dispersive metamaterials. For these layers we use the Drude model of 
equations (\ref{eqdrudea}) and (\ref{eqdrudeb}), with  $\omega_\alpha=\omega_\beta=\omega_0$ and $\Gamma_\alpha=\Gamma_\beta=\Gamma$. Hence,
$h_\alpha(\omega)=h_\beta(\omega)=h(\omega)=1-\frac{\omega_0^2}{\omega(\omega+i\Gamma)}$. For low frequencies ($\omega<<\omega_0$) this is approximated by $h(\omega)=-\frac{\omega_0^2}{\omega(\omega+i\Gamma)}$.
Using equations (\ref{eqalpha}) and (\ref{eqbeta}), we write
\begin{eqnarray}
\alpha(\omega)&=&\alpha_0 h(\omega)=\bar\alpha_0\bar h(\omega),\\
\beta(\omega)&=&\beta_0 h(\omega)=\bar\beta_0\bar h(\omega),
\end{eqnarray}
with
\begin{eqnarray}
\bar\alpha_0&=&-\alpha_0\frac{\omega_0^2}{\omega_c^2}<0,\\
\bar\beta_0&=&-\beta_0\frac{\omega_0^2}{\omega_c^2}<0,\\
\bar h(\omega)&=&-h(\omega)\frac{\omega_c^2}{\omega_0^2}=\frac{\omega_c^2}{\omega(\omega+i\Gamma)},
\end{eqnarray}
where $\omega_c$ is the central angular frequency of the wave fields that will be considered. For consistency with the low frequency assumption, we assume $\omega_c<<\omega_0$.
In Figure \ref{Fig3}, the parameters of the DNG layers are the mass density $\bar\beta_0$ and the phase velocity $c_0=-(\bar\alpha_0\bar\beta_0)^{-\frac{1}{2}}$. 
The parameter $\Gamma$ in the DNG layers is set to $\Gamma=\omega_c/1000$. 

\begin{figure}[t]
\centerline{\epsfxsize=7.5cm \epsfbox{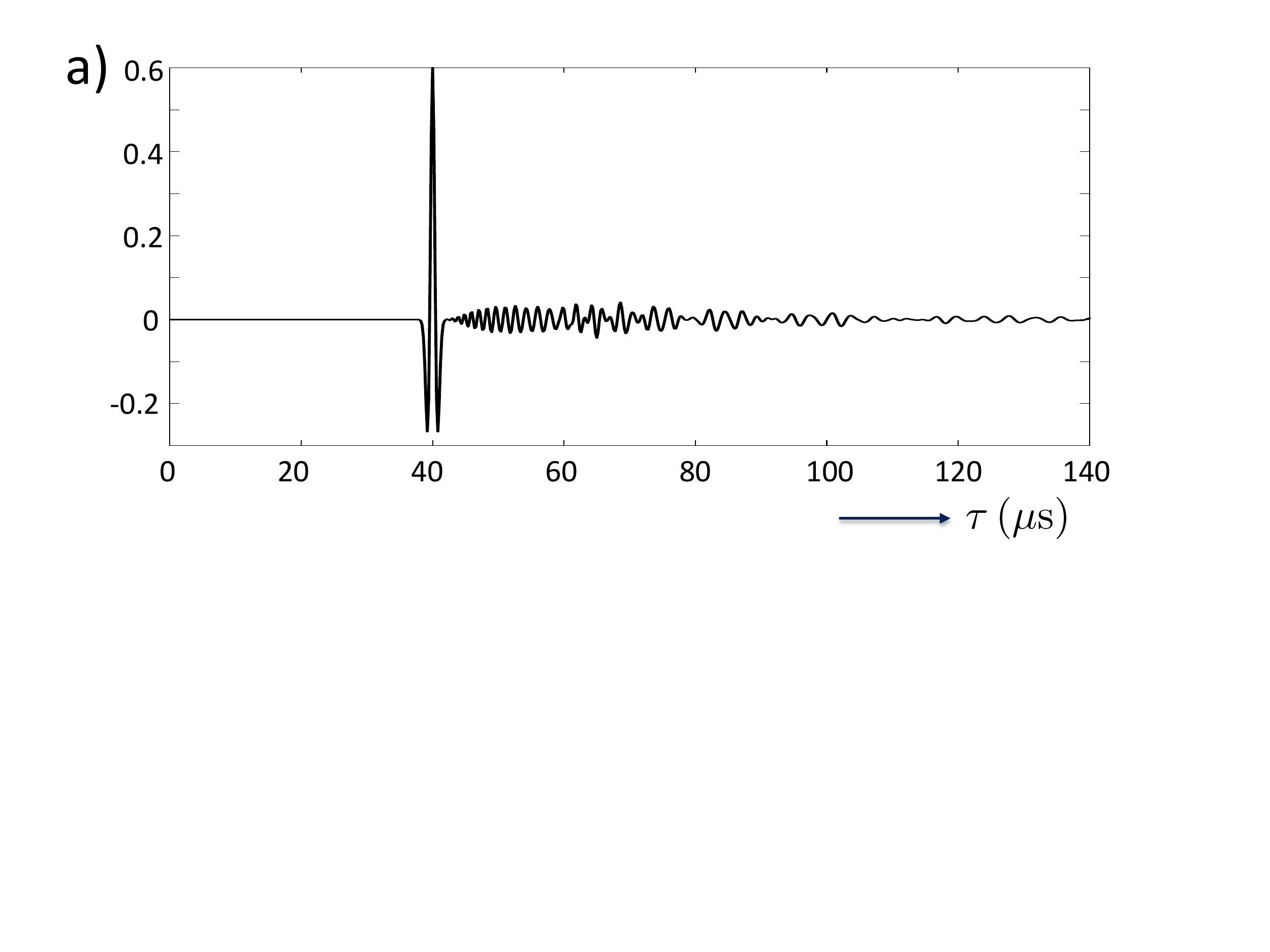}}
\vspace{-2.5cm}
\centerline{\epsfxsize=7.5cm \epsfbox{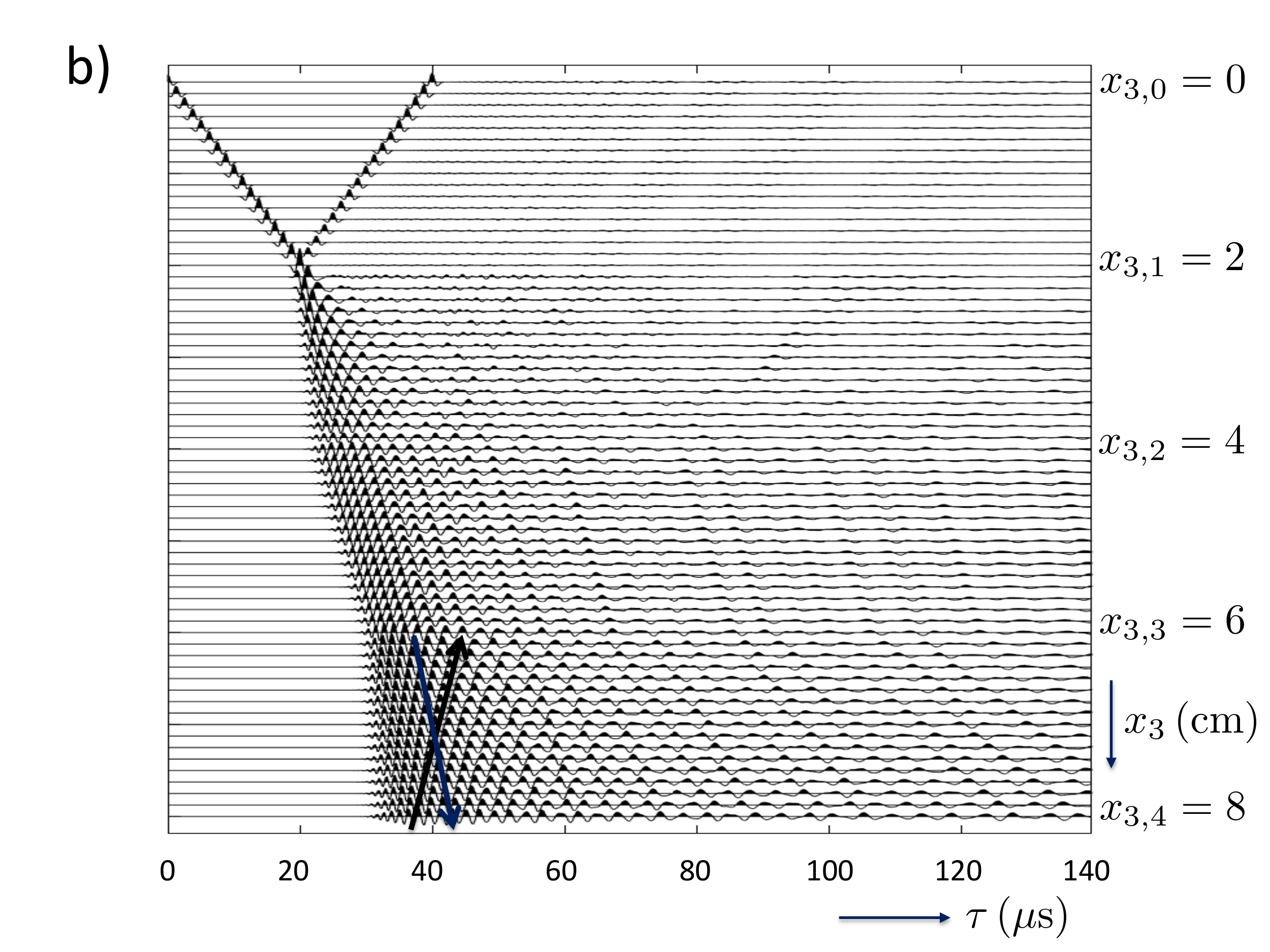}}
\centerline{\epsfxsize=7.5cm \epsfbox{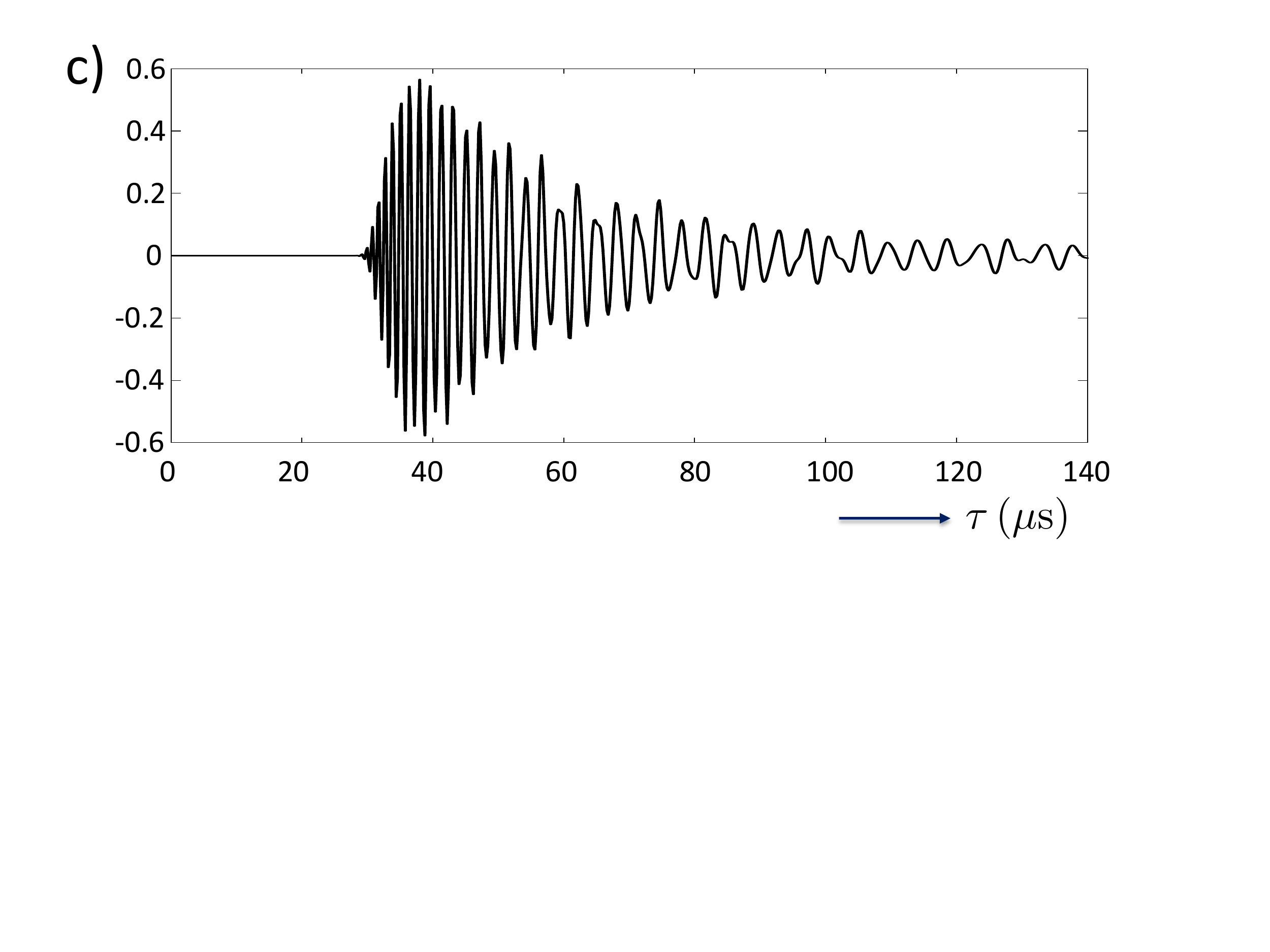}}
\vspace{-2.5cm}
\centerline{\epsfxsize=7.5cm \epsfbox{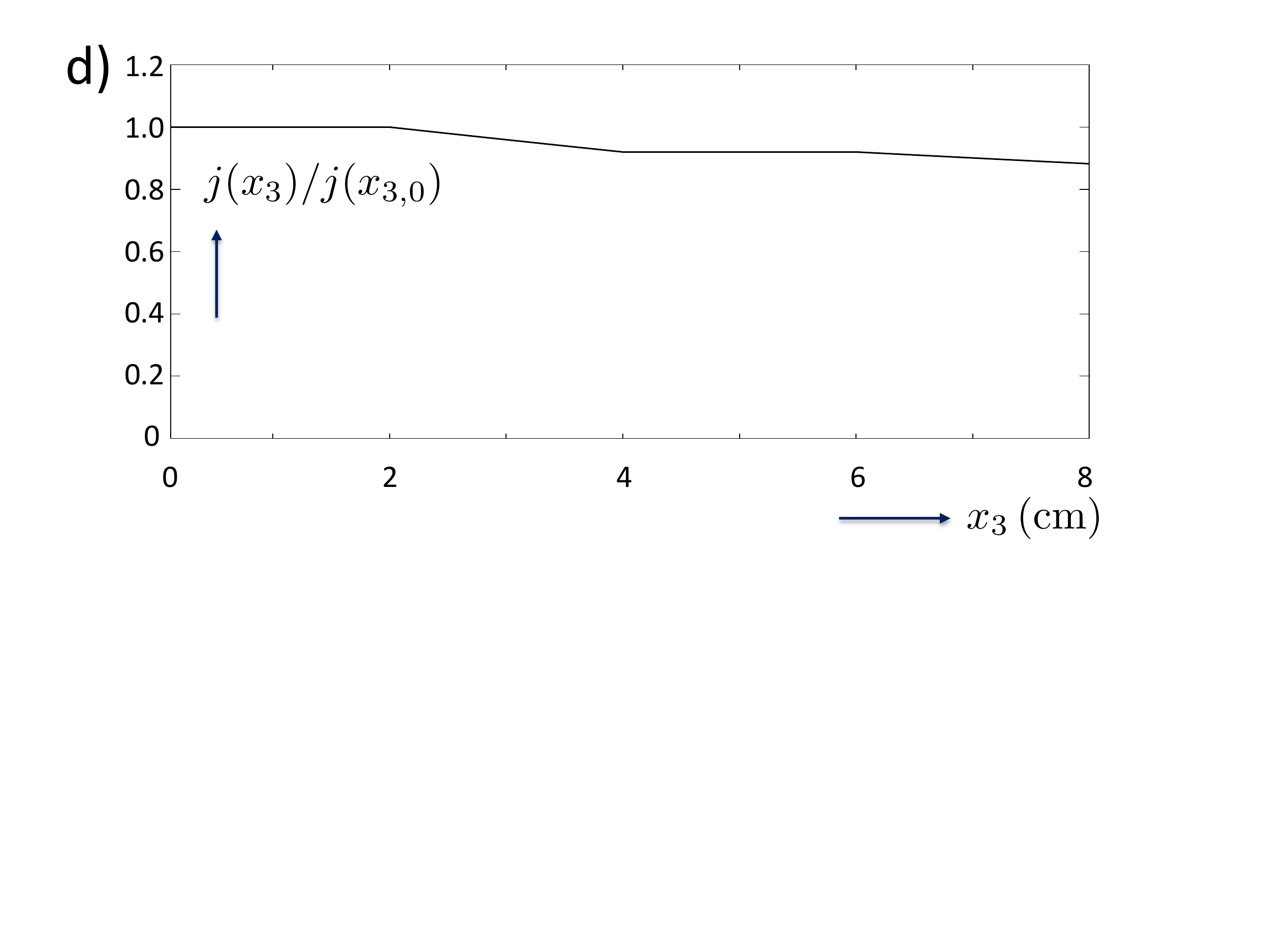}}
\vspace{-2.5cm}
\caption{Modelled wave field in the layered medium of Figure \ref{Fig3}. (a) Reflection response $R(s_1=0,x_{3,0},\tau)$.
(b) Green's function $G(s_1=0,x_3,x_{3,0},\tau)$. (c) Transmission response $T(s_1=0,x_{3,4},x_{3,0},\tau)$.
(d) Power flux density as a function of $x_3$.}\label{Fig4}
\end{figure}

\begin{figure}[t]
\centerline{\epsfxsize=7.5cm \epsfbox{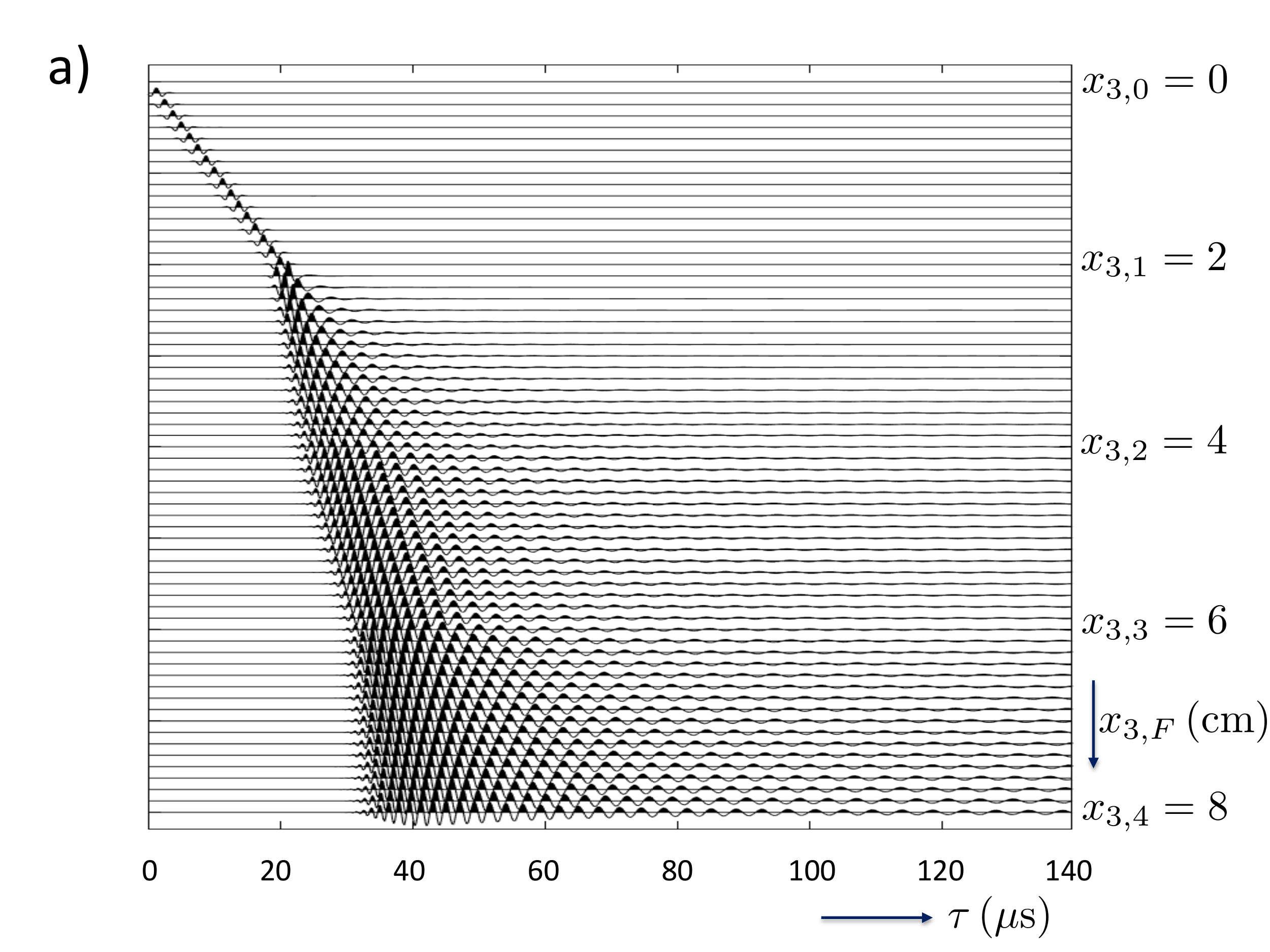}}
\centerline{\epsfxsize=7.5cm \epsfbox{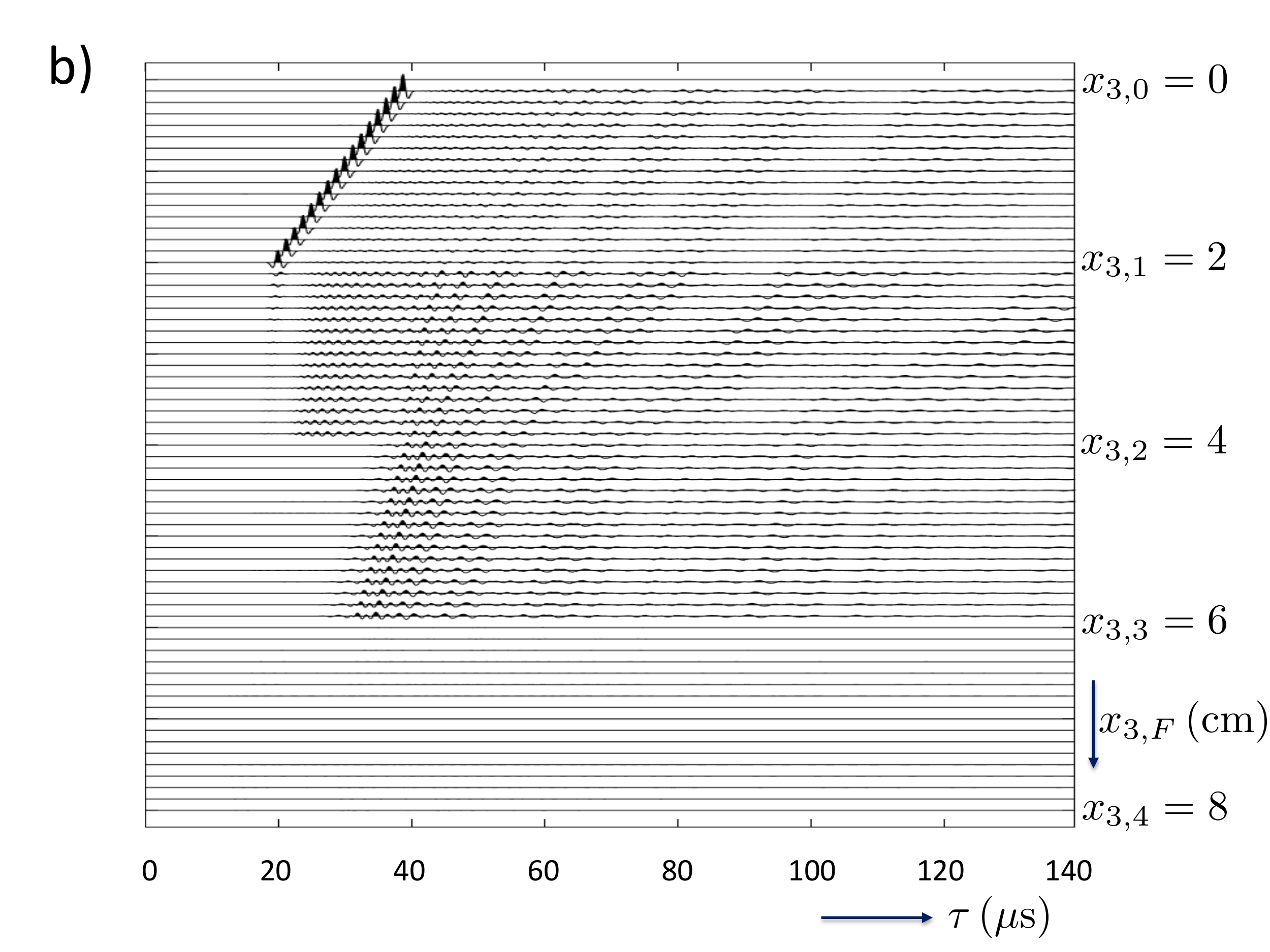}}
\centerline{\epsfxsize=7.5cm \epsfbox{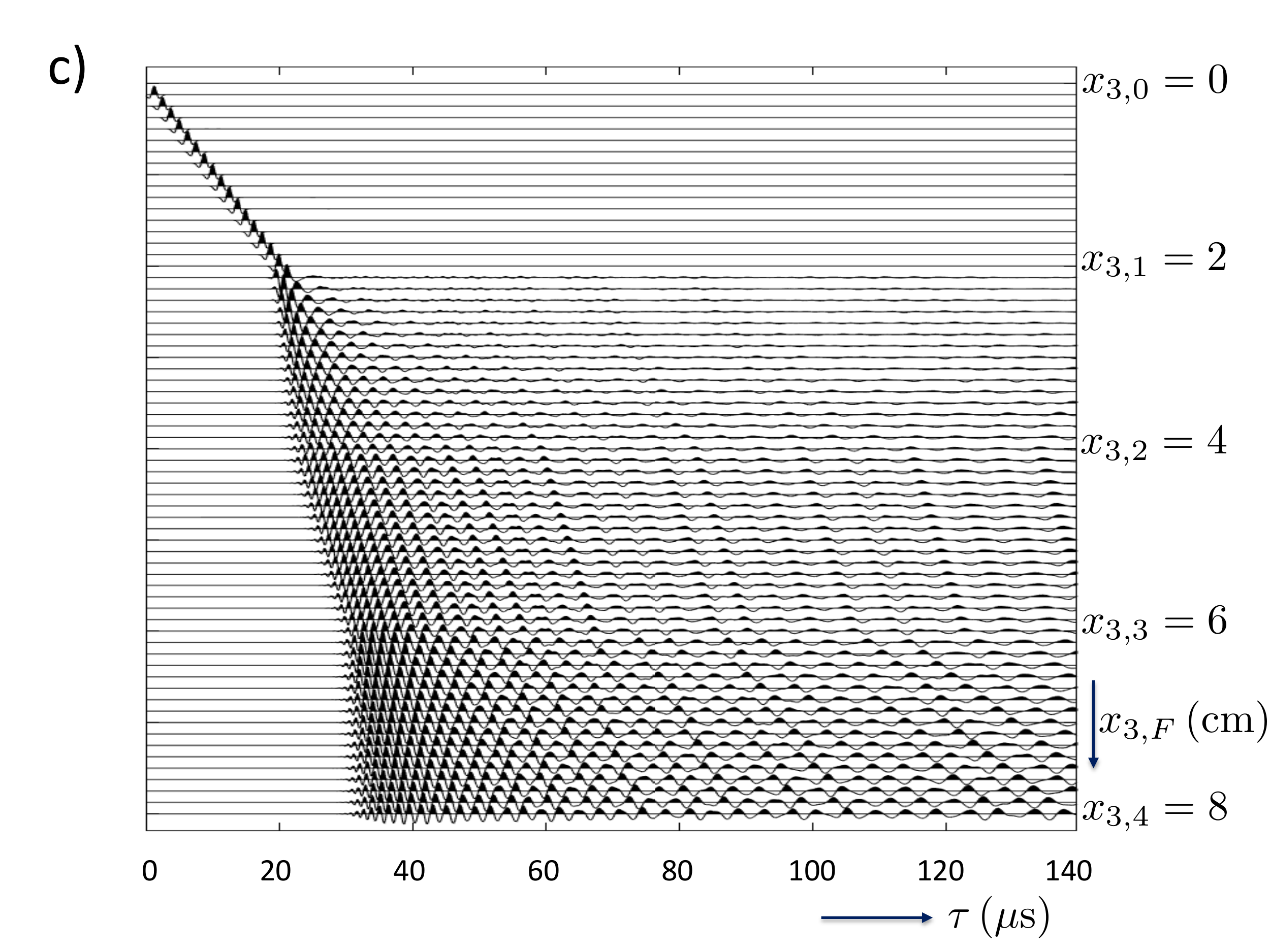}}
\caption{Wavefield retrieval with the Marchenko method. (a) Time-reversed direct arrival of the focusing function, $f_{1,{\rm d}}^+(s_1=0,x_{3,0},x_{3,F},-\tau)$.
(b) Retrieved upgoing Green's function $G^-(s_1=0,x_{3,F},x_{3,0},\tau)$.
(c) Retrieved downgoing Green's function $G^+(s_1=0,x_{3,F},x_{3,0},\tau)$.}\label{Fig5}
\end{figure}

We consider a band-limited ultrasonic downgoing plane wave, incident to the layered medium at $x_{3,0}=0$ cm.  
The source function of the incident wave is defined as 
$S(\tau)= (1-\omega_c^2\tau^2/2)\exp(-\omega_c^2\tau^2/4)$ (a so-called Ricker wavelet), with a central frequency $\omega_c/2\pi=500$ kHz.
{Note that this wavelet is symmetric in time.}
We use a wavenumber-frequency domain modelling method \citep{Kennett79GJRAS}, adjusted for metamaterials, to model the response to this plane wave. 
For the moment we consider vertically propagating plane waves, hence, we take $s_1=0$.
The modelled reflection response $R(s_1=0,x_{3,0},\tau)$, convolved with the wavelet $S(\tau)$, is shown in Figure \ref{Fig4}(a).
This figure clearly shows the non-dispersed reflection at 40 $\mu$s from the first layer interface at $x_{3,1}=2$ cm.
It also shows the dispersed reflections from deeper interfaces, including multiple reflections between the interfaces. 
Figure \ref{Fig4}(b) is the modelled  Green's function 
$G(s_1=0,x_3,x_{3,0},\tau)$ inside the medium, as a function of depth $x_3$ and time $\tau$. This serves as a reference for the results we will obtain with the Marchenko method.
To make the later arrivals visible, a time-dependent amplitude gain of $\exp(2.5\tau/\tau_{\rm max})$, with $\tau_{\rm max}=140\,\mu$s,  has been applied in this display.
This figure shows how the wave field propagates 
through the layers and scatters at the interfaces. 
The downward and upward pointing arrows in the deepest layer indicate the opposite group and phase propagation directions. 
The upward propagating part of the upper trace in this figure is proportional to the reflection response $R(s_1=0,x_{3,0},\tau)$ (see equation (\ref{eq54})), which is shown separately in Figure \ref{Fig4}(a).
Similarly, the lower trace  is proportional to the transmission response $T(s_1=0,x_{3,4},x_{3,0},\tau)$, with $x_{3,4}=8$ cm, which is shown separately in Figure \ref{Fig4}(c).
The trace at $x_3=5$ cm is equal to the superposition of Figures \ref{Figure4}(a) and (c).
Figure \ref{Fig4}(d) shows the power flux density $j(x_3)$, defined in equation (\ref{eq511}), divided by $j(x_{3,0})$. In the DNG layers it decreases because $\Gamma\ne 0$ in these layers.
In a lossless medium $j(x_3)$ would be constant, i.e., propagation invariant.
Its deviation from being constant implies that the Marchenko method, which is based on propagation invariants, cannot lead to exact results.
{Since in this example the losses are small, the effects on the results of the Marchenko method are limited.}

\begin{figure}[t]
\centerline{\epsfxsize=6.5cm \epsfbox{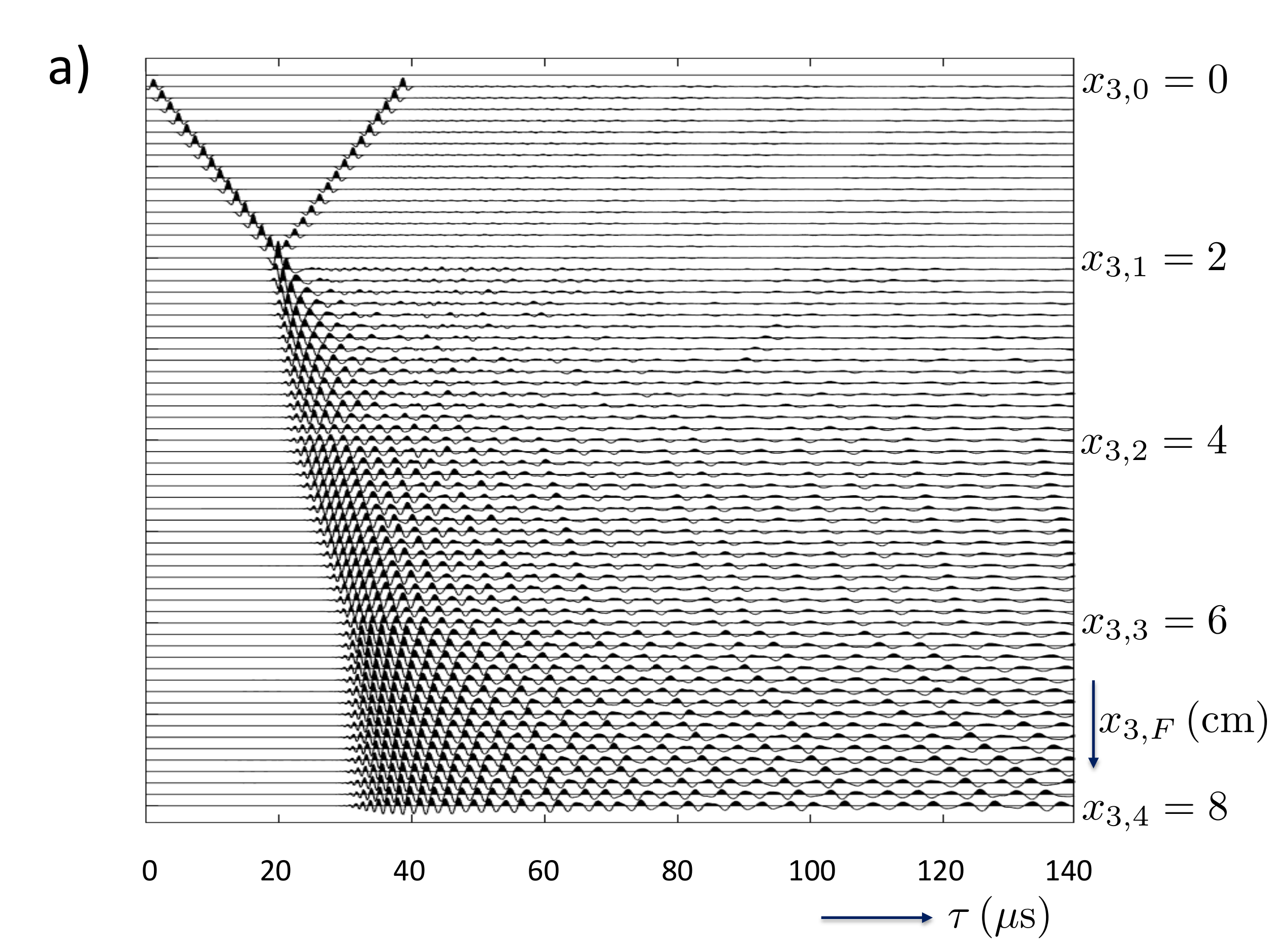}}
\centerline{\epsfxsize=6.5cm \epsfbox{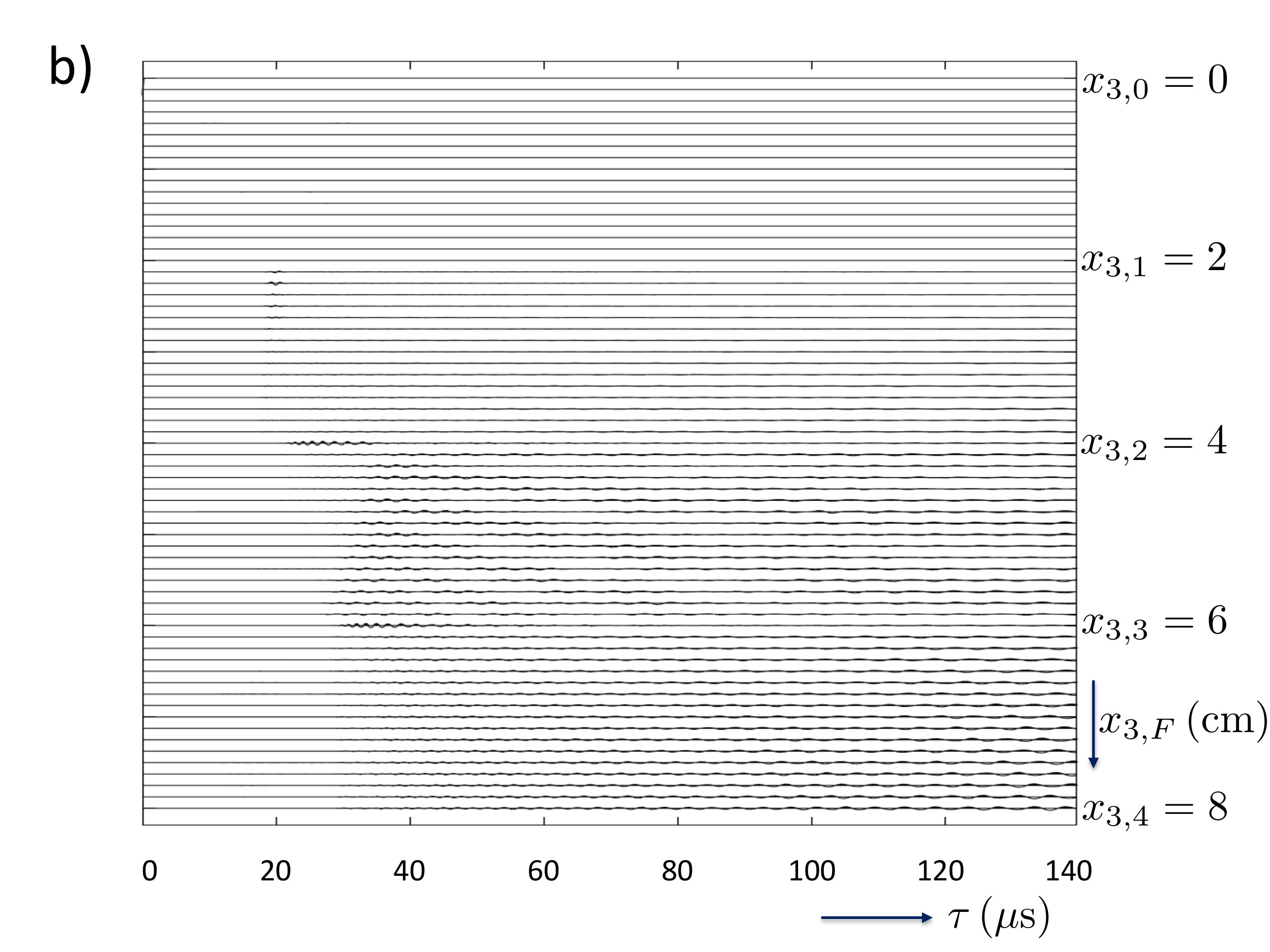}}
\centerline{\epsfxsize=6.5cm \epsfbox{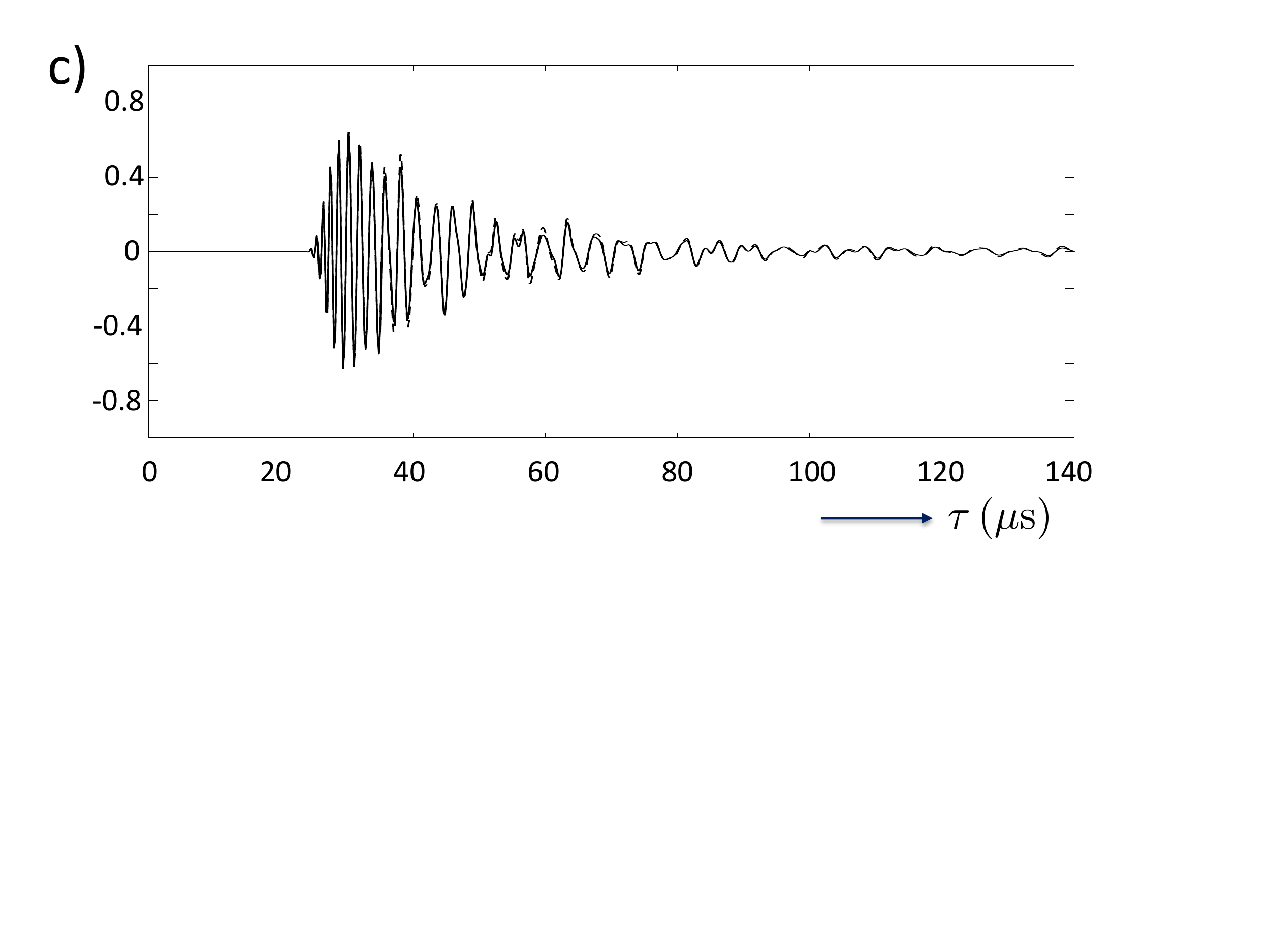}}
\vspace{-2.3cm}
\centerline{\epsfxsize=6.5cm \epsfbox{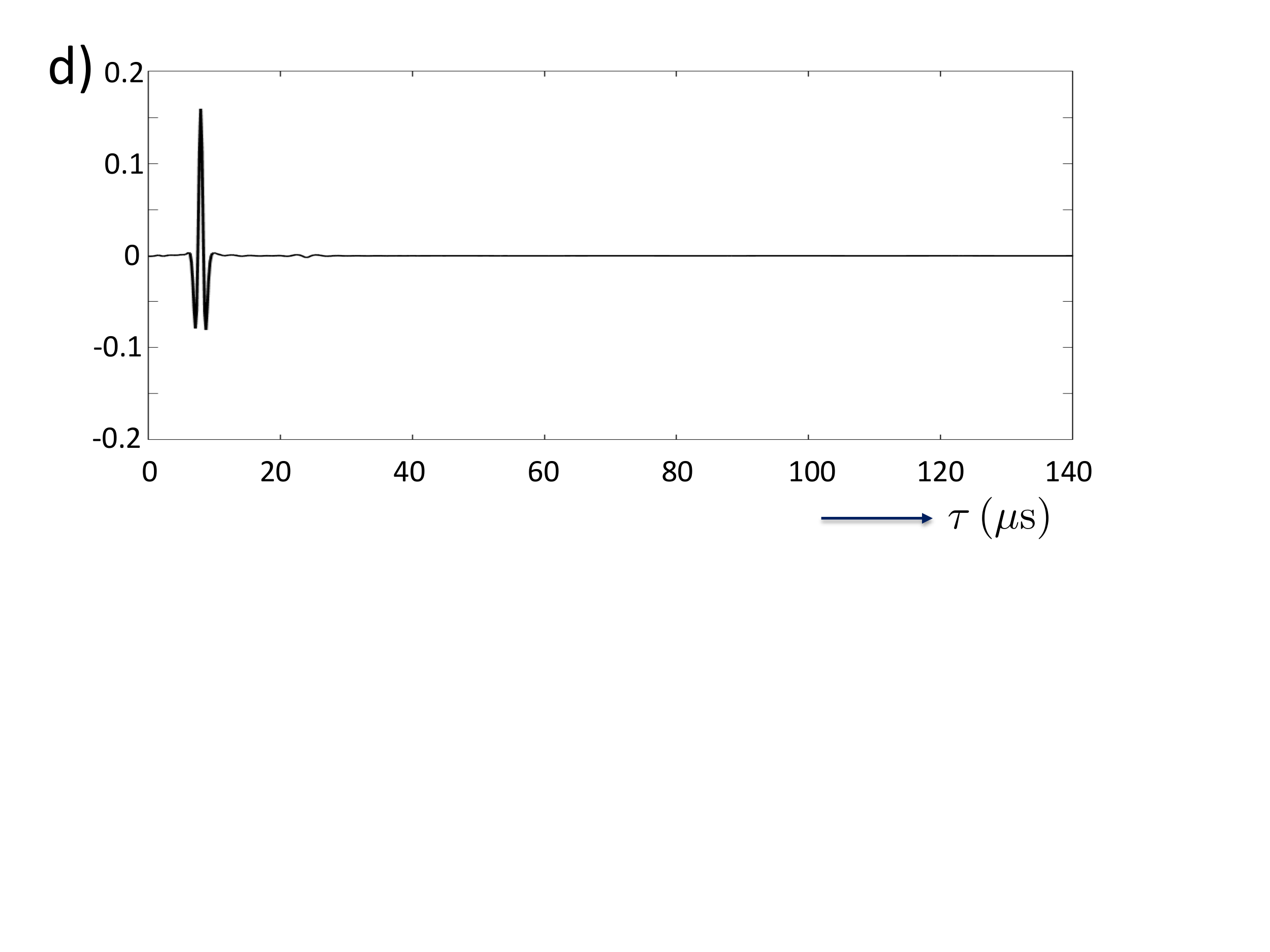}}
\vspace{-2.3cm}
\caption{Wavefield retrieval with the Marchenko method (continued). (a) Total  {retrieved} Green's function $G(s_1=0,x_{3,F},x_{3,0},\tau)=
G^+(s_1=0,x_{3,F},x_{3,0},\tau)+G^-(s_1=0,x_{3,F},x_{3,0},\tau)$.
(b) Difference of the retrieved Green's function with the directly modelled Green's function of Figure \ref{Fig4}(b). 
{(c) Overlay of directly modelled (solid) and retrieved (dashed) Green's functions at $x_{3,F}=5$ cm.
(d)} Retrieved reflection response {$R(s_1=0,x_{3,F},\tau)$} at $x_{3,F}=5$ cm of the interface at $x_{3,3}=6$ cm. 
}\label{Fig6}
\end{figure}

We use the Marchenko method discussed in section \ref{sec4} to retrieve the Green's function $G(s_1=0,x_{3,F},x_{3,0},\tau)$
 inside the medium from the reflection response $R(s_1=0,x_{3,0},\tau)$, shown in Figure \ref{Fig4}(a).
 Apart from the reflection response, we also need an estimate of the direct arrival of the focusing function, 
$f_{1,{\rm d}}^+(s_1=0,x_{3,0},x_{3,F},\tau)$, which, according to equation (\ref{eq71c}), follows from the inverse of the direct arrival of the transmission response.
For this direct arrival we need a background model of the medium. For the moment we use the exact model, but in a later example we replace it by an approximate model. 
Figure \ref{Fig5}(a) shows $f_{1,{\rm d}}^+(s_1=0,x_{3,0},x_{3,F},-\tau)$ (convolved with {the symmetric wavelet} $S(\tau)$) for variable $x_{3,F}$. 
The trace at $x_{3,F}=5$ cm is equal to {the direct arrival of the time-reversed focusing function in} Figure \ref{Figure4}(b).
Given the reflection response at the surface (Figure \ref{Fig4}(a)), the {time-reversed} direct arrival of the focusing function (Figure \ref{Fig5}(a)), 
and the depth-dependent onset times $\tau_{\rm on}^+$ and  $\tau_{\rm on}^-$,
we apply the iterative Marchenko scheme of equations (\ref{eq145tauwiniter}) $-$ (\ref{eq147tauwiniter}) for 64 focal depths, ranging from $x_{3,F}=$1.25 mm to $x_{3,F}=$8 cm, 
with steps  $\Delta x_{3,F}=$1.25 mm.
{For the window $w^-$ in equation (\ref{eq145tauwiniter}) we use a cosine-square taper with a length of $\frac{8\pi}{\omega_c}$ s
(except  in the upper DPS layer, where we replace this window by $w^+$). The length of the taper appears to have no strong effect on the results of the method.} 
For each focal depth $x_{3,F}$ we apply five iterations. 
{Actually for this relatively simple situation the method converges already after two iterations and it remains stable even after 100 iterations.}
The obtained focusing functions  $f_1^-(s_1=0,x_{3,0},x_{3,F},\tau)$ and
$f_1^+(s_1=0,x_{3,0},x_{3,F},\tau)$ are subsequently used in equations (\ref{eq145tau}) and (\ref{eq146tau}) to obtain the Green's functions
$G^-(s_1=0,x_{3,F},x_{3,0},\tau)$ and $G^+(s_1=0,x_{3,F},x_{3,0},\tau)$. These are shown in Figures \ref{Fig5}(b) and \ref{Fig5}(c) for variable $x_{3,F}$. 
Superposing these results yields the total Green's function  $G(s_1=0,x_{3,F},x_{3,0},\tau)$, see Figure \ref{Fig6}(a).
Figure \ref{Fig6}(b) shows the difference of this retrieved Green's function with the directly modelled Green's function in Figure \ref{Fig4}(b)
(the same time-dependent amplitude gain has been applied {in this display} as in the other figures).
Note that the difference is overall small. 
{This is also seen in Figure \ref{Fig6}(c), which shows a comparison of the directly modelled and retrieved Green's functions at $x_{3,F}=$ 5 cm. The phases match very well and the
 amplitudes deviate typically a few percent, with some outliers in the order of 10 $\%$.}
From the decomposed Green's functions we can retrieve the reflection response $R(s_1=0,x_{3,F},\tau)$ for any $x_{3,F}$ by inverting equation (\ref{eq888}).
The retrieved response for $x_{3,F}=5$ cm is shown in Figure \ref{Fig6}({d}).
This is the reflection response of the third interface at $x_{3,3}=6$ cm in Figure \ref{Fig3}, measured with a virtual source and a virtual receiver 1 cm above this interface.
We observe a single primary reflection event at 8.0 $\mu$s; the {dispersion effects of the overlying DNG layer and the} 
multiple reflections occurring in the medium above $x_{3,F}$ have been properly removed
(apart from a very small remnant of a multiple reflection at approximately 24.0 $\mu$s).
The amplitude of the reflection event at 8.0 $\mu$s is 0.159, which is a slight underestimation of the true reflection coefficient  $r_3=0.180$ for $\omega=\omega_c$. 
This discrepancy is due to the loss occurring in the DNG layer between $x_{3,1}=2$ cm and $x_{3,2}=4$ cm. 

\begin{figure}[t]
\centerline{\epsfxsize=7.5cm \epsfbox{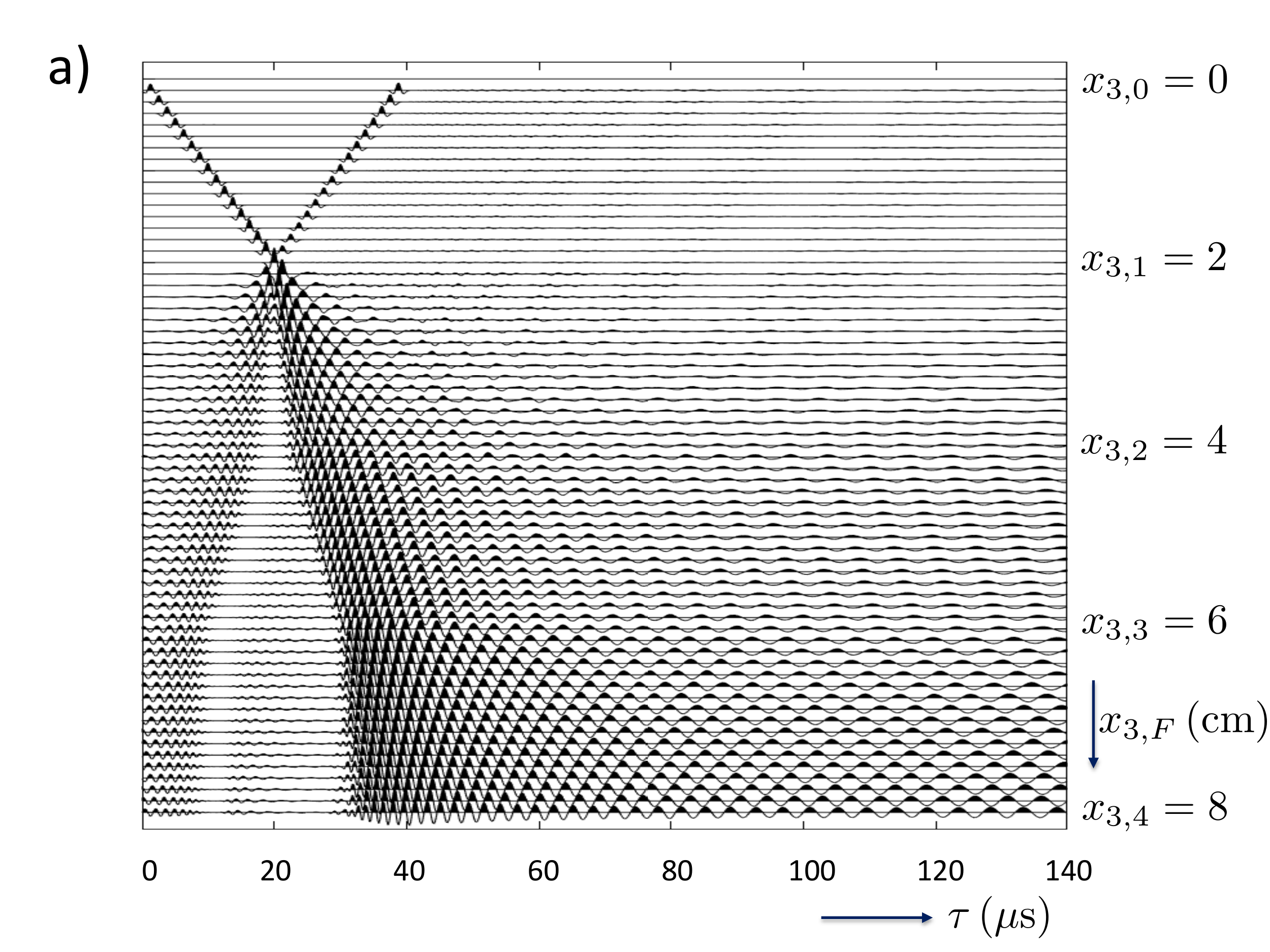}}
\centerline{\epsfxsize=7.5cm \epsfbox{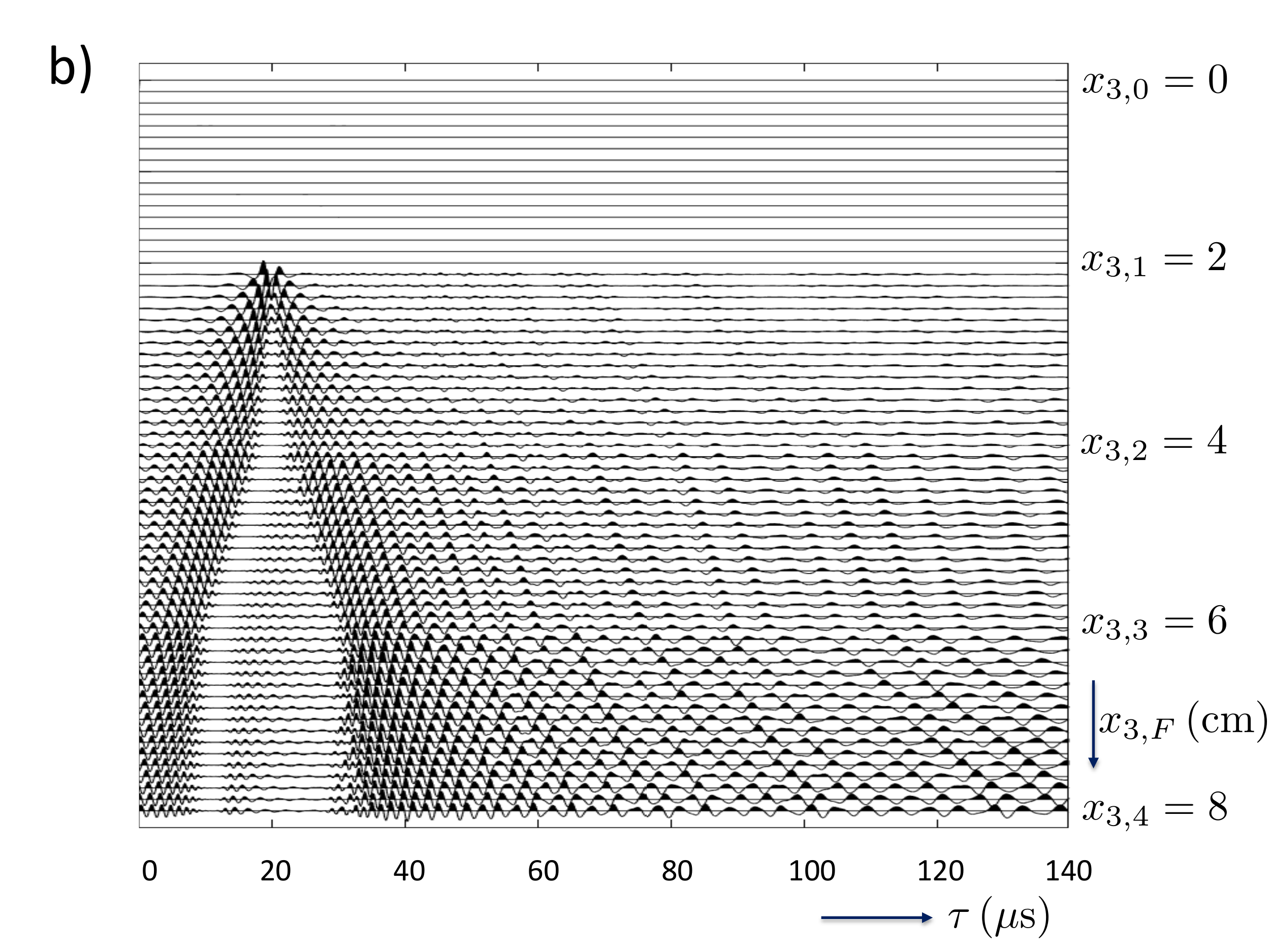}}
\centerline{\epsfxsize=7.5cm \epsfbox{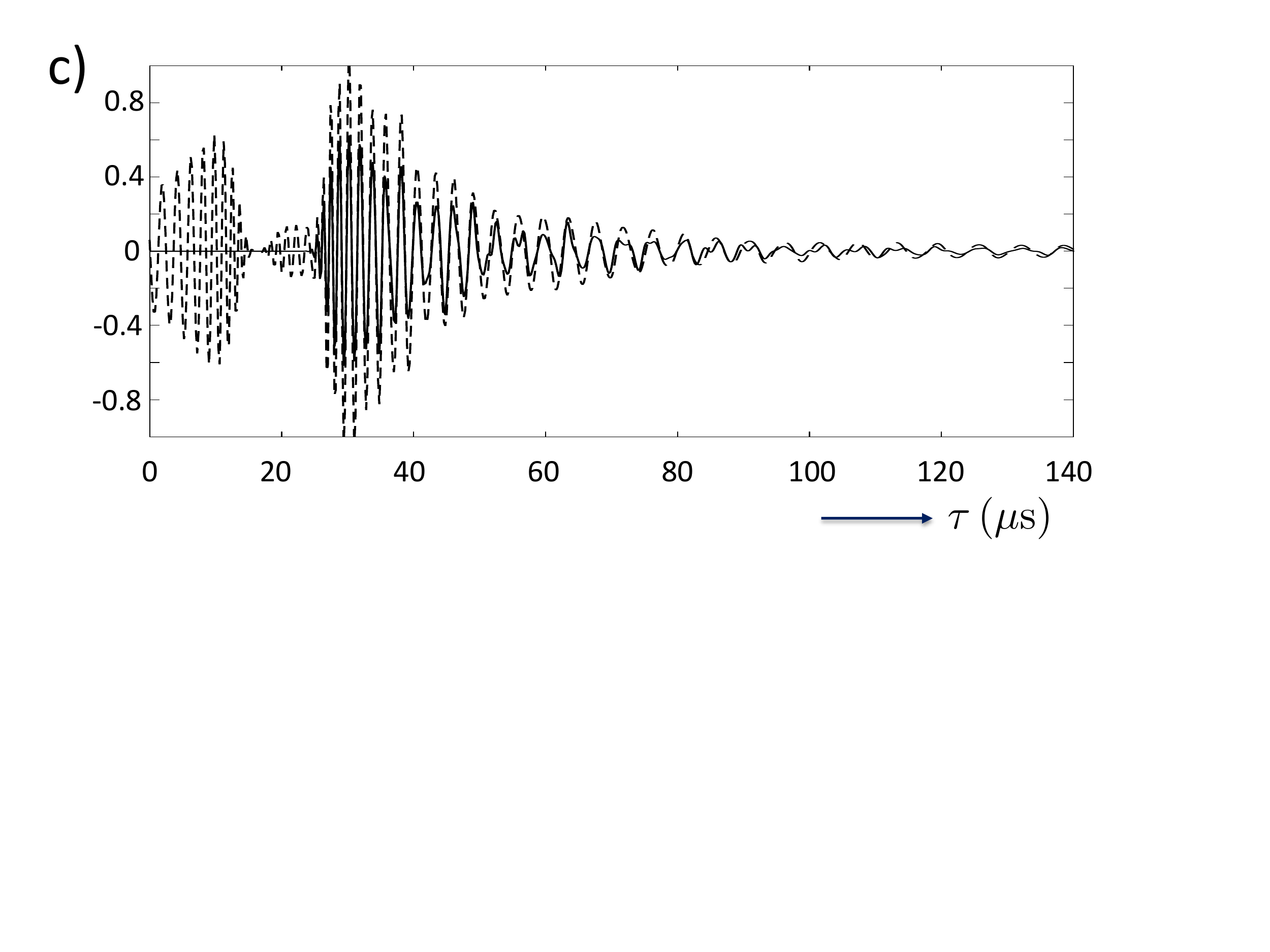}}
\vspace{-2.3cm}
\centerline{\epsfxsize=7.5cm \epsfbox{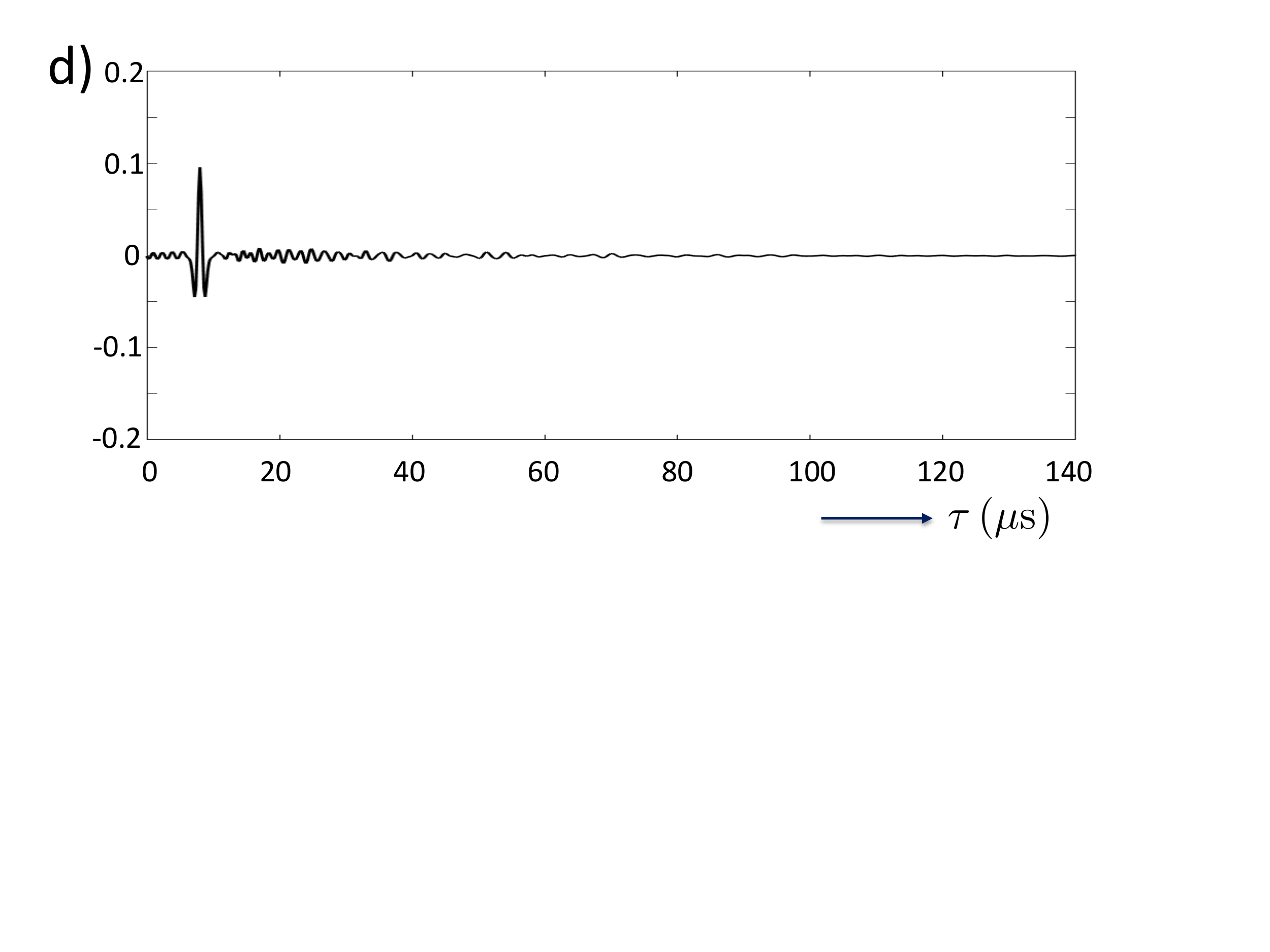}}
\vspace{-2.3cm}
\caption{As Figure \ref{Fig6}, but using  a method that handles primaries only.}\label{Figure9}
\end{figure}

\begin{figure}[t]
\centerline{\epsfxsize=7.5cm \epsfbox{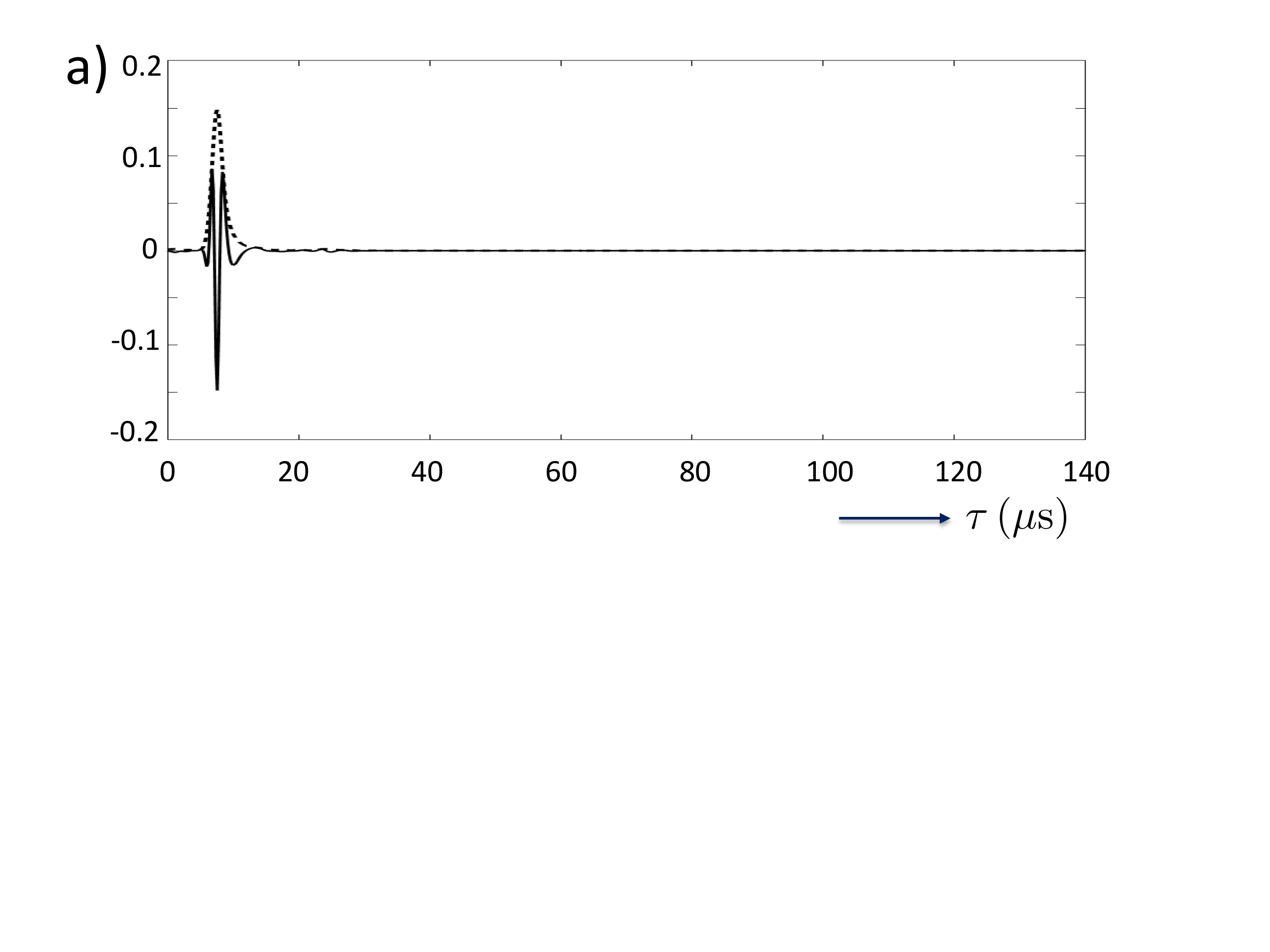}}
\vspace{-2.5cm}
\centerline{\epsfxsize=7.5cm \epsfbox{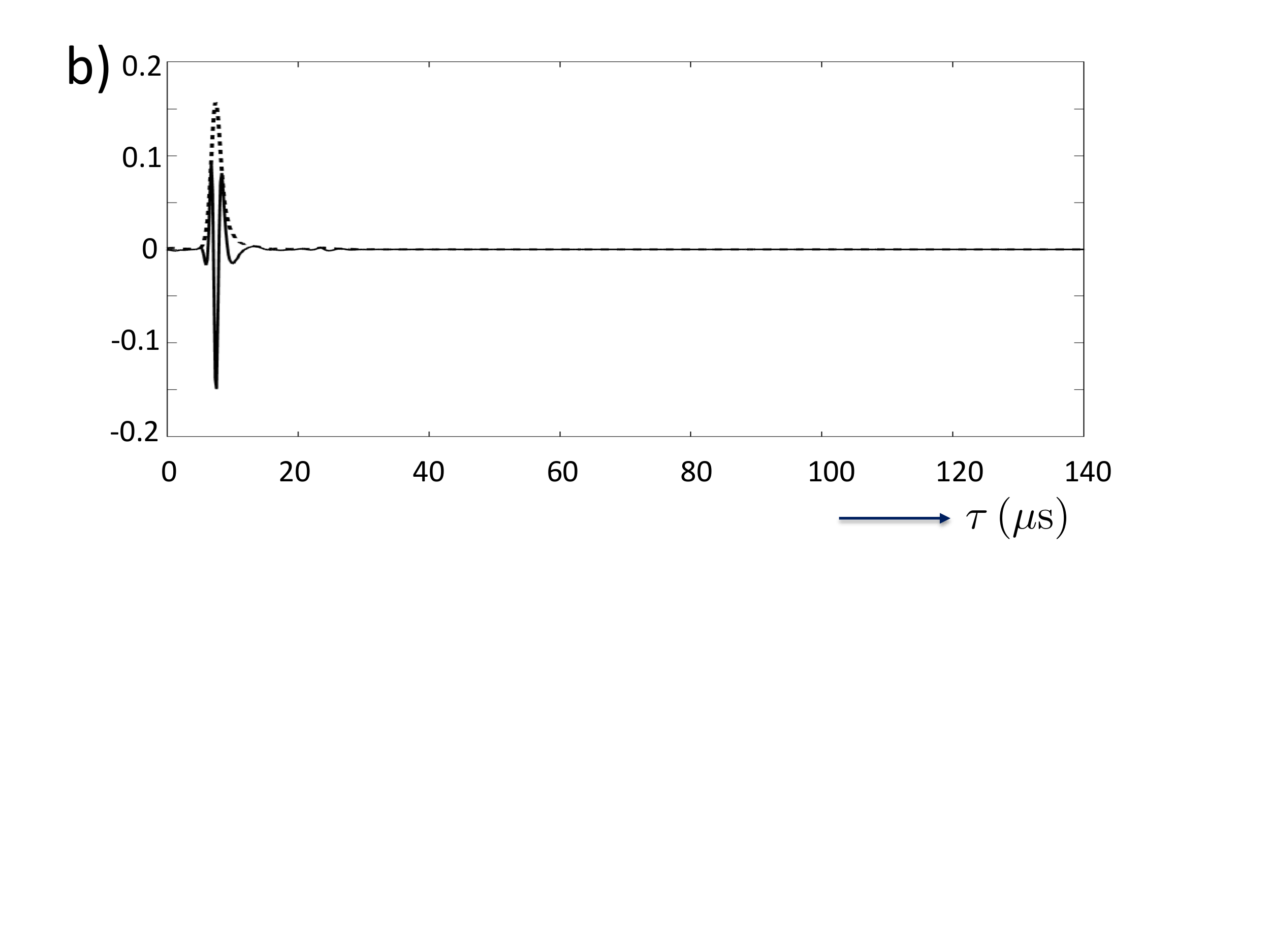}}
\vspace{-2.5cm}
\caption{(a) As Figure \ref{Fig6}({d}), but using erroneous phase velocities. (b) As Figure \ref{Fig6}({d}), but for $s_1=40\mu$s/m and using erroneous phase velocities.}\label{Fig7}
\end{figure}

To emphasize what we have achieved with the Marchenko method we repeat this numerical experiment with a method that handles primaries only 
(to this end, we use the same method as before, but apply zero iterations for each focal depth). Figure \ref{Figure9}(a) shows the 
retrieved Green's function  $G(s_1=0,x_{3,F},x_{3,0},\tau)$
and Figure \ref{Figure9}(b)  the difference with the directly modelled Green's function in Figure \ref{Fig4}(b). Note that this difference is significantly stronger than that in Figure \ref{Fig6}(b). 
{Figure \ref{Figure9}(c) shows again a comparison of the directly modelled and retrieved Green's functions at $x_{3,F}=$ 5 cm.}
Figure \ref{Figure9}({d}) shows the retrieved reflection response $R(s_1=0,x_{3,F},\tau)$ for $x_{3,F}=5$ cm, obtained by inverting equation (\ref{eq888}).
The amplitude of the retrieved reflection event at 8.0 $\mu$s is now 0.096, almost a factor two too low. 
Moreover, the events directly following this reflection event are caused by  multiple reflections in the medium above $x_{3,F}=$ 5 cm, which obviously have not been removed by this method.

In practice we do not know the exact model, so we can obtain only an estimate of the direct arrival of the focusing function and of the onset times $\tau_{\rm on}^+$ and  $\tau_{\rm on}^-$.
We apply the same Marchenko method as above (again with 5 iterations for each focal depth), 
but this time we use erroneous phase velocities $c_0$ of 975, $-2050$, 2550 and {$-2950$} m/s in the four layers (and the same numerical values for the mass density).
Because of the erroneous velocities, the traveltimes of the retrieved Green's functions are erroneous as well, but the multiple reflections are correctly handled \citep{Wapenaar2014JASA}.  
Figure \ref{Fig7}(a) shows the retrieved reflection response $R(s_1=0,x_{3,F},\tau)$ for $x_{3,F}=5$ cm.
We observe a single reflection event at 7.6 $\mu$s and hardly any remnants of multiples, which confirms that the multiple reflections occurring in the medium above $x_{3,F}$ have again been properly removed.
The phase of the reflection event is distorted, but the envelope (indicated by the dashed line) has a peak value of 0.149, which still approximates the true reflection coefficient $r_3=0.180$ reasonably well.

Finally, we repeat the numerical experiment, using the same erroneous phase velocities, 
for a dipping plane wave with a small non-zero horizontal slowness $s_1=40 \mu$s/m.  The propagation angle $\alpha$ for $\omega=\omega_c$ is related to the slowness $s_1$ and 
the phase velocities $c_0$ of the different layers via
$\alpha=\arcsin(s_1c_0)$.  For $s_1=40 \mu$s/m, the angles in the four layers of Figure \ref{Fig3} are $2.29^o$, $-4.59^o$,  $5.74^o$ and $-6.89^o$, respectively.
The retrieved reflection response $R(s_1,x_{3,F},\tau)$ for {$s_1=40 \mu$s/m and} $x_{3,F}=5$ cm is shown in Figure \ref{Fig7}(b).
The amplitude of the retrieved reflection response of the third interface is 0.157,
which is a reasonable approximation  of the true reflection coefficient $r_3=0.181$ for $\omega=\omega_c$ and $s_1=40 \mu$s/m. 

\section{Concluding remarks}\label{sec6}

We have shown that the Marchenko method, 
which retrieves the wavefield inside a medium from its reflection response at the surface, can be extended for metamaterials.
{The main modification is the use of a new window function, which better accounts for the strong dispersive behaviour of waves in metamaterials.}
The method holds, in the low frequency limit, for elastodynamic and electromagnetic waves in layered media, consisting of a mix of natural materials and metamaterials. 
Multiple scattering between the layer interfaces is properly taken into account.  We have shown with a numerical example that the method accurately retrieves
the response to a source at the surface, observed by virtual receivers inside the medium. By deconvolving the retrieved upgoing field by the retrieved downgoing field, 
we accurately obtain the reflection response between a virtual source and a virtual receiver, both inside the medium. 

The method works well for vertically propagating plane waves and for dipping plane waves with small horizontal slownesses, corresponding to propagation angles up to approximately 7$^o$.
For larger horizontal slownesses the method becomes unstable.
Due to the strong dispersive behaviour of the DNG layers, the propagation angle for a fixed horizontal slowness is frequency-dependent and can  become post-critical for high frequencies, which explains the unstable behaviour.
A possible remedy is to remove the high frequencies from the source spectrum, but this will go at the cost of resolution. Further research is needed to optimize the proposed method for a wider range of propagation angles.

{Whereas we only considered horizontally layered media, in principle the method can be extended for laterally varying metamaterials in a similar way as for 
natural materials \citep{Wapenaar2014JASA}. 
The strong dispersive character of
metamaterials will limit the maximum aperture angle of the space-time focusing operators and hence the obtainable lateral resolution. An interesting option to be investigated further
 is the virtual plane-wave Marchenko approach for laterally varying media \cite{Meles2018GJI}, modified for metamaterials.}

The proposed method can potentially be used in any application of metamaterials where knowledge of the wavefield inside the medium is required, 
 for example in non-destructive testing of layered materials, where anomalies of the retrieved reflectivity may be used to determine the  location of a delamination.

\section*{Acknowledgements}
{The constructive comments of Patrick Elison and an anonymous reviewer are highly appreciated.} 
This work has received funding from the European Union's Horizon 2020 research and innovation programme: European Research Council (grant agreement 742703).

\newpage
\centerline{{\Large References}}

\bibliography{/Users/cwapenaar/kees/lib/tex/bibliography/bibliography}

\end{document}